\font\fiverm=cmr5
\font\fivebf=cmbx5
\font\fivei=cmmi5

\font\fivesy=cmsy5

\font\sevenrm=cmr7
\font\sevenbf=cmbx7
\font\seveni=cmmi7

\font\sevensy=cmsy7

\font\ninerm=cmr9
\font\ninebf=cmbx9

\font\nineit=cmmi9
\font\ninei=cmmi9 
\font\ninesy=cmsy9  
\font\nineex=cmex10
\font\tenrm=cmr10
\font\tenbf=cmbx10
\font\tensl=cmsl10
\font\tenit=cmmi10
\font\teni=cmmi10 
 
\font\tensy=cmsy10
\font\tenex=cmex10
\font\twelverm=cmr12
\font\twelvebf=cmbx12
\font\twelvesl=cmsl12
\font\twelveit=cmmi12
\font\twelvei=cmmi12
\font\twelvesy=cmsy10 scaled\magstep1


%
%
 %
%
 \def\ninepoint{%
   \normalbaselineskip=11pt
   \def\rm{\fam0\ninerm}%
   \def\it{\fam0\nineit}%
   \def\bf{\fam\bffam\ninebf}%
   \def\bi{\fam\bffam\ninebf}%
   \def\rmit{\fam0\ninerm\def\it{\fam0\nineit}}%
   \def\bfit{\fam\bffam\ninebf\def\it{\bi}}%
   \def\bsl{\fam\bffam\ninebsl}
   \textfont0=\ninerm\scriptfont0=\sevenrm\scriptscriptfont0=\fiverm
   \textfont1=\ninei\scriptfont1=\seveni\scriptscriptfont1=\fivei
   \textfont2=\ninesy\scriptfont2=\sevensy\scriptscriptfont2=\fivesy
   \textfont3=\nineex \scriptfont3=\tenex \scriptscriptfont3=\tenex
   \textfont\bffam=\ninebf\scriptfont\bffam=\sevenbf\scriptscriptfont\bffam=
     \fivebf
   \normalbaselines\rm}%
%
 \def\tenpoint{%
   \normalbaselineskip=12pt
   \def\rm{\fam0\tenrm}%
   \def\it{\fam0\tenit}%
   \def\bf{\fam\bffam\tenbf}%
   \def\bi{\fam\bffam\tenbf}%
   \def\rmit{\fam0\tenrm\def\it{\fam0\tenit}}%
   \def\bfit{\fam\bffam\tenbf\def\it{\bi}}%
   \def\bsl{\fam\bffam\tenbsl}
   \textfont0=\tenrm\scriptfont0=\sevenrm\scriptscriptfont0=\fiverm
   \textfont1=\teni\scriptfont1=\seveni\scriptscriptfont1=\fivei
   \textfont2=\tensy\scriptfont2=\sevensy\scriptscriptfont2=\fivesy
   \textfont3=\tenex \scriptfont3=\tenex \scriptscriptfont3=\tenex
     \fivebf
   \textfont\bffam=\tenib\scriptfont\bffam=\sevenib\scriptscriptfont\bffam=
     \fiveib
   \normalbaselines\rm}%
%

%

     
\font\twelverm=cmr10 scaled 1200    \font\twelvei=cmmi10 scaled 1200
\font\twelvesy=cmsy10 scaled 1200   \font\twelveex=cmex10 scaled 1200
\font\twelvebf=cmbx10 scaled 1200   \font\twelvesl=cmsl10 scaled 1200
\font\twelvett=cmtt10 scaled 1200   \font\twelveit=cmti10 scaled 1200
     
\skewchar\twelvei='177   \skewchar\twelvesy='60
     
     
\def\twelvepoint{\normalbaselineskip=12.4pt
  \abovedisplayskip 12.4pt plus 3pt minus 9pt
  \belowdisplayskip 12.4pt plus 3pt minus 9pt
  \abovedisplayshortskip 0pt plus 3pt
  \belowdisplayshortskip 7.2pt plus 3pt minus 4pt
  \smallskipamount=3.6pt plus1.2pt minus1.2pt
  \medskipamount=7.2pt plus2.4pt minus2.4pt
  \bigskipamount=14.4pt plus4.8pt minus4.8pt
  \def\rm{\fam0\twelverm}          \def\it{\fam\itfam\twelveit}%
  \def\sl{\fam\slfam\twelvesl}     \def\bf{\fam\bffam\twelvebf}%
  \def\mit{\fam 1}                 \def\cal{\fam 2}%
  \def\tt{\twelvett}
  \textfont0=\twelverm   \scriptfont0=\tenrm   \scriptscriptfont0=\sevenrm
  \textfont1=\twelvei    \scriptfont1=\teni    \scriptscriptfont1=\seveni
  \textfont2=\twelvesy   \scriptfont2=\tensy   \scriptscriptfont2=\sevensy
  \textfont3=\twelveex   \scriptfont3=\twelveex  \scriptscriptfont3=\twelveex
  \textfont\itfam=\twelveit
  \textfont\slfam=\twelvesl
  \textfont\bffam=\twelvebf \scriptfont\bffam=\tenbf
  \scriptscriptfont\bffam=\sevenbf
  \normalbaselines\rm}
     
     
\def\tenpoint{\normalbaselineskip=12pt
  \abovedisplayskip 12pt plus 3pt minus 9pt
  \belowdisplayskip 12pt plus 3pt minus 9pt
  \abovedisplayshortskip 0pt plus 3pt
  \belowdisplayshortskip 7pt plus 3pt minus 4pt
  \smallskipamount=3pt plus1pt minus1pt
  \medskipamount=6pt plus2pt minus2pt
  \bigskipamount=12pt plus4pt minus4pt
  \def\rm{\fam0\tenrm}          \def\it{\fam\itfam\tenit}%
  \def\sl{\fam\slfam\tensl}     \def\bf{\fam\bffam\tenbf}%
  \def\smc{\tensmc}             \def\mit{\fam 1}%
  \def\cal{\fam 2}%
  \textfont0=\tenrm   \scriptfont0=\sevenrm   \scriptscriptfont0=\fiverm
  \textfont1=\teni    \scriptfont1=\seveni    \scriptscriptfont1=\fivei
  \textfont2=\tensy   \scriptfont2=\sevensy   \scriptscriptfont2=\fivesy
  \textfont3=\tenex   \scriptfont3=\tenex     \scriptscriptfont3=\tenex
  \textfont\itfam=\tenit
  \textfont\slfam=\tensl
  \textfont\bffam=\tenbf \scriptfont\bffam=\sevenbf
  \scriptscriptfont\bffam=\fivebf
  \normalbaselines\rm}
     

{\obeylines\gdef\
{}}
\def\singlespace{\baselineskip=\normalbaselineskip}

\def\doublespace{\baselineskip=\normalbaselineskip \multiply\baselineskip by 2}

\newcount\firstpageno
\firstpageno=2
\footline={\ifnum\pageno<\firstpageno{\hfil}\else{\hfil\twelverm\folio\hfil}\fi}
\let\rawfootnote=\footnote              
\def\footnote#1#2{{\rm\singlespace\parindent=0pt\rawfootnote{#1}{#2}}}

     
\hsize=6.5truein
\hoffset=0truein
\vsize=8.9truein
\voffset=0truein
\parskip=\medskipamount
\twelvepoint            
\doublespace            
\overfullrule=0pt       
     
     
\def\preprintno#1{
 \rightline{\rm #1}}    
     
     
\def\ref#1{Ref. #1}                     

\def\frac#1#2{{\textstyle{#1 \over #2}}}
\def\half{{\textstyle{ 1\over 2}}}

\def\sla{\raise.15ex\hbox{$/$}\kern-.57em}
\def\leaderfill{\leaders\hbox to 1em{\hss.\hss}\hfill}
\def\twiddle{\lower.9ex\rlap{$\kern-.1em\scriptstyle\sim$}}
\def\bigtwiddle{\lower1.ex\rlap{$\sim$}}
\def\gtwid{\mathrel{\raise.3ex\hbox{$>$\kern-.75em\lower1ex\hbox{$\sim$}}}}
\def\ltwid{\mathrel{\raise.3ex\hbox{$<$\kern-.75em\lower1ex\hbox{$\sim$}}}}
\def\square{\kern1pt\vbox{\hrule height 1.2pt\hbox{\vrule width 1.2pt\hskip 3pt
   \vbox{\vskip 6pt}\hskip 3pt\vrule width 0.6pt}\hrule height 0.6pt}\kern1pt}

\def\tablerule{\tablespace\noalign{\hrule}\tablespace}

\def\m@th{\mathsurround=0pt }
\def\leftrightarrowfill{$\m@th \mathord\leftarrow \mkern-6mu
 \cleaders\hbox{$\mkern-2mu \mathord- \mkern-2mu$}\hfill
 \mkern-6mu \mathord\rightarrow$}
\def\overleftrightarrow#1{\vbox{\ialign{##\crcr
     \leftrightarrowfill\crcr\noalign{\kern-1pt\nointerlineskip}
     $\hfil\displaystyle{#1}\hfil$\crcr}}}

\input psfig.sty
\singlespace
\preprintno{hep-ph/9602316}
\preprintno{CPTH-S422.1295}
\preprintno{CRETE-96-13}
\preprintno{UFIFT-HEP-96-5}
\preprintno{Revised June 1996}
\vskip 2cm
\centerline{\bf THE QUANTUM GRAVITATIONAL BACK-REACTION ON INFLATION}
\vskip 2cm
\centerline{\bf N. C. Tsamis$^{*}$}
\vskip .5cm
\centerline{\it Centre de Physique Th\'eorique, Ecole Polytechnique}
\centerline{\it Palaiseau 91128, FRANCE}
\vskip .5cm
\centerline{and}
\vskip .5cm
\centerline{\it Theory Group, FO.R.T.H.}
\centerline{\it Heraklion, Crete 71110, GREECE}
\vskip 1cm
\centerline{and}
\vskip 1cm
\centerline{\bf R. P. Woodard$^{\dagger}$}
\vskip .5cm
\centerline{\it Department of Physics, University of Florida}
\centerline{\it Gainesville, FL 32611, USA}
\vskip 2cm
\centerline{ABSTRACT}
\itemitem{}{\tenpoint We describe our recent calculation of two-loop 
corrections to the expansion rate of an initially inflating universe on the
manifold $T^3 \times \Re$. If correct, our result proves that quantum 
gravitational effects slow the rate of inflation by an amount which becomes 
non-perturbatively large at late times. In a preliminary discussion of basic
issues we show that the expansion rate is a gauge invariant, and that our
ultraviolet regulator does not introduce spurious time dependence. We also 
derive a sharp bound on the maximum strength of higher loop effects.}
\footnote{}{$^*$~~ \tenpoint e-mail: tsamis@iesl.forth.gr and 
tsamis@orphee.polytechnique.fr}
\footnote{}{$^{\dagger}$~~ \tenpoint e-mail: woodard@phys.ufl.edu}

\vfill\eject

\doublespace

\centerline{\bf {1. Introduction}}   

The purpose of this paper is to discuss the consistency, methodology and
accuracy of a very long calculation in perturbative quantum gravity. Although
such an exposition is necessary if the result is to win acceptance, we should 
best begin by explaining the motivation. This labor was undertaken to check the
suggestion [1,2] that corrections from the infrared of quantum gravity may 
extinguish inflation without the need for a special inflaton field. It will be 
seen that this is also a proposal for solving the problem of the cosmological 
constant. In our scheme the cosmological constant is not unnaturally small, it 
only appears so today on account of screening by infrared effects in quantum 
gravity. Inflation begins in Hubble-sized patchs of the early universe over
which the local temperature has fallen enough for the cosmological constant to
dominate the stress-energy. After a few e-foldings the temperature of such a
patch is sufficiently low to permit long range correlations, and the infrared 
screening effect begins to build up. This build-up requires a long time because
gravitational interactions are naturally weak; they can only become significant
through causal and coherent superposition over the past lightcone. Effective 
screening is therefore delayed until an enormous invariant volume has 
developed within the past lightcone of the observation point to the onset 
of inflation. This is why inflation lasts long enough to explain the 
homogeneity and isotropy of the observed universe.

Note that no fine tuning is necessary for our scheme, nor do we require new
matter fields or even new gravitational interactions. Of the phenomenologically 
viable quanta, our mechanism is unique to gravitons. The other known particles 
are either massive --- which precludes coherent superposition --- or else they 
possess conformal invariance on the classical level --- which means they cannot
exploit the enormous invariant volume in the past lightcone of a conformally 
flat, inflating universe.

Note also that causality, and the physically motivated initial condition of 
coherent inflation over a finite spatial region, preclude sensitivity to global 
issues such as whether or not the full de Sitter manifold is used. For
convenience we worked on the manifold $T^3 \times \Re$. What we computed is the
expectation value of the invariant element in the presence of a homogeneous and
isotropic state which is initially free de Sitter vacuum:
$$\Bigl\langle 0 \Bigl\vert \; g_{\mu \nu}(t,{\vec x}) \;
dx^{\mu} dx^{\nu} \; \Bigr\vert 0 \Bigr\rangle = 
-dt^2 + {\rm a}^2(t) \; d{\vec x} \cdot d{\vec x} 
\;\; . \eqno(1.1)$$
Inflation redshifts the temperature to zero so rapidly that we simply used zero
temperature quantum field theory. We were able to use the Lagrangian of
conventional general relativity:
$${\cal L} = {1 \over 16 \pi G} \Bigl(R - 2 \Lambda\Bigr) \; 
\sqrt{-g} \;\; , \eqno(1.2)$$
because our infrared mechanism is insensitive to the still unknown ultraviolet
sector of quantum gravity. We assumed only that the scale of inflation $M \sim 
(\Lambda / G)^{1/4}$ is at least a few orders of magnitude below the Planck 
mass $M_{\rm Pl} \sim G^{-(1/2)}$:
$$M \ltwid 10^{-3} \; M_{\rm Pl} \;\; . \eqno(1.3)$$ 
In this case ultraviolet modes should have plenty of time to reach their 
natural equilibria as they redshift down to scales at which the dynamics is 
described by quantum general relativity. Our mechanism can be shown to derive
from modes whose physical wavelength is about the Hubble radius at the time
that they contribute most strongly. It is also worth pointing out that our
mechanism is an inherently quantum mechanical effect, deriving ultimately from
the gravitational interaction between the zero point motions of the various 
modes. It in no way conflicts with the classical and semi-classical stability 
of locally de Sitter backgrounds [3].

We inferred the physical rate of expansion from the effective Hubble parameter:
$$H_{\rm eff}(t) \equiv {d \over dt} \; \ln ({\rm a}) 
\;\; . \eqno(1.4)$$
The first secular effect occurs at two loops and has the form:
$$H_{\rm eff}(t) = H \Biggl\{ 1 - \Bigl( {\kappa H \over 4 \pi} \Bigr)^4 \; 
\Bigl[ \; \frac16 \; (Ht)^2 + ({\rm subdominant}) \Bigr] \;
+ \; O(\kappa^6) \Biggr\} \;\; , \eqno(1.5)$$
where $\kappa^2 \equiv 16 \pi G$ is the usual loop counting parameter of 
quantum gravity and $3 H^2 \equiv \Lambda$ is the bare Hubble constant. We were
also able to show that the dominant contribution at $\ell$ loops has the form:
$$- \# \; (\kappa H)^{2\ell} \; (Ht)^{\ell} \;\; . \eqno(1.6)$$
This permits the following estimate of the number of e-foldings needed for the
extinction of inflation:
$$N \equiv Ht \sim (\kappa H)^{-2} \gtwid 10^{12} \;\; . \eqno(1.7)$$
Of course perturbation theory is no longer reliable when quantum corrections
become of order one so the valid conclusion is that quantum gravitational 
effects slow the rate of inflation by an amount which becomes 
non-perturbatively large at late times.

This completes our discussion of why we undertook the project and what we got.
A more complete discussion of the physical consequences can be found elsewhere
[4]. The remainder of this paper is devoted to an explanation of how we 
obtained the result and why we believe it is accurate. Section 2 deals with 
basic issues of the formalism. Crucial points here are the distinction between 
``in''-``out'' matrix elements and expectation values, the physical significance
of our ultraviolet regularization, the demonstration that $H_{\rm eff}(t)$ is 
independent of the choice of gauge, the reason why infrared logarithms are 
almost inevitable, and a derivation of our bound (1.6) for the dominant 
corrections at $\ell$ loops.  Sections 3 and 4 deal with our specific process. 
Section 3 is devoted to the class of diagrams which involve only 3-point 
vertices, while Section 4 describes the dominant diagram which has a 4-point 
vertex and a 3-point vertex. Section 5 assembles the final result and
discusses the many accuracy checks. We close the section, and the paper, 
with a brief consideration of the issues pertaining to the case of negative 
$\Lambda$.

\vfill\eject

\centerline{\bf {2. Basic Issues}} 

\noindent {\it 2.1 The Apparatus of Perturbation Theory.}

We take the onset of inflation to be $t=0$, and we work perturbatively 
around the classical background:
$${\rm a}_{\rm class}(t) = \exp(Ht) \;\; . \eqno(2.1)$$
It is simplest to perform the calculation in conformally flat coordinates, 
for which the invariant element of the background is:
$$-dt^2 + {\rm a}^2_{\rm class}(t) \; d{\vec x} \cdot d{\vec x} = 
\Omega^2 (-du^2 + d{\vec x} \cdot d{\vec x}) \;\; , \eqno(2.2a)$$
$$\Omega \equiv (Hu)^{-1} = \exp(Ht) \;\; . \eqno(2.2b)$$
Note the temporal inversion and the fact that the onset of inflation at 
$t=0$ corresponds to $u = H^{-1}$. Since $t \rightarrow \infty$ corresponds
to $u \rightarrow 0^+$, and since the spatial coordinates of $T^3$ fall 
within the region, $-\frac12 H^{-1} < x^i \leq \frac12 H^{-1}$, the range of 
conformal coordinates is rather small. This is why a conformally invariant 
field --- whose dynamics are locally the same as in flat space, except for 
ultraviolet regularization --- cannot induce a big infrared effect.

Perturbation theory is organized most conveniently in terms of a 
``pseudo-graviton'' field, $\psi_{\mu \nu}$, obtained by conformally 
re-scaling the metric:
$$g_{\mu \nu} \equiv \Omega^2 \; {\widetilde g}_{\mu \nu} \equiv 
\Omega^2 \; (\eta_{\mu \nu} + \kappa \psi_{\mu \nu}) 
\;\; . \eqno(2.3)$$
As usual, pseudo-graviton indices are raised and lowered with the Lorentz
metric, and the loop counting parameter is $\kappa^2 \equiv 16 \pi G$. 
After some judicious partial integrations the invariant part of the bare 
Lagrangian takes the following form [5]:
$$\eqalignno{{\cal L}_{\rm inv} =
\sqrt{-{\widetilde g}} \; &{\widetilde g}^{\alpha \beta} \;
{\widetilde g}^{\rho \sigma} \; {\widetilde g}^{\mu \nu} \; 
\Omega^{2} \; \Bigl[ 
\half \psi_{\alpha \rho , \mu} \; \psi_{\nu \sigma , \beta} - 
\half \psi_{\alpha \beta , \rho} \; \psi_{\sigma \mu , \nu} + 
\frac14 \psi_{\alpha \beta , \rho} \; \psi_{\mu \nu , \sigma} \cr 
&- \frac14 \psi_{\alpha \rho , \mu} \; \psi_{\beta \sigma , \nu}
\Bigr] - \half \sqrt{-{\widetilde g}} \; 
{\widetilde g}^{\rho \sigma} \; {\widetilde g}^{\mu \nu} \; 
(\Omega^{2})_{,\alpha} \; 
\psi_{\rho \sigma , \mu} \; \psi_{\nu}^{~\alpha} 
\;\; . &(2.4) \cr}$$
Gauge fixing is accomplished through the addition of $-\half \eta^{\mu \nu} 
F_{\mu} F_{\nu}$, where [5]:
$$F_{\mu} \equiv \Omega \; \Bigl( \psi^{\alpha}_{~\mu , \alpha} - \frac12 
\psi^{\alpha}_{~\alpha , \mu} + 2 \; \psi^{\alpha}_{~\mu} \; 
(\ln \Omega)_{,\alpha} \Bigr) \;\; . \eqno(2.5)$$
The associated ghost Lagrangian is [5]:
$$\eqalignno{{\cal L}_{\rm ghost} = -\Omega^2 \; 
&{\overline \omega}^{\mu , \nu} \; \Bigl[{\widetilde g}_{\rho \mu} \; 
\partial_{\nu} + {\widetilde g}_{\rho \nu} \; \partial_{\mu} + 
{\widetilde g}_{\mu \nu , \rho} + 2 {\widetilde g}_{\mu \nu} \; 
{(\ln \Omega)}_{, \rho} \Bigr] \; \omega^{\rho} \cr 
&+ {\Bigl( \Omega^2 \; {\overline \omega}^{\mu} \Bigr)}_{, \mu} 
\eta^{\rho \sigma} \; \Bigl[{\widetilde g}_{\nu \rho} \; 
\partial_{\sigma} + \frac12 {\widetilde g}_{\rho \sigma , \nu} + 
{\widetilde g}_{\rho \sigma} \; {(\ln \Omega)}_{, \nu} \Bigr] \; 
\omega^{\nu} \;\; . &(2.6) \cr}$$

In our gauge the pseudo-graviton kinetic operator has the form:
$$D_{\mu \nu}^{~~ \rho \sigma} \equiv \Bigl[
\frac12 {\overline \delta}_{\mu}^{~( \rho} \; 
{\overline \delta}_{\nu}^{~\sigma)} 
- \frac14 \eta_{\mu \nu} \; \eta^{\rho\sigma} 
- \frac12 t_{\mu} t_{\nu} t^{\rho} t^{\sigma} \Bigr] {\rm D}_A 
- t_{(\mu} {\overline \delta}_ {\nu)}^{~~(\rho} \; t^{\sigma)} \; {\rm D}_B 
+ t_{\mu} t_{\nu} t^{\rho} t^{\sigma} \; {\rm D}_C 
\;\; , \eqno(2.7)$$
where parenthesized indices are symmetrized. Two notational conventions reflect
the fact that the $0$ direction is special. First, we define $t_{\mu}$ as:
$$t_{\mu} \equiv \eta_{\mu 0} = - \delta_{\mu}^{~0} \;\; . \eqno(2.8a)$$
(Recall from (1.1) that our metric is spacelike.) Second, we define barred 
tensors to have their natural zero components nulled, for example:
$${\overline \delta}_{\nu}^{~\mu} \equiv \delta_{\nu}^{~\mu} - \delta_0^{~\mu}
\delta_{\nu}^{~0} = \delta_{\nu}^{~\mu} + t_{\nu} \; t^{\mu} 
\;\; . \eqno(2.8b)$$
Note also that ${\rm D}_A \equiv \Omega(\partial^2 + \frac2{u^2}) \Omega$ is 
the kinetic operator for a massless, minimally coupled scalar and ${\rm D}_B =
{\rm D}_C \equiv \Omega \> \partial^2 \Omega$ is the kinetic operator for a
conformally coupled scalar. 

The zeroth order action results in the following free field expansion [6]:
$$\psi_{\mu \nu}(u,{\vec x}) = 
\Biggl({{\rm Zero} \atop {\rm Modes}}\Biggr) + 
H^3 \sum_{\lambda, {\vec k}\neq 0} \Biggl\{ \Psi_{\mu \nu}\Bigl(u,{\vec x};
{\vec k},\lambda\Bigr) \; a({\vec k},\lambda) + \Psi_{\mu \nu}^*
\Bigl(u,{\vec x};{\vec k},\lambda\Bigr) \; a^{\dagger}({\vec k},\lambda) 
\Biggr\} \;\; . \eqno(2.9)$$
The spatial polarizations consist of ``A'' modes:
$$\Psi_{\mu \nu}\Bigl(u,{\vec x};{\vec k},\lambda\Bigr) = 
{Hu \over \sqrt{2 k}} \; \Bigl(1 + {i \over k u}\Bigr) \; 
\exp\Bigl[i k \Bigl(u - \frac1{H}\Bigr) + 
i {\vec k} \cdot {\vec x}\Bigr] \; \epsilon_{\mu \nu}({\vec k},\lambda) 
\qquad ; \qquad \forall \lambda \in A \eqno(2.10a)$$
while the space--time and purely temporal polarizations are associated, 
respectively, with ``B'' and ``C'' modes:
$$\Psi_{\mu \nu}\Bigl(u,{\vec x};{\vec k},\lambda\Bigr) = 
{Hu \over \sqrt{2 k}} \; \exp\Bigl[i k \Bigl(u - \frac1{H}\Bigr) + 
i {\vec k} \cdot {\vec x} \Bigr] \; \epsilon_{\mu \nu}({\vec k},\lambda) 
\qquad ; \qquad \forall \lambda \in B,C \eqno(2.10b)$$
In LSZ reduction one would integrate against and contract into $\Psi_{\mu 
\nu}(u,{\vec x};{\vec k},\lambda)$ to insert and ``in''-coming graviton of 
momentum ${\vec k}$ and polarization $\lambda$; the conjugate would be used 
to extract an ``out''-going graviton with the same quantum numbers. The 
zero modes evolve as free particles with time dependences $1$ and $u^3$ 
for the A modes, and $u$ and $u^2$ for the B and C modes. Since causality 
decouples the zero modes shortly after the onset of inflation, they play 
no role in screening and we shall not trouble with them further.

We define $\vert 0 \rangle$ as the Heisenberg state annihilated by 
$a({\vec k}, \lambda)$ --- and the analogous ghost operators --- at 
the onset of inflation. We can use this condition and expansion (2.9)
to express the free pseudo-graviton propagator as a mode sum [7]:
$$\eqalignno{i\Bigl[ {_{\mu \nu}}\Delta_{\rho \sigma}\Bigr](x;x') &\equiv
\Bigl\langle 0 \Bigl\vert T\Bigl\{\psi_{\mu \nu}(x) \; 
\psi_{\rho \sigma}(x')\Bigr\} \Bigr\vert 0 \Bigr\rangle_{\rm free} 
\qquad &(2.11a) \cr 
&= H^3 \sum_{\lambda, {\vec k} \neq 0} \Biggl\{ \theta(u'-u) \; \Psi_{\mu \nu}
\; {\Psi'}^*_{\rho \sigma} + \theta(u-u') \; \Psi^*_{\mu \nu} \; {\Psi'}_{\rho 
\sigma} \Biggr\} e^{- \epsilon \Vert {\vec k} \Vert} 
\;\; . \qquad \qquad &(2.11b) \cr}$$
Note that the convergence factor $e^{- \epsilon \Vert {\vec k} \Vert}$ serves 
as an ultraviolet mode cutoff. Although the resulting regularization is very 
convenient for this calculation, its failure to respect general coordinate 
invariance necessitates the use of non-invariant counterterms. These are 
analogous to the photon mass which must be added to QED when using a momentum 
cutoff. Just as in QED, these non-invariant counterterms do not affect long 
distance phenomena.

Because the propagator is only needed for small conformal coordinate 
separations, ${\Delta x} \equiv \Vert {\vec x}\ ' - {\vec x} \Vert$ and 
${\Delta u} \equiv u' - u$, the sum over momenta is well approximated as 
an integral whose lower limit is the momentum $k = H$ of the longest 
wavelength. When this is done the pseudo-graviton propagator becomes [6]: 
$$\eqalignno{
\int_H &{d^3k \over (2 \pi)^3} \Biggl\{ {H^2 u u' \over 2 k} 
\exp\Bigl[-i k \vert {\Delta u} \vert + i {\vec k} \cdot ({\vec x}\ ' 
- {\vec x}) - \epsilon k \Bigr] \; 
\Bigl[{2 \delta_{\mu}^{~ (\rho} \; \delta_{\nu}^{~ \sigma)}} - 
\eta_{\mu \nu} \; \eta^{\rho \sigma} \Bigr] \cr
&+ {H^2 (1 + i k \vert {\Delta u} \vert) \over 2 k^3} 
\exp\Bigl[-i k \vert {\Delta u} \vert + i {\vec k} \cdot ({\vec x}\ ' 
- {\vec x}) - \epsilon k \Bigr] \; 
\Bigl[ 2{\overline \delta_{\mu}^{~(\rho}} \; 
{\overline \delta_{\nu}^{~ \sigma)}} - 
2 {\overline \eta_{\mu \nu}} \; 
{\overline \eta^{\rho \sigma}} \Bigr] \; \Biggr\} 
\qquad\qquad &(2.12a) \cr}$$
so that:
$$\eqalignno{
i \Bigl[{_{\mu \nu}} \Delta^{\rho \sigma}\Bigr](x;x') \approx 
&{H^2 \over 8 {\pi}^2} \; \Biggl\{ 
{2u'u \over {\Delta x}^2 - {\Delta u}^2 + 
2 i \epsilon \vert {\Delta u} \vert + \epsilon^2} \;
\Bigl[{2 \delta_{\mu}^{~ (\rho} \; \delta_{\nu}^{~ \sigma)}} - 
\eta_{\mu \nu} \; \eta^{\rho \sigma} \Bigr] \cr
&- \ln \Bigl[ H^2 \Bigl(
{\Delta x}^2 - {\Delta u}^2 + 2 i \epsilon \vert {\Delta u} \vert + 
\epsilon^2\Bigr) \Bigr] \;
\Bigl[ 2{\overline \delta_{\mu}^{~(\rho}} \;
{\overline \delta_{\nu}^{~ \sigma)}} - 
2 {\overline \eta_{\mu \nu}} \;
{\overline \eta^{\rho \sigma}} \Bigr] \; \Biggr\} 
\qquad\qquad &(2.12b) \cr}$$
The same approximation gives the following result for the ghost propagator:
$$\eqalignno{i \Bigl[{_{\mu}} \Delta_{\nu}\Bigr](x;x') \approx 
\; {H^2 \over 8 {\pi}^2} \; \Biggl\{ &{2u'u \over {\Delta x}^2 - 
{\Delta u}^2 + 2 i \epsilon \vert {\Delta u} \vert + \epsilon^2} \;
\eta_{\mu\nu}\cr 
&- \ln\Bigl[H^2 \Bigl({\Delta x}^2 - {\Delta u}^2 + 2 i \epsilon \vert 
{\Delta u} \vert + \epsilon^2\Bigr)\Bigr] \; {\overline \eta_{\mu \nu}} 
\Biggr\} \;\; . \qquad &(2.13) \cr}$$
The decoupling between tensor indices and the functional dependence upon 
spacetime --- and the simplicity of each --- greatly facilitates calculations.
It is convenient to identify as the ``normal'' and ``logarithmic'' propagator 
functions as $i \Delta_{N}$ and $i \Delta_{L}$ respectively:
$$i \Delta_{N}(x,x') \equiv {H^2 \over 8 \pi^2} \; {2 u u' \over {\Delta x}^2
- {\Delta u}^2 + 2 i \epsilon \vert {\Delta u}\vert + \epsilon^2} \;\; , 
\eqno(2.14a)$$
$$i \Delta_{L}(x,x') \equiv {H^2 \over 8 \pi^2} \; \ln\Bigl[ H^2 \Bigl({\Delta
u}^2 - {\Delta u}^2 + 2 i \epsilon \vert {\Delta u}\vert + \epsilon^2 \Bigl)
\Bigr] \;\; . \eqno(2.14b)$$
In this notation we can write the pseudo-graviton and ghost propagators as:
$$\eqalignno{ i \Bigl[{_{\mu \nu}} \Delta^{\rho \sigma}\Bigr](x;x') = &
i \Delta_N(x;x') \; \Bigl[{2 \delta_{\mu}^{~ (\rho} \; \delta_{\nu}^{~ \sigma)}}- \eta_{\mu \nu} \; \eta^{\rho \sigma} \Bigr] \cr
& - i \Delta_L(x;x') \; \Bigl[ 2{\overline \delta_{ \mu}^{~(\rho}} \; 
{\overline \delta_{\nu}^{~ \sigma)}} - 2 {\overline \eta_{ \mu \nu}} \; 
{\overline \eta^{\rho \sigma}} \Bigr] \;\; , &(2.15a) \cr}$$
$$i \Bigl[{_{\mu}} \Delta_{\nu}\Bigr](x;x') = \; i \Delta_{N}(x,x') \; 
\eta_{\mu\nu} - i \Delta_L(x;x') \; {\overline \eta_{\mu \nu}} 
\;\; . \eqno(2.15b)$$

\vfill\eject

\noindent {\it 2.2 ``In''-``Out'' Matrix Elements and the S-Matrix.}

Perturbative quantum field theory is usually formulated to give ``in''-``out'' 
amplitudes and S-matrix elements. These quantities are not well suited for our
study because they require specification of the vacuum at asymptotically early 
and late times, which is precisely what we wish to determine. However, they can
at least be used to negate the hypothesis that the vacuum of an initially
inflating universe suffers only perturbatively small corrections. The procedure
is simply to assume the ``in'' and ``out'' vacua are both free de Sitter, and 
then do the computation. If the hypothesis is correct, the result should be 
free of infrared divergences.

What one actually finds is that both ``in''-``out'' amplitudes [1,2] and 
S-matrix elements [7] are infrared divergent. Since even 3-particle tree 
amplitudes are affected, there is no possibility for solving the problem by 
summing degenerate ensembles. There is simply nothing of lower order in 
perturbation theory that could cancel the problem. 

Why this happens can be readily understood from the previous section. S-matrix
elements consist of wavefunctions integrated against interaction vertices, 
which are linked by propagators. From (2.10a) we see that physical 
wavefunctions become constant at late times; while (2.14-15) shows that 
non-coincident propagators remain of order one as $u \rightarrow 0$.
But (2.4) reveals that vertices blow up at late times. A more physical way of 
understanding the phenomenon is that although the {\it coordinate} momentum of 
a graviton is unchanged by time evolution, its {\it physical} momentum is
redshifted to zero. Since all gravitons of fixed coordinate momentum approach 
the same physical momentum, their interaction becomes infinitely strong at late 
times. 

The correct interpretation of these infrared divergences is that the ``in''
vacuum is infinitely far from the ``out'' vacuum. Stated differently, the 
vacuum of an initially inflating universe suffers non-perturbatively large 
corrections at late times. However, it does not follow that these corrections 
must depend upon time in the same way that the ``in''-``out'' divergences 
depend upon some arbitrarily chosen infrared cutoff. To obtain a quantitative 
result one must actually follow the evolution.

\noindent {\it 2.3 How to Compute Expectation Values.}

The perturbative rules for calculating expectation values of operators
in field theory were developed by Schwinger [8] and have been adapted
to our particular problem and initial conditions in [2]. They are 
quite similar to the usual rules for computing ``in''-``out'' matrix 
elements since the propagators and vertices utilized are simple 
variations of the usual ones. The main idea is to evolve forward 
from the initial state with the action functionally integrated over
the dummy field $\psi_{\mu \nu}^{(+)}$ and, then, to evolve back to 
the initial state using the conjugate action functionally integrated
over the dummy field $\psi_{\mu \nu}^{(-)}$.

The Feynman rules are simple. An external line may be chosen as either ``$+$'' 
or ``$-$'' but one does not sum over both possibilities. Vertices have either 
all ``$+$'' or all ``$-$'' lines, and one sums the two possibilities for each 
vertex. The ``$+$'' vertices are identical to those of the ``in''-``out'' 
formalism, while the ``$-$'' vertices are complex conjugated. Propagators can 
link fields of either sign. All four possibilities can be obtained from 
(2.12a-b) by effecting the following substitutions for the quantity $({\Delta 
x}^2 - {\Delta u} + 2 i \epsilon \vert {\Delta u} \vert + \epsilon^2)$ in 
(2.14):
$$\eqalignno{ &(+ \;\; +) \qquad \Longrightarrow \qquad 
(\Delta x - \vert \Delta u \vert + i\epsilon) \; 
(\Delta x + \vert \Delta u \vert - i\epsilon) \;\; , &(2.16a)\cr
&(- \;\; +) \qquad \Longrightarrow \qquad 
(\Delta x - \Delta u + i\epsilon) \; 
(\Delta x + \Delta u - i\epsilon) \;\; , &(2.16b)\cr
&(+ \;\; -) \qquad \Longrightarrow \qquad 
(\Delta x - \Delta u - i\epsilon) \; 
(\Delta x + \Delta u + i\epsilon) \;\; , &(2.16c)\cr
&(- \;\; -) \qquad \Longrightarrow \qquad 
(\Delta x - \vert \Delta u \vert - i\epsilon) \;
(\Delta x + \vert \Delta u \vert + i\epsilon) \;\; . &(2.16d)\cr}$$

Two properties of Schwinger's formalism have significance for our computation.
First, the expectation value of a Hermitian operator such as the metric must be
real. Second, the interference between ``$+$'' and ``$-$'' vertices results in 
complete cancellation whenever an interaction strays outside the past lightcone
of the observation point. This means that no infrared regularization is needed.
The infrared divergences of the ``in''-``out'' formalism come from interactions
in the infinite future, and these drop out of ``in''-``in'' calculations. In 
their place one expects growth in the observation time.

\noindent {\it 2.4 Ultraviolet Regularization.}

The simplest way of regulating the ultraviolet is by keeping the parameter
$\epsilon$ non-zero in the ``$++$''propagators (2.14-15) and their variations 
(2.16). Consideration of the mode sums (2.11) reveals $\epsilon^{-1}$ to be 
an exponential cutoff on the coordinate 3-momentum. This would not be a very 
natural technique for an ``in''-``out'' calculation because there is nothing
unique about the $t=0$ surface of simultaneity. However, this surface has a
special significance in our modified ``in''-``in'' computation: it is where the
initial state is defined, and it is the point at which interactions begin. In
the context of expectation values, our method corresponds to weighting the 
usual Fock space inner product with a factor of $\exp(-\half \epsilon \Vert 
{\vec k} \Vert)$ for each creation and annihilation operator. In other words, 
{\it time evolution is unaffected by our method, only the inner product on the
initial value surface changes.} 

The point just made is crucial because spurious time dependence can be made to
reside on the ultraviolet regularization parameter. Consider, for example, a 
logarithmic divergence of the form $\ln(\epsilon)$. Had we instead regulated by
replacing $\epsilon$ everywhere with $\varepsilon Hu$, then we might have 
claimed to see a secular infrared effect:
$$\ln(\varepsilon Hu) = \ln(\varepsilon) - H t \eqno(2.17)$$
which is in fact of purely ultraviolet origin. We emphasize that the 
possibility for this sort of delusion has nothing to do with the perturbative
non-renormalizability of quantum general relativity. One can see it even in the
flat space, massless $\phi^3$ model which was our original paradigm for 
relaxation [2]. Nor does the phenomenon signify any fuzziness in the 
distinction between infrared and ultraviolet. The unambiguous signal for an
ultraviolet effect is propagators at or near coincidence.

The time dependence of the inflating background has led to much confusion about
ultraviolet regularization. Some researchers have thought it more natural to 
employ a mode cutoff which is invariant with respect to the background 
geometry. (Note that this method is no more invariant than ours with respect to
the full geometry.) This amounts to replacing our parameter $\epsilon$ with 
$\varepsilon \sqrt{H^2 u u'}$. One consequence is that the coincidence limit of
an undifferentiated propagator which has been so regulated grows linearly in
the co-moving time. By first taking the coincidence limit of our propagator and
then replacing $\epsilon$ with $\varepsilon H u$:
$$\eqalignno{\Bigl[{_{\mu \nu}} \Delta^{\rho \sigma}\Bigr](x;x) &= \; {H^2 
\over 8 {\pi}^2} \; \Biggl\{ {2u'u \over \epsilon^2} \; \Bigl[{2 \delta_{
\mu}^{~ (\rho} \; \delta_{\nu}^{~ \sigma)}} - \eta_{\mu \nu} \; \eta^{\rho 
\sigma} \Bigr] \cr
&\qquad \qquad  - \ln \Bigl[H^2 \epsilon^2 \Bigr] \; \Bigl[ 2{\overline 
\delta_{\mu}^{~(\rho}} \; {\overline \delta_{\nu}^{~ \sigma)}} - 2 {\overline 
\eta_{\mu \nu}} \; {\overline \eta^{\rho \sigma}} \Bigr] \;\Biggr\} &(2.18a)\cr
&\longrightarrow \; {H^2 \over 8 {\pi}^2} \; \Biggl\{ {2 \over H^2 
\varepsilon^2} \; \Bigl[{2 \delta_{\mu}^{~ (\rho} \; \delta_{\nu}^{~ \sigma)}} 
- \eta_{\mu \nu} \; \eta^{\rho \sigma} \Bigr] \cr
&\qquad \qquad \;\; + \Bigl(2 H t - \ln \Bigl[H^2 \varepsilon^2 \Bigr] \Bigr)\; 
\Bigl[ 2{\overline \delta_{\mu}^{~(\rho}}\; {\overline \delta_{\nu}^{~\sigma)}}
- 2 {\overline \eta_{\mu \nu}} \; {\overline \eta^{\rho \sigma}} \Bigr] \;
\Biggr\} \;\; , \qquad &(2.18b) \cr}$$
we see that this growth is identical to that displayed in (2.17), and hence of
purely ultraviolet provenance. One gets the same result with 
background-invariant point splitting. 

An erroneous argument is sometimes given that this spurious time dependence 
from the ultraviolet is actually a reliable effect from the infrared. One
first notes that on $\Re^3$ the propagator would be our integral expression 
(2.12a), but without the lower bound at $k=H$. The second term of the 
integrand grows rapidly enough near $k=0$ to give an infrared divergence.
According to the fallacious argument, one should regulate this infrared 
divergence by cutting off the integration at $k=(u u')^{-\frac12}$. Of course 
this replaces the factor of $H^2$ inside the logarithm term of our propagator
with $(u u')^{-\frac12}$, and one can see from (2.18a) that the coincidence 
limit then exhibits the same linear growth as (2.18b).

We stress that the introduction of a time dependent infrared cutoff prevents
the propagator from even inverting the kinetic operator. The rationale 
behind the cutoff is that coordinate momenta smaller than $k = u^{-1}$ 
correspond to physical wavelengths which have redshifted beyond the Hubble 
radius and should therefore decouple. This is correct physics but faulty 
mathematics. Modes indeed decouple when their physical wavelengths redshift 
beyond the Hubble radius, but this is accomplished by the causality of
interactions in Schwinger's formalism, not by the {\it ad hoc} imposition of
an infrared cutoff on the naive mode sum. The mode sum is actually regulated 
by the physically motivated restriction of coherent de Sitter vacuum to a 
spatial patch of finite extent. For us this was the Hubble radius, but any 
value would serve. The key point is that the infrared cutoff on the free 
mode sum is not time dependent. The only time dependence which appears in the 
coincidence limit of the propagator got there through a poor choice of the 
ultraviolet regulator.

Motivated by the faulty argument --- for which we emphasize that he bore no 
responsibility --- Ford [9] proposed a very interesting relaxation mechanism 
of which ours is, in some ways, a mirror image.\footnote{*}{\tenpoint We thank 
M. B. Einhorn for bringing Ford's work to our attention.} Among other 
differences, the most important diagrams for Ford were those with a coincident 
propagator attached to two legs of the same vertex. These are precisely the 
least important ones for us.
 
\noindent {\it 2.5 The Threshold for a Late Time Effect.}

We found it convenient to compute the amputated expectation value of the 
pseudo-graviton field and then attach the external line by solving an 
ordinary differential equation. The homogeneity and isotropy of the dynamics,
and of our initial state, allow us to express the amputated 1-point function 
in terms of two functions of $u$:
$$D_{\mu \nu}^{~~\alpha \beta} \; \Bigl\langle 0 \Bigl\vert \; 
\kappa \psi_{\alpha \beta}(u,{\vec x}) \; \Bigr\vert 0 \Bigr\rangle 
= a(u) \; {\overline \eta}_{\mu \nu} + 
c(u) \; \delta_{\mu}^{~0} \delta_{\nu}^{~0} \;\; . \eqno(2.19)$$
The full 1-point function must have the same form, although with different 
coefficient functions:
$$\Bigl\langle 0 \Bigl\vert \; \kappa \psi_{\mu \nu}(u,{\vec x}) 
\; \Bigr\vert 0 \Bigr\rangle = A(u) \; {\overline \eta}_{\mu \nu} + 
C(u) \; \delta_{\mu}^{~0} \delta_{\nu}^{~0} \;\; . \eqno(2.20)$$
Comparing these two expressions and taking account of the kinetic operator 
(2.7), we see that $A(u)$ and $C(u)$ can be expressed in terms of $a(u)$ and
$c(u)$ using the retarded Green's functions for the massless, minimally coupled
and conformally coupled scalars:
$$A(u) = - 4 \ G_A^{\rm ret}\Bigl[ a \Bigr](u) + G_C^{\rm ret}\Bigl[ 3a + c 
\Bigr](u) \;\; , \eqno(2.21a)$$
$$C(u) = G_C^{\rm ret}\Bigl[ 3a + c \Bigr](u) \;\; . \eqno(2.21b)$$
Although it is simple to give integral expressions for these Greens functions,
the more revealing form is to work them out for arbitrary powers of $u$ [2]:
$$G_A^{\rm ret} \Bigl[ u^{-4} \; (Hu)^{\zeta} \Bigr](u) = 
{H^2 \over 3 \zeta (1 - \frac13 \zeta)} \; 
\Biggl\{ (Hu)^{\zeta} - (1 - \frac13 \zeta) - 
\frac13 \zeta (Hu)^3 \Biggr\} \;\; , \eqno(2.22a)$$
$$G_C^{\rm ret} \Bigl[ u^{-4} \; (Hu)^{\zeta} \Bigr](u) =  
{H^2 \over 2 (1 -\zeta) (1 - \frac12 \zeta)} \;
\Biggl\{ - (Hu)^{\zeta} + 2 (1 - \frac12 \zeta) Hu - 
(1 - \zeta) (Hu)^2\Biggr\} \eqno(2.22b)$$
Note that $\zeta =0$ constitutes a sort of threshold. For larger values of
$\zeta$ the late time limits of $A(u)$ and $C(u)$ approach constants, whereas
smaller values of $\zeta$ give functions which grow as $u$ approaches zero.

To obtain $H_{\rm eff}(t)$ we compare with the invariant element in co-moving 
coordinates
$$- dt^2 + {\rm a}^2(t) \; d{\vec x} \cdot d{\vec x} = \Omega^2 \; \Bigl\{
-\Bigl[ 1 - C(u) \Bigr] \; du^2 + \Bigl[ 1 + A(u) \Bigr] \; d{\vec x} \cdot 
d{\vec x} \; \Bigr\} \;\; . \; \; \eqno(2.23)$$
Substituting into the definition (1.4) gives a result which is true even beyond
perturbation theory:
$$H_{\rm eff}(t) = {H \over \sqrt{1 - C(u)}} \; \Biggl\{ \; 
1 - {\frac12 u \; \frac{d}{du} A(u) \over 1 + A(u)} \; \Biggr\} 
\;\; . \eqno(2.24)$$
With relations (2.21) and (2.22), this shows that $a(u)$ and $c(u)$ must grow 
faster than $u^{-4}$ if quantum corrections are to overwhelm the classical 
result of $H_{\rm eff}(t) = H$ at late times.

\vfill\eject

\noindent {\it 2.6 Gauge Independence of $H_{\rm eff}(t)$.}

In demonstrating the gauge independence of $H_{\rm eff}(t)$ it is useful to
recall that any quantity can be made invariant by defining it in a unique 
coordinate system. In this case we can also exploit the homogeneity and 
isotropy of our special state, although we feel confident that a satisfactory
definition of the effective Hubble constant could be found for more general 
initial states. Since $H_{\rm eff}(t)$ was defined in co-moving coordinates, 
for a state whose scale factor obeys the initial conditions:
$$a(0) = 1 \qquad , \qquad {da(0) \over dt} = H \;\; ; \eqno(2.25)$$
our first step in any new coordinate system would be to restore these 
conditions by performing an appropriate coordinate transformation. We can 
therefore limit ourselves to consideration of gauge changes $F_{\mu} 
\rightarrow F_{\mu} + {\delta F}_{\mu}$ which preserve homogeneity, 
isotropy, and the initial conditions. Any such change can be parametrized 
as follows:
$$\delta F_{\mu} = \delta_{\mu}^{~0} \; \Omega^{-1} \; \varphi(u) \;\; , 
\eqno(2.26)$$ 
where preserving the initial condition requires $\varphi(H^{-1}) = 0$.

Changing our gauge condition (2.5) by (2.26) induces the following 1-point
interaction:
$$\eqalignno{\delta \Biggl( -\frac12 \int d^4x \; F_{\mu} \eta^{\mu\nu} F_{\nu}
\Biggr) &= \int d^4x \; F_0 \; \Omega^{-1} \varphi(u) &(2.27a) \cr
&= \int d^4x \; \Bigl\{ \psi_{00} \; \varphi_{,0} + \frac12 \; \psi \;
\varphi_{,0} + \frac2{u} \; \psi_{00} \; \varphi\Bigr\} \;\; . 
&(2.27b) \cr}$$
The coefficient functions of the amputated 1-point function therefore acquire
the following variations:
$$\delta a(u) = {\kappa \over 2} \; {d \over du} \; \varphi(u)
\qquad , \qquad 
\delta c(u) = {\kappa \over 2} \; {d \over du} \; \varphi(u) + 
{2\kappa \over u} \; \varphi(u) \;\; . \eqno(2.28)$$ 
Now consider variation by a general power, minus a constant to enforce the 
initial condition:
$$\varphi (u) = u^{-3} \; (Hu)^{\zeta} - H^3 \;\; . \eqno(2.29)$$
The induced variations on $a(u)$ and $c(u)$ are:
$$\delta a(u) = {\kappa \over 2} \; (\zeta -3) \; u^{-4} \; (Hu)^{\zeta} 
\;\; , \eqno(2.30a)$$
$$\delta c(u) = {\kappa \over 2} \; (\zeta + 1) \; u^{-4} \; (Hu)^{\zeta} - 
{2 \kappa H^3 \over u} \;\; . \eqno(2.30b)$$
The coefficient functions of the unamputated 1-point function suffer the
following variations:
$$\delta A(u) = 2 \kappa H^2 \Biggl\{ - {(Hu)^{\zeta} \over \zeta (\zeta -1)}
+ {\zeta - 3 \over 3 \zeta} - \Bigl({\zeta - 3 \over \zeta - 1}\Bigr) {H u 
\over 2} + {(Hu)^3 \over 6}\Biggr\} \;\; , \eqno(2.31a)$$
$$\delta C(u) = 2 \kappa H^2 \Biggl\{ - {(Hu)^{\zeta} \over \zeta -1} - 
\Bigl({\zeta - 3 \over \zeta - 1}\Bigr) {H u \over 2} + {(Hu)^3 \over 2}
\Biggr\} \;\; . \eqno(2.31b)$$
From (2.24) we see that the variation of $H_{\rm eff}(t)$ is:
$$\delta H_{\rm eff}(t) = H \; \Biggl\{ - {1 \over 2} \; u \; {d \over du} \; 
\delta A(u) + {1 \over 2} \; \delta C(u) \Biggr\} \;\; , \eqno(2.32)$$ 
and substitution of (2.31) gives $\delta H_{\rm eff}(t) = 0$. The proof is 
completed by noting that a general variation of the required form can be built
by superposing terms of the form (2.29).

\noindent {\it 2.7 The Genesis of Infrared Logarithms.}

It turns out that perturbative corrections to the amputated 1-point function 
can grow no faster than $u^{-4}$ times powers of $\ln(Hu)$ [2,10]. Since we
will need to extend the argument, it is useful to begin by restating it. 
Recall first that $\kappa$ has the dimensions of length, while $H$ goes like an
inverse length, and the amputated 1-point function has the dimensions of an 
inverse length squared. Diagrams which contribute to the amputated 1-point 
function at $\ell$ loops can involve up to $2\ell-1$ 3-point interactions and 
$3\ell-2$ propagators. Each 3-point vertex contributes a constant factor of 
$\kappa H^{-2}$, while each propagator contributes a factor of $H^2$. Including 
the single factor of $\kappa$ in our definition (2.19), we see that the 
$\ell$-loop contribution consists of $\kappa^{2\ell} H^{2\ell-2}$ times 
a function of $H$, $u$ and $\epsilon$ which has the dimensionality 
$({\rm length})^{-4}$. 

Since the ultraviolet regularization parameter goes to zero after 
renormalization, we can ignore positive powers of $\epsilon$. The constant $H$ 
can enter through undifferentiated $i\Delta_L$, and through the limits of 
integration on the conformal time integrals. Neither can contribute negative 
powers of $H$. This is obvious for $i\Delta_L$. To see that factors of $H$ from
the limits cannot contribute negative powers, note that performing the spatial 
integrations leaves $2\ell-2$ conformal time integrations of an integrand which
falls off like $S^{-2\ell-6}$ if all the conformal times are scaled by the 
common factor $S$. Since the integrand can contain at most two isolated factors
of $u^{-1}$, the conformal time integrations must converge even if the factors
of $H^{-1}$ in their upper limits are taken to infinity. Hence there can be 
no negative powers of $H$.

To complete the argument consider a dimension $({\rm length})^{-4}$ function of 
$H$, $u$ and $\epsilon$ which contains no negative powers of $H$ or positive 
powers of $\epsilon$. We can write any such function as $u^{-4}$ times a 
combination of the dimensionless parameters $Hu$ and $\epsilon u^{-1}$. Since 
there can be no negative powers of the first parameter, or positive powers of 
the second, the strongest growth possible as $u \rightarrow 0^+$ is $u^{-4}$ 
times powers of $\ln(Hu)$ and $\ln(\epsilon u^{-1}) = \ln(H\epsilon) - 
\ln(Hu)$. 

Because the amputated 1-point function must grow faster than $u^{-4}$ in order
for there to be a significant effect at late times, the appearance of infrared 
logarithms is essential for our scheme of relaxation. The preceding argument
can be extended to show that infrared logarithms are just about inevitable if
one accepts that there are logarithmic ultraviolet divergences. Since $\epsilon$ 
is a dimensionful quantity, logarithmic ultraviolet divergences can only come 
in the form of $\ln(\epsilon u^{-1})$ and $\ln(H\epsilon)$. As before, 
there are just two sources of $H$ dependence: an undifferentiated $i \Delta_L$
and the upper limits $H^{-1}$ of the conformal time integrations. The latter 
cannot provide factors of $\ln(H)$ for the same reason it gives no negative 
powers of $H$: the conformal time integrands fall off rapidly enough for large 
conformal times to make them converge even when the upper limits become 
infinite. The logarithm part of an undifferentiated propagator can indeed 
provide a factor of $\ln(H)$, but there are never enough such terms to pair 
with all the factors of $\ln(\epsilon)$. At $\ell$ loops one has to expect 
$\ell$ factors of $\ln(\epsilon)$ from ultraviolet divergences, but there 
is a maximum of $\ell-1$ logarithms from undifferentiated $i \Delta_L$'s.
\footnote{*}{ \tenpoint See the end of sub-section 2.9 for a proof.} 
Hence at least one of the ultraviolet logarithms must come in the form 
$\ln(\epsilon u^{-1})$. 

We emphasize that the association we have exploited between infrared logarithms
and logarithmic ultraviolet divergences in no way implies that factors of 
$\ln(u)$ are of ultraviolet origin. Some of them originate as factors of 
$\ln(Hu)$ from the logarithm parts of undifferentiated propagators.
\footnote{**}{\tenpoint The reader should not be misled by our argument that 
there are fewer intrinsic factors of $\ln(Hu)$ than of $\ln(\epsilon u^{-1})$. 
We noted this only to rule out the remote possibility that the two factors 
come in pairs so as to cancel the infrared logarithms: $\ln(\epsilon u^{-1}) 
+ \ln(Hu) = \ln(H \epsilon)$. In fact the two sources of infrared logarithms 
tend to add.} 
The physical origin of such terms is the increasing correlation of the free 
graviton vacuum as inflation proceeds. This effect is the casual analog of the 
problem discussed in sub-section 2.4 when one assumes the existence of 
coherent de Sitter vacuum over an infinite surface of simultaneity.

Even the factors of $\ln(u)$ which originate as $\ln(\epsilon u^{-1})$ are 
infrared effects. They derive from the coherent superposition of interactions 
throughout the invariant volume of the past lightcone. In the absence of a mass,
ultraviolet divergences must always be associated with the infrared in this 
way. See, for example, the result we obtained for massless, $\phi^3$ theory 
in flat space [2].

\noindent {\it 2.8 A Bound on the Number of Infrared Logarithms.}

We argued in the preceding sub-section that infrared logarithms are all but
inevitable, and that they derive jointly from the infrared regions of loop
integrals which harbor logarithmic ultraviolet divergences and from an
undifferentiated $i\Delta_L$. However, we did not explain how many of the $\ell
-1$ undifferentiated propagator logarithms can go to reinforce the $\ell$ 
ultraviolet logarithms that one expects in an $\ell$-loop diagram. The answer 
turns out to be ``none:'' there can be at most $\ell$ infrared logarithms at 
$\ell$ loops. All the undifferentiated factors of $i\Delta_L$ do is to sometimes 
make up the difference for diagrams which would otherwise contribute less than 
$\ell$ infrared logarithms. 

To see why, consider the way a single undifferentiated propagator might enhance
the number of infrared logarithms. There can be $\ell$ logarithmic ultraviolet
divergences at $\ell$ loops so what we are looking for is:
$$\ln^{\ell}(\epsilon u^{-1}) \times \ln(Hu) = 
\ln^{\ell}(\epsilon u^{-1}) \times \ln(H) +
\ln^{\ell}(\epsilon u^{-1}) \times \ln(u) \;\; . \eqno(2.33)$$
In other words, if we replace the propagator in question with $\ln(H)$, the
resulting diagram must contribute $\ell$ logarithmic ultraviolet divergences.
But replacing a propagator by a constant cuts the leg and results in a diagram
containing only $\ell-1$ loops. Such a diagram can contribute at most $\ell-1$ 
factors of $\ln(\epsilon u^{-1})$. Therefore, an undifferentiated $i\Delta_L$ 
cannot increase the number of infrared logarithms in an $\ell$-loop diagram 
beyond $\ell$.

It is worth mentioning that this argument applies to diagrams, not to portions
of diagrams. However, the only practical way of performing the computation was 
to break each diagram up into many pieces. At two loops some of these pieces
produce three infrared logarithms; it is only their sums which are limited 
to two. On the first run of this calculation we did not appreciate that the
triple log terms must cancel. This argument was only discovered after noting
what was to us then, a surprising and disturbing cancellation. In retrospect
the cancellation stands as a powerful check on the consistency and accuracy of
our work.

\noindent {\it 2.9 Comments on One and Two Loops.}

The first quantum corrections to the amputated 1-point function are the 
one-loop graphs shown in Fig.~1. Since the external line has been amputated, 
there is no integration over the single interaction vertex at $x^{\mu} = 
(u,{\vec x})$. One obtains just the coincidence limit of a pseudo-graviton 
or ghost propagator, acted upon by the appropriate vertex operator. It turns 
out that there can be no undifferentiated ``logarithmic'' parts $i\Delta_L$, 
but even if there were, they would only contribute factors of $\ln(H\epsilon)$. One needs to {\it integrate} something over a large invariant volume in order 
to see an effect. This is one way that our mechanism differs from Ford's [9]. 

\vskip -1.5cm

\centerline{\psfig{figure=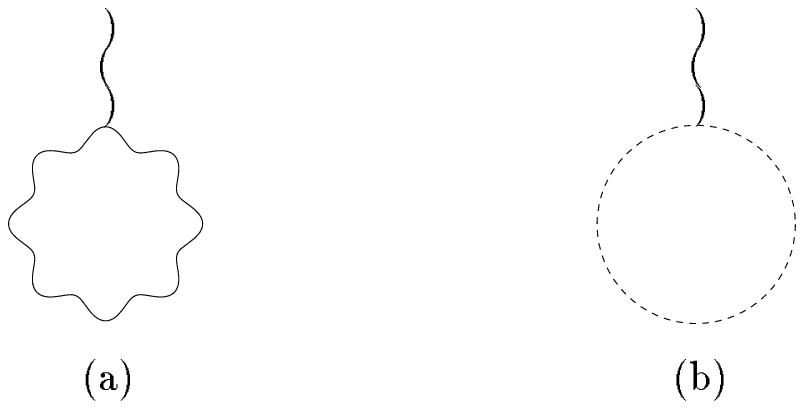,height=21cm,bbllx=0bp
,bblly=0bp,bburx=596bp,bbury=843bp,rheight=5.5cm,rwidth=12.7cm}}

{\bf Fig.~1:} {\ninepoint One-loop contributions to the background 
geometry. Gravitons reside on wavy lines and ghosts 
\vskip -16pt \noindent \hglue 2.50truecm on segmented lines.}

\vskip 0.3cm

\noindent
The one-loop graphs of Fig.~1 contribute at most terms of order $u^{-2}$, 
which is well below the threshold for producing an effect at late times.

\vskip -1.5cm

\centerline{\psfig{figure=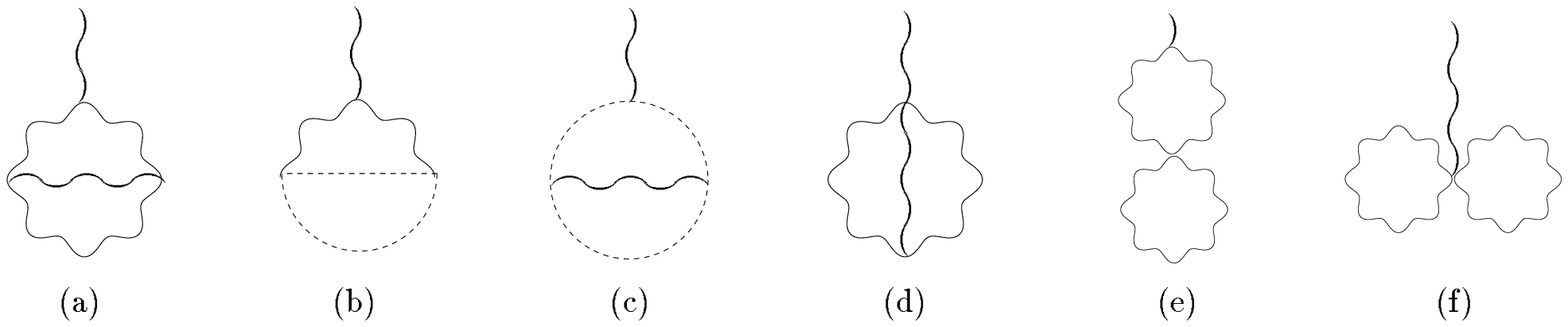,height=21cm,bbllx=0bp
,bblly=0bp,bburx=596bp,bbury=843bp,rheight=5.5cm,rwidth=12.7cm}}

{\bf Fig.~2:} {\ninepoint Two-loop contributions to the background 
geometry. Gravitons reside on wavy lines and ghosts 
\vskip -16pt \noindent \hglue 2.50truecm on segmented lines.}

\vskip 0.3cm

The graphs which contribute at two loops are displayed in Fig.~2. Diagram (2f)
is another ultra-local coincidence limit, so it contributes no infrared 
logarithms. Diagram (2e) contains a single free interaction vertex, but the
entire diagram is canceled by the counterterm needed to renormalize its 
coincident lower loop. Daigram (2d) also contains a single free interaction, 
and an undifferentiated $i\Delta_L$ allows it to contribute two infrared 
logarithms. We can this the ``4-3'' diagram because it consists of a 4-point 
and a 3-point vertex. 
What we call ``3-3-3'' diagrams are shown in (2a-c). They contain two free 
interaction vertices and would give two infrared logarithms if the calculation 
were done in flat space. By the argument of the preceding sub-section we know 
that undifferentiated logarithms from the propagators do not change this.

\vskip 1.3cm

\centerline{\psfig{figure=fg3.eps,height=7cm}}

\vskip 0.2cm

{\bf Fig.~3:} {\ninepoint In the 4-3 diagram (a) and the generic 
3-3-3 diagram (b), solid lines can represent gravitons or  
\vskip -16pt \noindent \hglue 2.40truecm
ghosts and the arrows indicate the action of the vertex 
derivatives.}

\vskip 0.3cm

Fig.~3 is useful in understanding why two-loop contributions to the 1-point
function have at most a single undifferentiated $i\Delta_L$. One can see from 
(2.4) and (2.6) that every interaction vertex contains two fields which are 
either differentiated or else contain a $0$-index. From (2.15) one can see that
a $0$-index precludes coupling to the logarithm term in the propagator. We call
a line which is either differentiated or forced to carry a $0$-index, 
``contaminated.'' 

To treat the case for a general loop $\ell$, note that a graph of this order 
can have $2\ell-1$ 3-point vertices and $3\ell-2$ internal propagators. One of 
the vertices is the external one, whose external leg may be contaminated, but 
the other vertices must each contribute two contaminated legs. This makes for a
total number of either $4\ell-3$ or $4\ell-4$ contaminating factors 
(derivatives or 0-indices) to distribute among the $3\ell -2$ internal 
propagators. The smallest number of ``spoiled'' propagators is therefore $2\ell
-1$, leaving $\ell-1$ which can harbor undifferentiated logarithms. The result 
is the same for $\ell$-loop graphs constructed from higher point vertices 
because the number of contaminating factors falls by two for each internal line
which is lost.

\vskip 1cm
\centerline{\bf {3. The 3--3--3 Diagrams}} 

The three 3-3-3 diagram consists of an outer 3-point vertex joined to two 
freely integrated 3-point vertices (see Fig.~4). 

\vskip 1.3cm

\centerline{\psfig{figure=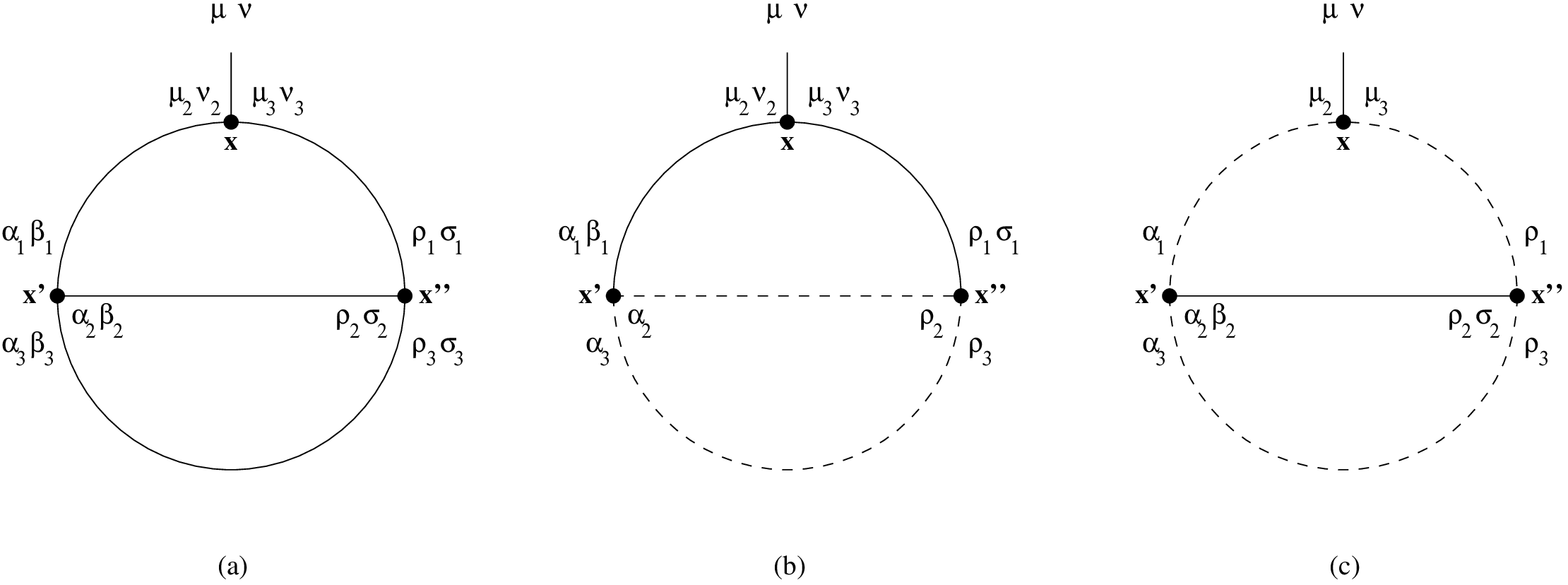,width=16truecm,angle=-90}}

\vskip 0.7truecm 

{\bf Fig.~4:}{\ninepoint The tensor structure of the 3-3-3 
diagrams. Gravitons reside on solid lines and ghosts on 
\vskip -16pt \noindent \hglue 2.30truecm segmented lines.}

\vskip 0.4cm 

\noindent The fixed vertex is taken to be at $x^{\mu}= (u,{\vec x})$, and we 
can identify the locations of the two free vertices as ${x'}^{\mu} = (u',{\vec 
x}~')$ and ${x''}^{\mu} = (u'',{\vec x}~'')$. One can form three differences 
from these positions:
$$w^{\mu} \equiv (x-x')^{\mu} \qquad , \qquad 
y^{\mu} \equiv (x' - x'')^{\mu} \qquad , \qquad 
z^{\mu} \equiv (x'' - x)^{\mu} \;\; . \eqno(3.1)$$
The fixed vertex is taken to be of ``$+$'' type, and there are a total of four 
variations when each of the two free vertices is summed over ``$+$'' and 
``$-$''. These variations have no effect until quite late in the reduction 
process because they control only the overall sign of the diagram and the order
$\epsilon$ terms in the four propagators. However, since the variations 
interfere destructively whenever ${x'}^{\mu}$ or ${x''}^{\mu}$ is outside the 
rather small past lightcone of $x^{\mu}$, we will make no error by extending 
the spatial integrations from $T^3$ to $\Re^3$.

The Lorentz squares of the vectors (3.1) are understood to include terms of
order $\epsilon$ according to the following scheme:
$$\eqalignno{ &w^2 = (r' - d_w + i \epsilon) \; 
(r' + d_w - i \epsilon) \qquad , \qquad 
{\vec r}\ ' \equiv \vec x - {\vec x}\ ' \;\; ; &(3.2a) \cr 
&z^2 = (r'' - d_z + i \epsilon) \; 
(r'' + d_z - i \epsilon) \qquad , \qquad 
{\vec r}\ '' \equiv {\vec x}\ '' - \vec x \;\; ; &(3.2b) \cr 
&y^2 = (\Vert {\vec r}\ ' + {\vec r}\ '' \Vert - d_y + i \epsilon) 
\; (\Vert {\vec r}\ ' + {\vec r}\ '' \Vert + d_y - i \epsilon) 
\;\; . &(3.2c) \cr}$$ 
Here $r' \equiv \Vert {\vec r}\ ' \Vert$ , $r'' \equiv \Vert {\vec r}\ '' 
\Vert$ , and the three $d$'s are conformal time differences which depend upon 
the four ``$\pm$'' variations. For example, when ${x'}^{\mu}$ is ``$+$'' and 
${x''}^{\mu}$ is ``$-$'' we have:
$$d_w = \vert u'-u \vert \qquad , \qquad d_z = u-u'' \qquad , \qquad
d_y = u' -u'' \;\; . \eqno(3.3)$$
We will eventually show that the order $u \leq u' \leq u'' \leq H^{-1}$ can be
enforced, in which case the $d$'s have the form given in Table~1.

\vskip 1cm 

\vbox{\tabskip=0pt \offinterlineskip
\def\tablerule{\noalign{\hrule}}
\halign to450pt {\strut#& \vrule#\tabskip=1em plus2em& 
\hfil#& \vrule#& \hfil#\hfil& \vrule#& \hfil#& \vrule#& \hfil#\hfil& 
\vrule#\tabskip=0pt\cr
\tablerule
\omit&height4pt&\omit&&\omit&&\omit&&\omit&\cr
&&\omit\hidewidth {$V(x') , V(x'')$}\hidewidth&&
\omit\hidewidth {$d_w$}\hidewidth&& 
\omit\hidewidth {$d_z$}\hidewidth&& 
\omit\hidewidth {$d_y$} 
\hidewidth&\cr
\omit&height4pt&\omit&&\omit&&\omit&&\omit&\cr
\tablerule
\omit&height2pt&\omit&&\omit&&\omit&&\omit&\cr
&& $+$ , $+$ && $u'-u$ && $u''-u$ && $u''-u'$ &\cr
\omit&height2pt&\omit&&\omit&&\omit&&\omit&\cr
\tablerule
\omit&height2pt&\omit&&\omit&&\omit&&\omit&\cr
&& $+$ , $-$ && $u'-u$ && $u-u''$ && $u'-u''$ &\cr
\omit&height2pt&\omit&&\omit&&\omit&&\omit&\cr
\tablerule
\omit&height2pt&\omit&&\omit&&\omit&&\omit&\cr
&& $-$ , $+$ && $u-u'$ && $u''-u$ && $u''-u'$ &\cr
\omit&height2pt&\omit&&\omit&&\omit&&\omit&\cr
\tablerule
\omit&height2pt&\omit&&\omit&&\omit&&\omit&\cr
&& $-$ , $-$ && $u-u'$ && $u-u''$ && $u'-u''$ &\cr
\omit&height2pt&\omit&&\omit&&\omit&&\omit&\cr
\tablerule}}

\vskip 0.2cm

{\bf Table~1:} {\ninepoint The values of the $d$'s for the four possible 
variations of the vertices at $x'$ and $x''$, 
\vskip -16pt \noindent \hglue 2.85truecm assuming $u \leq u' \leq u'' \leq 
H^{-1}$.}

\vskip 0.3cm

The ghost-graviton interactions can be obtained from expression (2.6):
$$\eqalignno{ {\cal L}_{\rm ghost}^{(3)} = 
\; \kappa \; \Omega^2 &\; \Bigl\{ 
- \psi_{\mu \nu} \; {\overline \omega}^{\mu , \alpha} \; 
\omega^{\nu}_{~, \alpha} 
- \psi_{\mu \nu} \; {\overline \omega}^{\alpha , \mu} \; 
\omega^{\nu}_{~, \alpha} 
- \psi_{\mu \nu , \alpha} \; {\overline \omega}^{\mu , \nu} \; 
\omega^{\alpha} 
+ \psi_{\mu \nu} \; {\overline \omega}^{\alpha}_{~, \alpha} \; 
\omega^{\mu , \nu} \qquad \cr 
&+ \frac12 \psi_{, \mu} \; {\overline \omega}^{\alpha}_{~, \alpha} \; 
\omega^{\mu}    
-\frac2u \psi_{\mu \nu} \; {\overline \omega}^{\mu , \nu} \; 
\omega^{\alpha} \; t_{\alpha} 
+\frac1u \psi \; {\overline \omega}^{\mu}_{~, \mu} \; 
\omega^{\nu} \; t_{\nu} \cr 
&+\frac2u \psi_{\mu \nu} \; {\overline \omega}^{\alpha} \; 
\omega^{\mu , \nu} \; t_{\alpha} 
+\frac1u \psi_{, \mu} \; {\overline \omega}^{\nu} \; 
\omega^{\mu} \; t_{\nu}
+ \frac{2}{u^2} \psi \; {\overline \omega}^{\mu} \; \omega^{\nu} \; 
t_{\mu} \; t_{\nu} 
\Bigr\} \;\; . &(3.4) \cr}$$
(Recall that $t_{\mu} \equiv \eta_{\mu 0}$ and $\psi \equiv \eta^{\mu\nu} 
\psi_{\mu\nu}$.) The associated vertex operators $V_i^{\alpha_1 \beta_1\mu\nu}$
are defined by the relation:
$${\cal L}^i_{\rm ghost} \equiv \eta_{\alpha_1 \beta_1} \; \eta_{\alpha_2 
\alpha_3} \;\; V_i^{\alpha_1 \beta_1 \alpha_2 \alpha_3} (x \; ; \; \partial_1 ,
\partial_2 , \partial_3) \qquad , \qquad i=1,...,10 \;\; ; \eqno(3.5)$$ 
and are explicitly displayed in Table 2. 

\vskip1cm

\vbox{\tabskip=0pt \offinterlineskip
\def\tablerule{\noalign{\hrule}}
\halign to450pt {\strut#& \vrule#\tabskip=1em plus2em& 
\hfil#& \vrule#& \hfil#\hfil& \vrule#& \hfil#& \vrule#& \hfil#\hfil& 
\vrule#\tabskip=0pt\cr
\tablerule
\omit&height4pt&\omit&&\omit&&\omit&&\omit&\cr
&&\omit\hidewidth \# 
&&\omit\hidewidth {\rm Vertex Operator}\hidewidth&& 
\omit\hidewidth \#\hidewidth&& 
\omit\hidewidth {\rm Vertex Operator}
\hidewidth&\cr
\omit&height4pt&\omit&&\omit&&\omit&&\omit&\cr
\tablerule
\omit&height2pt&\omit&&\omit&&\omit&&\omit&\cr
&& 1 && $- \eta^{\alpha_2 (\alpha_1} \; \eta^{\beta_1) \alpha_3}
\; \partial_2 \cdot \partial_3$ 
&& 6 && $\frac12 \eta^{\alpha_1 \beta_1} \; \partial_2^{\alpha_2} 
\; \partial_1^{\alpha_3}$ &\cr
\omit&height2pt&\omit&&\omit&&\omit&&\omit&\cr
\tablerule
\omit&height2pt&\omit&&\omit&&\omit&&\omit&\cr
&& 2 && $- \eta^{\alpha_3 (\alpha_1} \; \partial_2^{\beta_1)} 
\; \partial_3^{\alpha_2}$ 
&& 7 && $\frac1{u} \eta^{\alpha_1 \beta_1} \; \partial_2^{\alpha_2} 
\; t^{\alpha_3}$ &\cr
\omit&height2pt&\omit&&\omit&&\omit&&\omit&\cr
\tablerule
\omit&height2pt&\omit&&\omit&&\omit&&\omit&\cr
&& 3 && $- \eta^{\alpha_2 (\alpha_1} \; \partial_2^{\beta_1)} 
\; \partial_1^{\alpha_3}$ 
&& 8 && $\frac2{u} \eta^{\alpha_3 (\alpha_1} \; \partial_3^{\beta_1)} 
\; t^{\alpha_2}$ &\cr
\omit&height2pt&\omit&&\omit&&\omit&&\omit&\cr
\tablerule
\omit&height2pt&\omit&&\omit&&\omit&&\omit&\cr
&& 4 && $- \frac2{u} \eta^{\alpha_2 (\alpha_1} \; \partial_2^{\beta_1)}
\; t^{\alpha_3}$ 
&& 9 && $\frac1{u} \eta^{\alpha_1 \beta_1} \; \partial_1^{\alpha_3} 
\; t^{\alpha_2}$ &\cr
\omit&height2pt&\omit&&\omit&&\omit&&\omit&\cr
\tablerule
\omit&height2pt&\omit&&\omit&&\omit&&\omit&\cr
&& 5 && $\eta^{\alpha_3 (\alpha_1} \; \partial_3^{\beta_1)} 
\; \partial_2^{\alpha_2}$ 
&& 10 && $\frac2{u^2} \eta^{\alpha_1 \beta_1} \; t^{\alpha_2} 
\; t^{\alpha_3}$ &\cr
\omit&height2pt&\omit&&\omit&&\omit&&\omit&\cr
\tablerule}}

\vskip -0.3cm

{\bf Table~2:} {\ninepoint Vertex operators of ghost-pseudograviton 
interactions without the factor of $\kappa \Omega^2$.}

\vskip 0.3cm

We can read the graviton 3-point interaction off from expression (2.4):
$$\eqalignno{ {\cal L}_{\rm inv}^{(3)} =  
\; \kappa \; \Omega^{2} &\; \Bigl\{
\frac14 \psi \; \psi^{\alpha \beta , \mu} \; \psi_{\mu \alpha , \beta} -
\psi^{\alpha \beta} \; \psi_{\alpha \mu , \nu} \; 
\psi^{\mu \nu}_{~~, \beta} -
\frac12 \psi^{\alpha \beta} \; \psi_{\alpha \mu}^{~~, \nu} \;
\psi_{\beta \nu}^{~~, \mu} \cr
&- \frac14 \psi \; \psi_{, \mu} \; \psi^{\mu \nu}_{~~, \nu} +
\frac12 \psi^{\alpha \beta} \; \psi_{\alpha \beta , \mu} \;
\psi^{\mu \nu}_{~~, \nu} +
\frac12 \psi^{\alpha \beta} \; \psi_{, \alpha} \; 
\psi_{\mu \beta}^{~~, \mu} +
\frac12 \psi^{\alpha \beta} \; \psi_{\alpha \mu , \beta} \;
\psi^{, \mu} \cr
&+ \frac18 \psi \; \psi^{, \mu} \; \psi_{, \mu} -
\frac12 \psi^{\alpha \beta} \; \psi_{\alpha \beta , \mu} \; 
\psi^{, \mu} -
\frac14 \psi^{\alpha \beta} \; \psi_{,\alpha} \; \psi_{, \beta} \cr
&- \frac18 \psi \; \psi^{\alpha \beta , \mu} \;
\psi_{\alpha \beta , \mu} +
\frac12 \psi^{\alpha \beta} \; \psi_{\alpha \mu , \nu} \;
\psi_{\beta}^{~~\mu , \nu} +
\frac14 \psi^{\alpha \beta} \; \psi_{\mu \nu , \alpha} \;
\psi^{\mu \nu}_{~~, \beta} \cr
&- \frac{1}{2u} \psi \; \psi_{, \mu} \; 
\psi^{\mu \nu} \; t_{\nu} + 
\frac1u \psi^{\alpha \beta} \; \psi_{\alpha \beta , \mu} \; 
\psi^{\mu \nu} \; t_{\nu} + 
\frac1u \psi_{\alpha \mu} \; \psi^{, \alpha} \; 
\psi^{\mu \nu} \; t_{\nu}  
\Bigr\} \;\; . &(3.6) \cr}$$ 
It is worth noting that all but the last three terms can be checked against the
previously published, flat space 3-point interactions [11] by merely omitting
the factor of $\Omega^2$ and regarding $\psi_{\mu\nu}$ as the graviton.

In deriving the associated vertex operators we must account for the 
indistinguishability of gravitons. This would ordinarily be accomplished by 
fully symmetrizing each interaction, which turns out to give 75 distinct terms.
For the pure graviton diagram (4a) it is wasteful to first sum over the 
$(75)^3 = 421,875$ possibilities for the three vertices and then divide by 
the symmetry factor of 4 to compensate for overcounting. The more efficient 
strategy is to symmetrize the vertices only on line \#1 and then sum over 
the interchange of lines \#2 and \#3 for only the vertex at ${x''}^{\mu}$, 
dispensing with the symmetry factor. One saves over a factor of 3 this way.

To obtain the partially symmetrized vertices one first takes any of the terms
from (3.6) and permutes graviton \#1 over the three possibilities. As an 
example, consider the first term in (3.6). Denoting graviton \#1 by a breve, we
obtain the following three terms:
$$\frac14 \; \kappa \Omega^2 \; {\breve \psi} \; \psi^{\alpha \beta , \mu} \; 
\psi_{\mu \alpha , \beta} \; + \;  \frac14 \; \kappa \Omega^2 {\breve \psi_{\mu
\alpha , \beta}} \; \psi \; \psi^{\alpha \beta , \mu} \; + \; \frac14 \; \kappa
\Omega^2 {\breve \psi^{\alpha \beta , \mu}} \; \psi_{\mu \alpha , \beta} \; 
\psi \;\; . \eqno(3.7)$$  
One then assigns the remaining two gravitons in each term as \#2 and \#3 in any
way. For example, from (3.7) we could infer the following three vertex 
operators:
$$\frac14 \; \kappa \Omega^2 \eta^{\alpha_1 \beta_1} \; \partial_3^{(\alpha_2} 
\; \eta^{\beta_2)(\alpha_3} \; \partial_2^{\beta_3)} \;\; , \eqno(3.8a)$$
$$\frac14 \; \kappa \Omega^2 \eta^{\alpha_2 \beta_2} \; \partial_1^{(\alpha_3} 
\; \eta^{\beta_3)(\alpha_1} \; \partial_3^{\beta_1)} \;\; , \eqno(3.8b)$$
$$\frac14 \; \kappa \Omega^2 \eta^{\alpha_3 \beta_3} \; \partial_2^{(\alpha_1} 
\; \eta^{\beta_1)(\alpha_2} \; \partial_1^{\beta_2)} \;\; . \eqno(3.8c)$$ 
The 43 operators which comprise the full vertex are given in Table 3.

The various 3-3-3 diagrams can be written in terms of the vertex operators of 
Tables~2-3 and the propagators (2.15). The diagram of Fig.~4a has gravitons on 
both loops and results in the following expression:
$$\eqalignno{ \kappa \;\; &\int_u^{H^{-1}} du' \int d^3r' \; 
\int_{u}^{H^{-1}} du'' \int d^3r'' \;\; \sum_{i,j,k=1}^{43} \; \cr 
&  V_i^{\mu \nu \mu_2 \nu_2 \mu_3 \nu_3} (x \; ; \; 
\partial_1 , \partial_2 , \partial_3) \;\; 
i\Bigl[ {_{\mu_2 \nu_2}}\Delta_{\alpha_1 \beta_1} \Bigr](x \; ; x') \;\; 
i\Bigl[ {_{\mu_3 \nu_3}}\Delta_{\rho_1\sigma_1}\Bigr](x \; ; x'')\cr 
& \Biggl\{ \; V_j^{\alpha_1 \beta_1 \alpha_2 \beta_2 \alpha_3 \beta_3}
(x'; \; \partial_1' , \partial_2' , \partial_3') \;\;  
i\Bigl[ {_{\alpha_2 \beta_2}}\Delta_{\rho_2 \sigma_2} \Bigr] (x' ; x'') \;\; 
i\Bigl[ {_{\alpha_3 \beta_3}}\Delta_{\rho_3 \sigma_3} \Bigr](x' ; x'') \cr 
& \Bigl( \; V_k^{\rho_1 \sigma_1 \rho_2 \sigma_2 \rho_3 \sigma_3} 
(x''; \; \partial_1'' , \partial_2'' , \partial_3'') \; + \; 
V_k^{\rho_1 \sigma_1 \rho_3 \sigma_3 \rho_2 \sigma_2}
(x''; \; \partial_1'' , \partial_3'' , \partial_2'') \; \Bigr) \; \Biggr\} 
\;\; . &(3.9a) \cr}$$ 

\vfill\eject

\vbox{\tabskip=0pt \offinterlineskip
\def\tablerule{\noalign{\hrule}}
\halign to450pt {\strut#& \vrule#\tabskip=1em plus2em& 
\hfil#& \vrule#& \hfil#\hfil& \vrule#& \hfil#& \vrule#& \hfil#\hfil& 
\vrule#\tabskip=0pt\cr
\tablerule
\omit&height4pt&\omit&&\omit&&\omit&&\omit&\cr
&&\omit\hidewidth \# &&\omit\hidewidth {\rm Vertex Operator}\hidewidth&& 
\omit\hidewidth \#\hidewidth&& \omit\hidewidth {\rm Vertex Operator}
\hidewidth&\cr
\omit&height4pt&\omit&&\omit&&\omit&&\omit&\cr
\tablerule
\omit&height2pt&\omit&&\omit&&\omit&&\omit&\cr
&& 1 && $-\frac1{2u} \eta^{\alpha_1 \beta_1} \; \eta^{\alpha_2 \beta_2}
\; \partial_2^{(\alpha_3} \; t^{\beta_3)}$ 
&& 22 && $\frac12 \eta^{\alpha_1 (\alpha_2} \; \eta^{\beta_2) \beta_1} 
\; \partial_2^{(\alpha_3} \; \partial_3^{\beta_3)}$ &\cr
\omit&height2pt&\omit&&\omit&&\omit&&\omit&\cr
\tablerule
\omit&height2pt&\omit&&\omit&&\omit&&\omit&\cr
&& 2 && $-\frac1{2u} \eta^{\alpha_2 \beta_2} \; \eta^{\alpha_3 \beta_3}
\; \partial_3^{(\alpha_1} \; t^{\beta_1)}$ 
&& 23 && $\frac12 \eta^{\alpha_2 (\alpha_3} \; \eta^{\beta_3) \beta_2} 
\; \partial_3^{(\alpha_1} \; \partial_1^{\beta_1)}$ &\cr
\omit&height2pt&\omit&&\omit&&\omit&&\omit&\cr
\tablerule
\omit&height2pt&\omit&&\omit&&\omit&&\omit&\cr
&& 3 && $-\frac1{2u} \eta^{\alpha_3 \beta_3} \; \eta^{\alpha_1 \beta_1}
\; \partial_1^{(\alpha_2} \; t^{\beta_2)}$ 
&& 24 && $\frac12 \eta^{\alpha_3 (\alpha_1} \; \eta^{\beta_1) \beta_3} 
\; \partial_1^{(\alpha_2} \; \partial_2^{\beta_2)}$ &\cr
\omit&height2pt&\omit&&\omit&&\omit&&\omit&\cr
\tablerule
\omit&height2pt&\omit&&\omit&&\omit&&\omit&\cr
&& 4 && $\frac1{u} \eta^{\alpha_1 (\alpha_2} \; \eta^{\beta_2) \beta_1}
\; \partial_2^{(\alpha_3} \; t^{\beta_3)}$ 
&& 25 && $\frac12 \partial_2^{(\alpha_1} \; \eta^{\beta_1) (\alpha_3} 
\; \partial_3^{\beta_3)} \; \eta^{\alpha_2 \beta_2}$ &\cr
\omit&height2pt&\omit&&\omit&&\omit&&\omit&\cr
\tablerule
\omit&height2pt&\omit&&\omit&&\omit&&\omit&\cr
&& 5 && $\frac1{u} \eta^{\alpha_2 (\alpha_3} \; \eta^{\beta_3) \beta_2}
\; \partial_3^{(\alpha_1} \; t^{\beta_1)}$ 
&& 26 && $\frac12 \partial_3^{(\alpha_2} \; \eta^{\beta_2) (\alpha_1} 
\; \partial_1^{\beta_1)} \; \eta^{\alpha_3 \beta_3}$ &\cr
\omit&height2pt&\omit&&\omit&&\omit&&\omit&\cr
\tablerule
\omit&height2pt&\omit&&\omit&&\omit&&\omit&\cr
&& 6 && $\frac1{u} \eta^{\alpha_3 (\alpha_1} \; \eta^{\beta_1) \beta_3}
\; \partial_1^{(\alpha_2} \; t^{\beta_2)}$ 
&& 27 && $\frac12 \partial_1^{(\alpha_3} \; \eta^{\beta_3) (\alpha_2} 
\; \partial_2^{\beta_2)} \; \eta^{\alpha_1 \beta_1}$ &\cr
\omit&height2pt&\omit&&\omit&&\omit&&\omit&\cr
\tablerule
\omit&height2pt&\omit&&\omit&&\omit&&\omit&\cr
&& 7 && $\frac1{u} t^{(\alpha_3} \; \eta^{\beta_3) (\alpha_1}
\; \partial_2^{\beta_1)} \; \eta^{\alpha_2 \beta_2}$ 
&& 28 && $\frac12 \partial_2^{(\alpha_1} \; \eta^{\beta_1) (\alpha_2} 
\; \partial_3^{\beta_2)} \; \eta^{\alpha_3 \beta_3}$ &\cr
\omit&height2pt&\omit&&\omit&&\omit&&\omit&\cr
\tablerule
\omit&height2pt&\omit&&\omit&&\omit&&\omit&\cr
&& 8 && $\frac1{u} t^{(\alpha_1} \; \eta^{\beta_1) (\alpha_2} 
\; \partial_3^{\beta_2)} \; \eta^{\alpha_3 \beta_3}$ 
&& 29 && $\frac12 \partial_3^{(\alpha_2} \; \eta^{\beta_2) (\alpha_3} 
\; \partial_1^{\beta_3)} \; \eta^{\alpha_1 \beta_1}$ &\cr
\omit&height2pt&\omit&&\omit&&\omit&&\omit&\cr
\tablerule
\omit&height2pt&\omit&&\omit&&\omit&&\omit&\cr
&& 9 && $\frac1{u} t^{(\alpha_2} \; \eta^{\beta_2) (\alpha_3} 
\; \partial_1^{\beta_3)} \; \eta^{\alpha_1 \beta_1}$ 
&& 30 && $\frac12 \partial_1^{(\alpha_3} \; \eta^{\beta_3) (\alpha_1} 
\; \partial_2^{\beta_1)} \; \eta^{\alpha_2 \beta_2}$ &\cr
\omit&height2pt&\omit&&\omit&&\omit&&\omit&\cr
\tablerule
\omit&height2pt&\omit&&\omit&&\omit&&\omit&\cr
&& 10 && $\frac14 \eta^{\alpha_1 \beta_1} \; \partial_3^{(\alpha_2} 
\; \eta^{\beta_2) (\alpha_3} \; \partial_2^{\beta_3)}$ 
&& 31 && $\frac18 \eta^{\alpha_1 \beta_1} \; \eta^{\alpha_2 \beta_2} 
\; \eta^{\alpha_3 \beta_3} \; \partial_2 \cdot \partial_3$ &\cr
\omit&height2pt&\omit&&\omit&&\omit&&\omit&\cr
\tablerule
\omit&height2pt&\omit&&\omit&&\omit&&\omit&\cr
&& 11 && $\frac14 \eta^{\alpha_2 \beta_2} \; \partial_1^{(\alpha_3} 
\; \eta^{\beta_3) (\alpha_1} \; \partial_3^{\beta_1)}$ 
&& 32 && $\frac14 \eta^{\alpha_1 \beta_1} \; \eta^{\alpha_2 \beta_2} 
\; \eta^{\alpha_3 \beta_3} \; \partial_3 \cdot \partial_1$ &\cr
\omit&height2pt&\omit&&\omit&&\omit&&\omit&\cr
\tablerule
\omit&height2pt&\omit&&\omit&&\omit&&\omit&\cr
&& 12 && $\frac14 \eta^{\alpha_3 \beta_3} \; \partial_2^{(\alpha_1} 
\; \eta^{\beta_1) (\alpha_2} \; \partial_1^{\beta_2)}$ 
&& 33 && $-\frac12 \eta^{\alpha_1 (\alpha_2} \; \eta^{\beta_2) \beta_1} 
\; \eta^{\alpha_3 \beta_3} \; \partial_2 \cdot \partial_3$ &\cr
\omit&height2pt&\omit&&\omit&&\omit&&\omit&\cr
\tablerule
\omit&height2pt&\omit&&\omit&&\omit&&\omit&\cr
&& 13 && $-\partial_3^{(\alpha_1} \; \eta^{\beta_1) (\alpha_2} 
\; \eta^{\beta_2) (\alpha_3} \; \partial_2^{\beta_3)}$ 
&& 34 && $-\frac12 \eta^{\alpha_2 (\alpha_3} \; \eta^{\beta_3) \beta_2}
\; \eta^{\alpha_1 \beta_1} \; \partial_3 \cdot \partial_1$ &\cr
\omit&height2pt&\omit&&\omit&&\omit&&\omit&\cr
\tablerule
\omit&height2pt&\omit&&\omit&&\omit&&\omit&\cr
&& 14 && $-\partial_1^{(\alpha_2} \; \eta^{\beta_2) (\alpha_3} 
\; \eta^{\beta_3) (\alpha_1} \; \partial_3^{\beta_1)}$ 
&& 35 && $-\frac12 \eta^{\alpha_3 (\alpha_1} \; \eta^{\beta_1) \beta_3}
\; \eta^{\alpha_2 \beta_2} \; \partial_1 \cdot \partial_2$ &\cr
\omit&height2pt&\omit&&\omit&&\omit&&\omit&\cr
\tablerule
\omit&height2pt&\omit&&\omit&&\omit&&\omit&\cr
&& 15 && $-\partial_2^{(\alpha_3} \; \eta^{\beta_3) (\alpha_1} 
\; \eta^{\beta_1) (\alpha_2} \; \partial_1^{\beta_2)}$ 
&& 36 && $-\frac14 \partial_2^{(\alpha_1} \; \partial_3^{\beta_1)} \;
\eta^{\alpha_2 \beta_2} \; \eta^{\alpha_3 \beta_3}$ &\cr
\omit&height2pt&\omit&&\omit&&\omit&&\omit&\cr
\tablerule
\omit&height2pt&\omit&&\omit&&\omit&&\omit&\cr
&& 16 && $-\frac12 \partial_3^{(\alpha_2} \; \eta^{\beta_2) (\alpha_1} 
\; \eta^{\beta_1) (\alpha_3} \; \partial_2^{\beta_3)}$ 
&& 37 && $-\frac12 \partial_3^{(\alpha_2} \; \partial_1^{\beta_2)} \;
\eta^{\alpha_3 \beta_3} \; \eta^{\alpha_1 \beta_1}$ &\cr
\omit&height2pt&\omit&&\omit&&\omit&&\omit&\cr
\tablerule
\omit&height2pt&\omit&&\omit&&\omit&&\omit&\cr
&& 17 && $-\frac12 \partial_1^{(\alpha_3} \; \eta^{\beta_3) (\alpha_2} 
\; \eta^{\beta_2) (\alpha_1} \; \partial_3^{\beta_1)}$ 
&& 38 && $-\frac18 \eta^{\alpha_1 \beta_1} \; \eta^{\alpha_2 (\alpha_3} 
\; \eta^{\beta_3) \beta_2} \; \partial_2 \cdot \partial_3$ &\cr
\omit&height2pt&\omit&&\omit&&\omit&&\omit&\cr
\tablerule
\omit&height2pt&\omit&&\omit&&\omit&&\omit&\cr
&& 18 && $-\frac12 \partial_2^{(\alpha_1} \; \eta^{\beta_1) (\alpha_3} 
\; \eta^{\beta_3) (\alpha_2} \; \partial_1^{\beta_2)}$ 
&& 39 && $-\frac14 \eta^{\alpha_2 \beta_2} \; \eta^{\alpha_3 (\alpha_1} 
\; \eta^{\beta_1) \beta_3} \; \partial_3 \cdot \partial_1$ &\cr
\omit&height2pt&\omit&&\omit&&\omit&&\omit&\cr
\tablerule
\omit&height2pt&\omit&&\omit&&\omit&&\omit&\cr
&& 19 && $-\frac14 \eta^{\alpha_1 \beta_1} \; \eta^{\alpha_2 \beta_2}
\; \partial_2^{(\alpha_3} \; \partial_3^{\beta_3)}$ 
&& 40 && $\frac12 \eta^{\alpha_1) (\alpha_2} \; \eta^{\beta_2) (\alpha_3} 
\; \eta^{\beta_3) (\beta_1} \; \partial_2 \cdot \partial_3$ &\cr
\omit&height2pt&\omit&&\omit&&\omit&&\omit&\cr
\tablerule
\omit&height2pt&\omit&&\omit&&\omit&&\omit&\cr
&& 20 && $-\frac14 \eta^{\alpha_2 \beta_2} \; \eta^{\alpha_3 \beta_3} 
\; \partial_3^{(\alpha_1} \; \partial_1^{\beta_1)}$ 
&& 41 && $\eta^{\alpha_1) (\alpha_2} \; \eta^{\beta_2) (\alpha_3} \; 
\eta^{\beta_3) (\beta_1} \; \partial_3 \cdot \partial_1$ &\cr
\omit&height2pt&\omit&&\omit&&\omit&&\omit&\cr
\tablerule
\omit&height2pt&\omit&&\omit&&\omit&&\omit&\cr
&& 21 && $-\frac14 \eta^{\alpha_3 \beta_3} \; \eta^{\alpha_1 \beta_1} 
\; \partial_1^{(\alpha_2} \; \partial_2^{\beta_2)}$ 
&& 42 && $\frac14 \partial_2^{(\alpha_1} \; \partial_3^{\beta_1)} \;
\eta^{\alpha_2 (\alpha_3} \; \eta^{\beta_3) \beta_2}$ &\cr
\omit&height2pt&\omit&&\omit&&\omit&&\omit&\cr
\tablerule
\omit&height2pt&\omit&&\omit&&\omit&&\omit&\cr
&& \omit && \omit 
&& 43 && $\frac12 \partial_3^{(\alpha_2} \; \partial_1^{\beta_2)} 
\; \eta^{\alpha_3 (\alpha_1} \; \eta^{\beta_1) \beta_3}$ &\cr 
\omit&height2pt&\omit&&\omit&&\omit&&\omit&\cr
\tablerule}}

\vskip 0.5cm 

{\bf Table~3:} {\ninepoint Partially symmetrized cubic pseudo-graviton 
vertex operators with vertex line \#1 
\vskip -0.65truecm \noindent \hglue 2.8truecm distinguished and without 
the factor of $\kappa \Omega^2$.}

\vfill\eject

\noindent The diagram of Fig.~4b has a ghost inner loop and gives:
$$\eqalignno{ - \kappa \;\; &\int_u^{H^{-1}} du' \int d^3r' \; \int_{u}^{H^{-1
}} du'' \int d^3r'' \;\; \sum_{i=1}^{43} \; \sum_{j,k=1}^{10} \; \cr 
& V_i^{\mu \nu \mu_2\nu_2 \mu_3\nu_3}(x\; ;\; \partial_1,\partial_2,\partial_3)
\;\; i\Bigl[ {_{\mu_2 \nu_2}}\Delta_{\alpha_1 \beta_1} \Bigr](x \; ; x') \;\; 
i\Bigl[ {_{\mu_3 \nu_3}}\Delta_{\rho_1 \sigma_1} \Bigr](x \; ; x'') \cr 
& \Biggl\{ \; V_j^{\alpha_1 \beta_1 \alpha_2 \alpha_3}(x'; \; \partial_1' , 
\partial_2' , \partial_3') \;\;  i\Bigl[ {_{\alpha_2}}\Delta_{\rho_2} \Bigr]
(x' ; x'') \;\; i\Bigl[ {_{\alpha_3}}\Delta_{\rho_3} \Bigr](x' ; x'') \cr 
& V_k^{\rho_1 \sigma_1 \rho_2 \rho_3}(x''; \; \partial_1'' , \partial_2'' , 
\partial_3'')  \; \Biggr\} \;\; . &(3.9b) \cr}$$ 
And the diagram of Fig.~4c has ghosts on the outer legs and takes the form:
$$\eqalignno{ - \kappa \;\; &\int_u^{H^{-1}} du' \int d^3r' \; \int_{u}^{
H^{-1}} du'' \int d^3r'' \;\; \sum_{i,j,k=1}^{10} \; \cr 
& V_i^{\mu \nu \mu_2 \mu_3}(x \; ; \; \partial_1 , \partial_2 , \partial_3) 
\;\;  i\Bigl[ {_{\mu_2}}\Delta_{\alpha_1} \Bigr](x \; ; x') \;\; 
i\Bigl[ {_{\mu_3}}\Delta_{\rho_1} \Bigr](x \; ; x'') \cr 
& \Biggl\{ \; V_j^{\alpha_1 \alpha_2 \beta_2 \alpha_3}(x'; \; \partial_1' , 
\partial_2' , \partial_3') \;\;  i\Bigl[ {_{\alpha_2 \beta_2}}\Delta_{\rho_2 
\sigma_2} \Bigr](x' ; x'') \;\; \cr 
& i\Bigl[ {_{\alpha_3}}\Delta_{\rho_3} \Bigr](x' ; x'') \;\; V_k^{\rho_1 \rho_2
\sigma_2 \rho_3}(x''; \; \partial_1'' , \partial_2'' , \partial_3'')  
\; \Biggr\} \;\; . &(3.9c) \cr}$$ 
The meaning of the various derivatives is determined by the subscripts --- 
indicating which propagator is being acted upon --- and by the number of primes
--- telling with respect to which coordinate the derivative is being taken.
For example, the derivative $\partial_2'$ of $V_j^{\alpha_1\beta_1\alpha_2
\beta_2\alpha_3\beta_3}(x';\partial_1',\partial_2',\partial_3')$ in (3.9a) 
differentiates the propagator $i[{_{\alpha_2\beta_2}}\Delta_{\rho_2\sigma_2}]
(x';x'')$ with respect to ${x'}^{\mu}$. The overall factor of $\kappa$ in each 
expression is the one appearing in our definition (2.19) of the amputated 
1-point function. The sign of (3.9a) is positive because each vertex comes with
a factor of $+i$ and there is an additional factor of $+i$ from amputation:  
$$D^{\mu \nu \kappa \lambda} \; 
i \Bigl[{_{\kappa \lambda}} \Delta_{\mu_1 \nu_1} \Bigr] (x ; x') = 
i \; \delta^{\mu}_{~(\mu_1} \; \delta_{\nu_1)}^{~~~\nu} \;\; 
\delta^{(4)} (x-x') \;\; . \eqno(3.10)$$ 
Expressions (3.9b) and (3.9c) are negative on account of the same factors with
the Fermi minus sign of their single ghost loops.

The various 3-3-3 diagrams were computed in six steps using the symbolic
manipulation program Mathematica [12]. For diagrams (4a) and (4b) we first
performed the tensor algebra and acted the derivatives of the inner loop, then 
did the tensor algebra and derivatives of the outer loop and, finally, 
performed the integrations. The sums in diagram (4c) were short enough that we 
could do the contractions and act the derivatives for both loops at the same
time. The tensor algebra was performed using FeynCalc [13] loaded in 
Mathematica.\footnote{*}{\tenpoint It is not very efficient to use FeynCalc 
this way because most of the 2 megabytes memory of the program goes to define 
operations we never need. After completing the calculation we wrote a very 
short contraction program in collaboration with G. G. Huey which reproduces 
the tiny portion of FeynCalc we actually used.}

\noindent {\it 3.1 Inner Loop Tensor Algebra and Derivatives.}

The inner loop is defined by the propagators which link ${x'}^{\mu}$ and 
${x''}^{\mu}$, and by the inner loop derivatives $\partial_{2-3}'$ and 
$\partial_{2-3}''$.  There are eight inner loop indices to contract in 
diagram (4a) and four in (4b), and there can be anywhere from zero to four 
inner loop derivatives to act. The outer derivatives, $\partial_1'$ and 
$\partial_1''$ are retained as free vector operators at this stage. 

What one gets after contracting the inner loop indices and acting the inner 
loop derivatives is a series of scalar functions times the 4-index objects 
which we denote by the symbol, $\Pi_A^{\alpha_1\beta_1\rho_1\sigma_1}$. These 
are the 79 independent 4-index objects, with the appropriate symmetries, 
which can be formed from $\eta^{\mu\nu}$, $y^{\mu}$, $t^{\mu}$, and up to 
one of each outer derivative. It is useful to categorize the scalar functions 
by which, if any, contracted outer derivatives they possess. What remains 
are functions of $u'$, $u''$ and $y^2$. We call these coefficient functions, 
$C_{mA}(u',u'',y^2)$, where the index $m$ stands for which of the ten 
contracted external derivatives multiplies the function and the index $A$ 
stands for which of the 79 4-index objects the combination multiplies. We 
therefore reach expressions of the following form for both (4a) and (4b):
$$\eqalignno{ \kappa \;\; &\int_u^{H^{-1}} du' \int d^3r' \; 
\int_{u}^{H^{-1}} du'' \int d^3r'' \;\; \sum_{i=1}^{43} \; 
V_i^{\mu \nu \mu_2 \nu_2 \mu_3 \nu_3}
(x;\partial_1,\partial_2,\partial_3) \cr
&i\Bigl[ {_{\mu_2 \nu_2}}\Delta_{\alpha_1 \beta_1} \Bigr](x;x') \;\; 
i\Bigl[ {_{\mu_3 \nu_3}}\Delta_{\rho_1 \sigma_1} \Bigr](x;x'') \; 
\sum_{A=1}^{79} \Pi_A^{\alpha_1 \beta_1\rho_1 \sigma_1} \cr
&\Biggl\{ C_{1A} + t \cdot \partial_1' \; C_{2A} + 
y \cdot \partial_1' \; C_{3A} + t \cdot \partial_1'' \; C_{4A} + 
y \cdot \partial_1'' \; C_{5A} \cr
&\;\; + (t \cdot \partial_1') \; (t \cdot \partial_1'') \; C_{6A} +
(t \cdot \partial_1') \; (y \cdot \partial_1'') \; C_{7A} + 
(y \cdot \partial_1') \; (t \cdot \partial_1'') \; C_{8A} \cr
&\;\; + (y \cdot \partial_1') \; (y \cdot \partial_1'') \; C_{9A} + 
\partial_1' \cdot \partial_1'' \; C_{10A} \Biggr\} 
\;\; . &(3.11) \cr}$$
What we call ``the inner loop'' part begins with the sum over $A$.

As an example, consider the contribution to the inner loop of (4a) from the
vertex $j=k= 41$ of Table~3:
$$\eqalignno{ {\rm (inner)_{41,41}} \equiv & 
\; \Bigl\{ \; 
V_{41}^{\alpha_1 \beta_1 \alpha_2 \beta_2 \alpha_3 \beta_3}
(x'; \; \partial_1' , \partial_2' , \partial_3') \;\;  
i\Bigl[ {_{\alpha_2 \beta_2}}\Delta_{\rho_2 \sigma_2} \Bigr]
(x' ; x'') \;\; 
i\Bigl[ {_{\alpha_3 \beta_3}}\Delta_{\rho_3 \sigma_3} \Bigr]
(x' ; x'') \cr 
& \qquad V_{41}^{\rho_1 \sigma_1 \rho_2 \sigma_2 \rho_3 \sigma_3}
(x''; \; \partial_1'' , \partial_2'' , \partial_3'') 
\; \Bigr\} \cr 
= & \; \kappa^2  \; {\Omega'}^2 \; {\Omega''}^2 \; 
\partial_1' \cdot \partial_3' \enskip 
\partial_1'' \cdot \partial_3'' \enskip 
\eta^{\alpha_1) (\alpha_2} \; \eta^{\beta_2) (\alpha_3} \; 
\eta^{\beta_3) (\beta_1} \enskip 
\eta^{\rho_1) (\rho_2} \; \eta^{\sigma_2) (\rho_3} \; 
\eta^{\sigma_3) (\sigma_1} \cr
&\;\;\; \Bigl\{ i\Delta_{2N} \; 
\Bigl[ 2 \eta_{\alpha_2 (\rho_2} \; \eta_{\sigma_2) \beta_2} 
- \eta_{\alpha_2 \beta_2} \; \eta_{\rho_2 \sigma_2} \Bigr] \cr 
&\;\;\;\; - i\Delta_{2L} \; 
\Bigl[ 2 {\overline \eta}_{\alpha_2 (\rho_2} \; 
{\overline \eta}_{\sigma_2) \beta_2} 
- 2 {\overline \eta}_{\alpha_2 \beta_2} \; 
{\overline \eta}_{\rho_2 \sigma_2} \Bigr] 
\Bigr\} \cr
&\;\;\; \Bigl\{ i\Delta_{3N} \; 
\Bigl[ 2 \eta_{\alpha_3 (\rho_3} \; \eta_{\sigma_3) \beta_3} 
- \eta_{\alpha_3 \beta_3} \; \eta_{\rho_3 \sigma_3} \Bigr] \cr 
&\;\;\;\; - i\Delta_{3L} \; 
\Bigl[ 2 {\overline \eta}_{\alpha_3 (\rho_3} \; 
{\overline \eta}_{\sigma_3) \beta_3} 
- 2 {\overline \eta}_{\alpha_3 \beta_3} \; 
{\overline \eta}_{\rho_3 \sigma_3} \Bigr] 
\Bigr\} 
\;\; . \qquad &(3.12) \cr}$$ 
After the contractions we obtain:
$$\eqalignno{ {\rm (inner)_{41,41}} = & 
\; \kappa^2  \; {\Omega'}^2 \; {\Omega''}^2 \;
\partial_1' \cdot \partial_3' \enskip 
\partial_1'' \cdot \partial_3'' 
\; \Biggl\{
i\Delta_{2N} \;\; i\Delta_{3N} \; 
\Bigl[ 2 \eta^{\alpha_1\beta_1} \; \eta^{\rho_1 \sigma_1} + 
2 \eta^{\alpha_1 (\rho_1} \; \eta^{\sigma_1) \beta_1} \Bigr] \cr 
& + \Bigl( i\Delta_{2N} \;\; i\Delta_{3L} + 
i\Delta_{2L} \;\; i\Delta_{3N} \Bigr) \;
\Bigl[ -3 \eta^{\alpha_1\beta_1} \; \eta^{\rho_1\sigma_1}
+ \eta^{\alpha_1 (\rho_1} \; \eta^{\sigma_1) \beta_1} \cr 
& \;\;\;\; - 3 \eta^{\alpha_1 \beta_1} t^{\rho_1} t^{\sigma_1}  
- 3 t^{\alpha_1} t^{\beta_1} \; \eta^{\rho_1 \sigma_1} 
+ 4 t^{(\alpha_1} \eta^{\beta_1) (\rho_1} t^{\sigma_1)} \Bigr] \cr
& + i\Delta_{2L} \;\; i\Delta_{3L} \; 
\Bigl[ 5 \eta^{\alpha_1\beta_1} \; \eta^{\rho_1\sigma_1} 
- 3 \eta^{\alpha_1 (\rho_1} \; \eta^{\sigma_1) \beta_1} 
+ 5 \eta^{\alpha_1 \beta_1} t^{\rho_1} t^{\sigma_1} \cr 
& \;\;\;\; + 5 t^{\alpha_1} t^{\beta_1} \eta^{\rho_1 \sigma_1} 
- 6 t^{(\alpha_1} \eta^{\beta_1) (\rho_1} t^{\sigma_1)} 
+ 2 t^{\alpha_1} t^{\beta_1} t^{\rho_1} t^{\sigma_1} \Bigr] 
\Biggr\} 
\;\; . \qquad &(3.13) \cr}$$ 
At this stage we recognize six of the 4-index objects:
$$\Pi_1^{\alpha_1\beta_1\rho_1\sigma_1} \equiv 
\eta^{\alpha_1\beta_1} \; \eta^{\rho_1\sigma_1} \qquad , \qquad 
\Pi_4^{\alpha_1\beta_1\rho_1\sigma_1} \equiv 
t^{\alpha_1} \; t^{\beta_1} \; \eta^{\rho_1\sigma_1} 
\;\; ; \eqno(3.14a)$$
$$\Pi_2^{\alpha_1\beta_1\rho_1\sigma_1} \equiv 
\eta^{\alpha_1(\rho_1} \; \eta^{\sigma_1)\beta_1} \qquad , \qquad 
\Pi_9^{\alpha_1\beta_1\rho_1\sigma_1} \equiv 
t^{(\alpha_1} \; \eta^{\beta_1)(\rho_1} \; t^{\sigma_1)} 
\;\; ; \eqno(3.14b)$$
$$\Pi_3^{\alpha_1\beta_1\rho_1\sigma_1} \equiv 
\eta^{\alpha_1\beta_1} \; t^{\rho_1} \; t^{\sigma_1} \qquad , \qquad 
\Pi_{13}^{\alpha_1\beta_1\rho_1\sigma_1} \equiv 
t^{\alpha_1} \; t^{\beta_1} \; t^{\rho_1} \; t^{\sigma_1} 
\;\; . \eqno(3.14c)$$
There does not seem to be any point in giving the entire list of the 79 
objects since we will not use them further. Dropping the indices and 
acting the derivatives gives:
$$\eqalignno{ {\rm (inner)_{41,41}} & = 
{\kappa^2 \over 2^6 \pi^4} \; \Biggl\{ \Bigl( 
- \frac{32 x\cdot\partial_1' \; x\cdot\partial_1''}{x^8} 
+ \frac{8 \partial_1'\cdot\partial_1''}{x^6} 
+ \frac{8 x\cdot\partial_1' \; t\cdot\partial_1''}{u'' x^6} 
- \frac{8 t\cdot\partial_1' \; x\cdot\partial_1''}{u' x^6} 
+ \frac{4 t\cdot\partial_1' \; t\cdot\partial_1''}{u' u'' x^4}
\Bigr) 
\Bigl[ 2 \Pi_1 + 2 \Pi_2 \Bigr] \cr
&\qquad\qquad + \Bigl[ \Bigl( 
- \frac{16 x\cdot\partial_1' \; x\cdot\partial_1''}{u' u'' x^6} 
+ \frac{4 \partial_1'\cdot\partial_1''}{u' u'' x^4} 
+ \frac{4 x\cdot\partial_1' \; t\cdot\partial_1''}{u' {u''}^2 x^4} 
- \frac{4 t\cdot\partial_1' \; x\cdot\partial_1''}{{u'}^2 u'' x^4} 
+ \frac{2 t\cdot\partial_1' \; t\cdot\partial_1''}{{u'}^2 {u''}^2 x^2}
\Bigr) \ln (H^2 x^2) \cr
&\qquad \qquad\qquad + \Bigl(
- \frac{4 \partial_1'\cdot\partial_1''}{u' u'' x^4} 
+ \frac{8 x\cdot\partial_1' \; x\cdot\partial_1''}{u' u'' x^6}
\Bigr) \Bigl] \Bigl[
-3 \Pi_1 + \Pi_2 - 3 \Pi_3 - 3 \Pi_4 + 4 \Pi_9 
\Bigr] \cr
&\qquad\qquad + \Bigl(
- \frac{2 \partial_1'\cdot\partial_1''}{{u'}^2 {u''}^2 x^2} 
+ \frac{4 x\cdot\partial_1' \; x\cdot\partial_1''}{{u'}^2 {u''}^2 x^4}
\Bigr) \ln (H^2 x^2) \cr 
&\qquad\qquad\qquad \Bigl[ 
5 \Pi_1 - 3 \Pi_2 + 5 \Pi_3 + 5 \Pi_4 - 6 \Pi_9 + 2 \Pi_{13} 
\Bigr] 
\Biggl\} \;\; , \qquad &(3.15) \cr}$$
from which we recognize 27 non-zero coefficient functions. A few examples are:
$$C_{61} (u',u'',y^2) = 
- \frac{64}{y^8} - \frac{24}{u' u'' y^6} + 
\frac{48 \ln(H^2 y^2)}{u' u'' y^6} + 
\frac{20 \ln(H^2 y^2)}{{u'}^2 {u''}^2 y^4}
\;\; , \eqno(3.16a)$$
$$C_{82} (u',u'',y^2) = 
\frac{16}{u'' y^6} + \frac{4 \ln(H^2 y^2)}{ u' {u''}^2 y^4} 
\;\; , \eqno(3.16b)$$
$$C_{1013} (u',u'',y^2) = 
- \frac{4 \ln(H^2 y^2)}{{u'}^2 {u''}^2 y^2}
\;\; . \eqno(3.16c)$$

\noindent {\it 3.2 Outer Loop Tensor Algebra and Derivatives.} 

What we mean by the outer loop tensor algebra is the contractions over 
$\alpha_1$, $\beta_1$; $\rho_1$, $\sigma_1$; $\mu_{2-3}$ and $\nu_{2-3}$. What 
we mean by the outer loop derivatives are $\partial_1'$, $\partial_1''$ and
$\partial_{2-3}$. The single possible derivative on the external leg acts
backwards on the entire expression, including the limits of integration. Only 
its 0-component survives:
$$\partial_1^{\mu} \longrightarrow t^{\mu} \partial_u \;\; . \eqno(3.17)$$
The great advantage of the representation (3.11) we use for the inner loop is
that it isolates the relatively simple dependence upon derivatives and indices
from the very complicated coefficient functions $C_{mA}(u',u'',y^2)$. We can 
contract the indices and act the various derivatives for each coefficient 
function without having to worry about its functional form. An additional
simplification is that we can contract the external indices, $\mu$ and $\nu$, 
into the two independent tensors --- $\eta_{\mu\nu}$ and $t_{\mu} \; t_{\nu}$ 
--- which survive the integrations. So the outer loop result can be expressed 
in terms of $2 \times (10 \times 79) = 1580$ outer coefficient functions:
$$\eqalignno{\alpha_{mA} (u,u',u'',w^2,y^2,z^2;\partial_u) 
\equiv &\eta_{\mu\nu} \; \sum_{i=1}^{43} \; 
V_i^{\mu\nu\mu_2\nu_2\mu_3\nu_3}(x; \partial_1, \partial_2, \partial_3) \cr
&{\cal D}_m(y,\partial_1',\partial_1'') \; 
\Pi_A^{\alpha_1\beta_1\rho_1 \sigma_1}(y,\partial_1',\partial_1'') \cr
&i\Bigl[ {_{\mu_2 \nu_2}} \Delta_{\alpha_1 \beta_1}\Bigr](x;x') \;\; 
i\Bigl[ {_{\mu_3 \nu_3}} \Delta_{\rho_1 \sigma_1} \Bigr](x;x'') 
\;\; , \qquad &(3.18a) \cr }$$
$$\eqalignno{\gamma_{mA} (u,u',u'',w^2,y^2,z^2;\partial_u) 
\equiv &t_{\mu} \; t_{\nu} \; \sum_{i=1}^{43} \; 
V_i^{\mu\nu\mu_2\nu_2\mu_3\nu_3}(x; \partial_1, \partial_2, \partial_3) \cr
&{\cal D}_m(y,\partial_1',\partial_1'') \; 
\Pi_A^{\alpha_1\beta_1\rho_1 \sigma_1}(y,\partial_1',\partial_1'') \cr
&i\Bigl[ {_{\mu_2 \nu_2}} \Delta_{\alpha_1 \beta_1}\Bigr](x;x') \;\; 
i\Bigl[ {_{\mu_3 \nu_3}} \Delta_{\rho_1 \sigma_1} \Bigr](x;x'') 
\;\; . \qquad &(3.18b) \cr }$$
The ten contracted derivatives ${\cal D}_m(y,\partial_1',\partial_1'')$ are
defined in expression (3.11). Because no derivative can appear more than once,
the inner coefficient function $C_{mA}$ is automatically zero for those 
combinations of $m$ and $A$ where either $\partial_1'$ or $\partial_1''$ 
appears in both the contracted derivative and the 4-index object, $\Pi_A^{
\alpha_1\beta_1\rho_1 \sigma_1}(y,\partial_1',\partial_1'')$. This means that 
only 696 of the coefficients (3.18) were actually used, so of course only these
were computed.

We should comment on how the outer coefficient functions (3.18) can be reduced
to dependence upon only the three conformal times --- $u$, $u'$, and $u''$ ---
and the three Lorentz squares of (3.2) --- $w^2$, $y^2$ and $z^2$. First, one 
can re-express contractions of spatial vectors in terms of Lorentz contractions
and 0-components:
$${\vec a} \cdot {\vec b} = a \cdot b + a^0 b^0 \;\; . \eqno(3.19)$$
Second, contractions involving $t_{\mu} \equiv \eta_{\mu 0}$ can be expressed
as follows:
$$t \cdot t = -1 \;\; ; \eqno(3.20a)$$
$$t \cdot w = u' - u \qquad , \qquad t \cdot y = u'' - u' \qquad , \qquad
t \cdot z = u - u'' \;\; . \eqno(3.20b)$$
Finally, dot products between the three coordinate differences can always be
expressed in terms of the Lorentz squares:
$$\eqalignno{ 
& w \cdot y = \frac12 (-w^2 - y^2 + z^2) \;\; , \cr 
& y \cdot z = \frac12 (w^2 - y^2 - z^2) \;\; , \cr 
& z \cdot w = \frac12 (-w^2 + y^2 - z^2) \;\; . 
&(3.21) \cr}$$

Because the outer coefficient functions involve a sum over the 43 outer vertex
operators, they often have lengthy expressions. One of reasonable size is
$\gamma_{113}$:
$$\eqalignno{ \gamma_{113} (u,u',u'',w^2,y^2,z^2;\partial_u) 
=& \frac{\kappa H^2}{2^6 \pi^4}\;\partial_u\; \Bigl\{ 
\frac{12 u' u''}{u w^2 z^2} + \frac{8 (u'-u) u' u''}{w^4 z^2} 
- \frac{8 u' (u''-u) u''}{w^2 z^4} \Bigr\} \cr
&+ \frac{\kappa H^2}{2^6 \pi^4} \; \Bigl\{
- \frac{3 u' u''}{u^2 w^2 z^2} - \frac{10 (u'-u) u' u''}{u w^4 z^2} 
+ \frac{22 u' (u''-u) u''}{u w^2 z^4} \cr
&\qquad\quad\;\; - \frac{24 (u'-u) u' (u''-u) u''}{w^4 z^4} + 
\frac{6 u' u'' (-w^2 + y^2 - z^2)}{w^4 z^4} \Bigr\} 
\;\; . \qquad\qquad &(3.22) \cr}$$
Terms containing single factors of $\ln(H^2 w^2)$ and/or $\ln(H^2 z^2)$ can
also appear but in these cases the outer loop coefficient functions are too
lengthy to display. Note that the external leg derivative $\partial_u$ is
really still a free operator at this stage since it can also act on the 
$u$-dependence in the limits of the conformal time integrations.

The $i=34$ contribution to $\alpha_{51}$ is in some ways more representative 
of what is involved in computing an outer loop coefficient function:
$$\eqalignno{ \alpha_{51} =& \eta_{\mu\nu} \; 
V_{34}^{\mu\nu\mu_2\nu_2\mu_3\nu_3}
(x; \partial_1,\partial_2,\partial_3) \;\; 
{\cal D}_5(y,\partial_1',\partial_1'') \;\;
\Pi_1^{\alpha_1\beta_1 \rho_1\sigma_1}(y,\partial_1',\partial_1'') \cr
&i\Bigl[ {_{\mu_2 \nu_2}}\Delta_{\alpha_1 \beta_1}\Bigr](x;x') \;\; 
i\Bigl[ {_{\mu_3 \nu_3}}\Delta_{\rho_1 \sigma_1}\Bigr](x;x'') \cr 
=& -2 \kappa \; \Omega^2 \; \eta^{\mu_2 (\mu_3} \; 
\eta^{\nu_3) \nu_2} \; \partial_u \;\; t\cdot\partial_3 \;\; 
y\cdot\partial_1'' \; \eta^{\alpha_1 \beta_1} \; \eta^{\rho_1\sigma_1} \cr
&\Bigl\{ i\Delta_{2N} \; \Bigl[ 
2 \; \eta_{\mu_2 (\alpha_1} \; \eta_{\beta_1) \nu_2} 
- \eta_{\mu_2 \nu_2} \; \eta_{\alpha_1 \beta_1} \Bigr] \cr
&\; - i\Delta_{2L} \; \Bigl[ 
2 \; {\overline \eta}_{\mu_2 (\alpha_1} \; 
{\overline \eta}_{\beta_1) \nu_2} 
- 2 \; {\overline \eta}_{\alpha_1 \beta_1} \; 
{\overline \eta}_{\alpha_1 \beta_1} \Bigr] \Bigr\} \cr
&\Bigl\{ i\Delta_{3N} \; \Bigl[ 
2\; \eta_{\mu_3 (\rho_1} \; \eta_{\sigma_1) \nu_3} 
- \eta_{\mu_3 \nu_3} \; \eta_{\rho_1 \sigma_1} \Bigr] \cr
&\; - i\Delta_{3L} \; \Bigl[ 
2 \; {\overline \eta}_{\mu_3 (\rho_1} \; 
{\overline \eta}_{\sigma_1) \nu_3} 
- 2 \; {\overline \eta}_{\mu_3 \nu_3} \; 
{\overline \eta}_{\rho_1 \sigma_1} \Bigr] \Bigr\} &(3.23a) \cr
=& -2 \kappa \; \Omega^2 \; \partial_u \;\; 
t\cdot\partial_3 \;\; y\cdot\partial_1'' \;
\Bigl\{ 16 \; i\Delta_{2N} \; i\Delta_{3N} \cr
&- 48 \; \Bigl( i\Delta_{2N} \; i\Delta_{3L} + 
i\Delta_{2L} \; i\Delta_{3N} \Bigr) + 
48 \; i\Delta_{2L} \; i\Delta_{3L}\Bigr\} &(3.23b) \cr
=& \frac{\kappa H^2}{2^6 \pi^4} \; \Biggl\{
-\frac{256 u' (u''-u')}{u w^2 z^2} 
- \frac{256 u' (u''-u)}{w^2 z^4} 
+ \frac{128 u' u'' (w^2-y^2-z^2)}{u w^2 z^4} \cr
&\qquad\quad - \frac{384 u' (u''-u)(w^2-y^2-z^2)}{u w^2 z^4} 
- \frac{512 u' (u''-u) u'' (w^2-y^2-z^2)}{w^2 z^6} \cr
&\qquad\quad + \partial_u \; \Bigl[
\frac{384 (u''-u') }{u^2 z^2} + \frac{384 u'(u''-u)}{u z^4}
- \frac{192 u'' (w^2-y^2-z^2)}{u^2 z^4} \cr
&\qquad\quad + \frac{192 (u''-u) (w^2-y^2-z^2)}{u^2 z^4} 
+ \frac{768 (u'' -u) u'' (w^2-y^2-z^2)}{u z^6} \Bigr] \; \ln(H^2 w^2) 
\Biggr\} \;\; . \qquad &(3.23c) \cr}$$
Note again that $\partial_u$ is a free operator at this stage. It will 
eventually act on all $u$'s, including those in the limits of integration.

\noindent {\it 3.3 Re-Organization of the Integrals.}

By adding to a generic integrand $I(x,x',x'')$ its reflection under 
${x'}^{\mu} \leftrightarrow {x''}^{\mu}$, we can make the integrand 
symmetric. We can then enforce a canonical time ordering, $u \leq u' 
\leq u'' \leq H^{-1}$, by splitting the conformal time integrations 
into halves with $u'$ before and after $u''$, and changing variables:
$$\eqalignno{ \int_u^{H^{-1}} &du' \int d^3r' \; 
\int_u^{H^{-1}} du'' \int d^3r''\; I(x,x',x'') \cr
&= \frac12 \int_u^{H^{-1}} du' \int d^3r' \; 
\int_u^{H^{-1}} du'' \int d^3r'' \;
\Bigl[ I(x,x',x'') + I(x,x'',x') \Bigr] \qquad &(3.24a)\cr 
&= \int_u^{H^{-1}} du' \int d^3r' \; 
\int_{u'}^{H^{-1}} du'' \int d^3r'' \; 
\Bigl[ I(x,x',x'') + I(x,x'',x') \Bigr] \;\; . \qquad &(3.24b) \cr}$$
The advantage of ordering the conformal times is that the three $d$'s can be 
written without using absolute value symbols. With the ordering chosen, they
are as given on Table~1.

The generic integrands for each 3-3-3 diagram can be obtained by contracting 
the integrands of expressions (3.9a-c) alternately with $\eta_{\mu\nu}$ (to 
give the $\alpha$-contraction) and with $t_{\mu} \; t_{\nu}$ (to give the 
$\gamma$-contraction). We can express the integrands for diagrams (4a) and (4b)
by summing over the inner and outer loop coefficient functions:
$$\eqalignno{ \Biggl[ {\alpha \atop \gamma} \Biggr]
(u,u',&u'',w^2,y^2,z^2;\partial_u) \cr
&= \kappa \; \sum_{m=1}^{10} \sum_{A=1}^{79} \; 
{\Biggl[ {\alpha \atop \gamma} \Biggr]}_{mA}
( u,u',u'',w^2,y^2,z^2;\partial_u) \;\; 
C_{mA} (u',u'',y^2) \;\; . \qquad &(3.25) \cr}$$
The analogous integrands for diagram (4c) are simple enough to compute without 
distinguishing between inner and outer loop tensor algebra and derivatives. 
Performing the integrals gives each diagram's contribution to the $\alpha$ and 
$\gamma$ contractions of the amputated 1-point function:
$$\eqalignno{ \Biggl[ {\alpha \atop \gamma} \Biggr] (u) \equiv
\int_u^{H^{-1}} &du' \int d^3x' \; 
\int_{u'}^{H^{-1}} du'' \int d^3x'' \cr
&\Biggl\{ \; \Biggl[ {\alpha \atop \gamma} \Biggr]
(u,u',u'',w^2,y^2,z^2,\partial_u) + 
\Biggl[ {\alpha \atop \gamma} \Biggr]
(u,u'',u',z^2,y^2,w^2;\partial_u) \Biggr\} 
\;\; . \qquad\qquad &(3.26) \cr}$$
(Note that the free operator $\partial_u$ acts {\it outside} the integrals.)
And the diagram's contributions to the coefficients $a(u)$ and $c(u)$ of (2.19)
are:
$$a(u) = \frac13 \Bigl[ \alpha(u) + \gamma(u) \Bigr]
\;\; , \eqno(3.27a)$$
$$c(u) = \gamma(u) \;\; . \eqno(3.27b)$$

It is useful to extract the universal constant factors which result from the 
initial $\kappa$, the three vertices, and the four propagators:
$$\kappa \times \Bigl({\kappa \over H^2}\Bigr)^3 
\times \Bigl({H^2 \over 2^3 \pi^2}\Bigr)^4 = 
{\kappa^4 H^2 \over 2^{12} \pi^8} \;\; . \eqno(3.28)$$
It is also useful is to split the various integrands up according to which, if 
any, of the three possible logarithms they contain. Recall from sub-section 2.9
that at most one logarithm from the propagators can survive differentiation. We
therefore have four possibilities: a single factor of $\ln(H^2 w^2)$, a single 
factor of $\ln(H^2 y^2)$, a single factor of $\ln(H^2 z^2)$, or no logarithms 
at all. A final re-organization is to break the integrands up into monomials 
of the three conformal times and the three Lorentz squares. What results is a 
dauntingly diverse series of integrals having the general form:
$$\eqalignno{ \# \; {\kappa^4 H^2 \over 2^{12} \pi^8} \; 
(\partial_u)^{0,1} \int_u^{H^{-1}} &du' \int_{u'}^{H^{-1}} du'' 
{1 \over (u)^i \; (u')^j \; (u'')^k} \cr
& \int d^3r' \int d^3r'' \;\; 
{\ln^{0,1} \Bigl[ H^2 \; (w^2,y^2,z^2) \Bigr] \over 
(w^2)^{\ell} \; (y^2)^m \; (z^2)^n} \;\; . &(3.29) \cr}$$
The exponents of the various conformal times may be negative, and are always
$\leq 2$. The exponents of the Lorentz squares may also be negative. Those
for $w^2$ and $z^2$ can range from $+3$ to $-2$, while the one for $y^2$ can
range from $+6$ to $-2$. It should also be noted that dimensional analysis
constrains the sum $[i+j+k+2(\ell+m+n)]$ to be either eleven, if there is an
external $\partial_u$, or twelve if there is not.

\noindent {\it 3.4 The Angular Integrations.} 

When the vectors ${\vec r}\ '$ and ${\vec r}\ ''$ are written using a polar 
coordinate system in which the $z$ axis of ${\vec r}\ ''$ is along 
${\vec r}\ '$, the three Lorentz squares have the following simple expressions:
$$w^2 = {r'}^2 - (d_w - i \epsilon)^2 = 
(r'-d_w+i\epsilon) \; (r'+d_w-i\epsilon)
\;\; , \eqno(3.30a)$$
$$z^2 = {r''}^2 - (d_z - i \epsilon)^2 = 
(r''-d_z+i\epsilon) \; (r''+d_z-i\epsilon)
\;\; , \eqno(3.30b)$$
$$y^2 = {r'}^2 - 2 r' r'' \cos(\theta'') + {r''}^2 - (d_y - i \epsilon)^2
\;\; . \eqno(3.30c)$$
Since the various integrands depend only on the ${\vec r}\ ''$ zenith angle, we 
can perform the other angular integrations trivially:
$$\int_0^{2\pi} d\phi' \; \int_0^{\pi} d\theta' \; \sin(\theta') \; \int_0^{2
\pi} d\phi'' = 8 \pi^2 \;\; . \eqno(3.31)$$
This factor is universal and we multiply it into the other universal factors
(3.28). When there is no $\ln(H^2 y^2)$ and $m\neq 1$ the integral over 
$\theta''$ gives:
$$\eqalignno{ 
\int_0^{\pi} d\theta'' \; {\sin(\theta'') \over (y^2)^m} &= 
{-1 \over 2 (m-1) r' r''} \Bigl[
{r'}^2 - 2 r' r'' \cos(\theta'') + (r'')^2 - (d_y - i\epsilon)^2
\Bigr]^{1-m} \qquad &(3.32a) \cr
&={-1 \over 2 (m-1) r' r''} \Biggl\{ 
\Bigl[(r' + r'')^2 - (d_y-i \epsilon)^2 \Bigr]^{1-m} \cr
&\qquad\qquad\qquad\qquad
- \Bigl[ (r' - r'')^2 - (d_y - i\epsilon)^2 \Bigr]^{1-m} 
\Biggr\} \;\; . \qquad &(3.32b) \cr}$$
When there is a $\ln(H^2 y^2)$ and $m\neq 1$ the integral over $\theta''$ 
gives:
$$\eqalignno{\int_0^{\pi} &d\theta'' \; \sin(\theta'') \; 
{\ln(H^2 y^2) \over (y^2)^m} \cr
&= {-1 \over 2 (m-1) r' r''} \Biggl\{ 
{\ln\Bigl[ \Bigl((r'+r'')^2 - (d_y - i\epsilon)^2\Bigr) \Bigr] 
+ \frac1{m-1} \over [(r'+r'')^2 - (d_y - i\epsilon)^2]^{m-1}} 
- \Bigl( r'' \rightarrow -r''\Bigr) \Biggr\} 
\;\; . \qquad &(3.33) \cr}$$
When $m=1$ one gets a new logarithm:
$$\int_0^{\pi} d\theta'' \; \sin(\theta'') \; {1 \over y^2} = 
{-1 \over 2 r' r''} \Biggl\{ 
\ln\Bigl[(r' + r'')^2 - (d_y-i \epsilon)^2 \Bigr] - 
\Bigl( r'' \rightarrow -r''\Bigr) 
\Biggr\} \;\; , \eqno(3.34a)$$
$$\int_0^{\pi} d\theta'' \; \sin(\theta'') \; {\ln(H^2 y^2) \over y^2} = 
{-1 \over 4 r' r''} \Biggl\{ 
\ln^2\Bigl[(r' + r'')^2 - (d_y-i \epsilon)^2 \Bigr] - 
\Bigl( r'' \rightarrow -r''\Bigr)
\Biggr\} \eqno(3.34b)$$
In this case we always eliminate the extra logarithm by partially integrating 
on one of the radial variables or conformal times.

It is useful at this stage to absorb the lower limit of the angular integration
--- the $r'' \rightarrow -r''$ term in (3.32b), (3.33) and (3.34a-b) --- into
an extension in the ranges of the radial integrations. For the case of a $\ln(
H^2 y^2)$ and $m \neq 1$ the extension goes as follows:
$$\eqalignno{ &\int_0^{\infty} dr' \; {r'}^2 \; 
\int_0^{\infty} dr'' \; {r''}^2 \; 
{1 \over (w^2)^{\ell} \; (z^2)^n} \times {-1 \over 2(m-1) r' r''} \cr
&\qquad\qquad\qquad\quad \Biggl\{ 
{\ln\Bigl[H^2 \Bigl( (r'+r'')^2 - (d_y-i\epsilon)^2\Bigr)\Bigr] + 
\frac1{m-1} \over [(r'+r'')^2 - (d_y-i\epsilon)^2]^{m-1}} - 
\Bigl(r'' \rightarrow - r''\Bigr) \Biggr\} \cr
&=- {1 \over 2(m-1)} \int_0^{\infty} dr' \; r' \; 
\int_{-\infty}^{\infty} dr'' \; r'' \; 
{\ln\Bigl[H^2 \Bigl((r'+r'')^2 - (d_y-i\epsilon)^2\Bigr)\Bigr] + 
\frac1{m-1} \over (w^2)^{\ell} \; 
[(r'+r'')^2 - (d_y-i\epsilon)^2]^{m-1} \; (z^2)^n} 
&(3.35a) \cr
&=- {1 \over 4(m-1)} \int_{-\infty}^{\infty} dr' \; r' \; 
\int_{-\infty}^{\infty} dr'' \; r'' \; 
{\ln\Bigl[H^2 \Bigl((r'+r'')^2 - (d_y-i\epsilon)^2\Bigr) \Bigr] + 
\frac1{m-1} \over (w^2)^{\ell} \; 
[(r'+r'')^2 - (d_y-i\epsilon)^2]^{m-1} \; (z^2)^n} 
\;\; . \qquad\qquad &(3.35b) \cr}$$
The reduction is similar for the other cases. 

We denote by $Y^2$ the combination of terms which descends from $y^2$ after
performing the angular integrations and extending the radial ranges:
$$Y^2 \equiv (r' + r'')^2 - (d_y - i\epsilon)^2 = 
(r'+r''-d_y+i\epsilon) \; (r'+r''-d_y-i\epsilon) 
\;\; . \eqno(3.36)$$
At this stage the answer is a long series of integrals of the form:
$$\eqalignno{ \# \; {\kappa^4 H^2 \over 2^{9} \pi^6} \; 
(\partial_u)^{0,1} \int_u^{H^{-1}} &du' \int_{u'}^{H^{-1}} du'' \;
{1 \over(u)^i\; (u')^j \; (u'')^k} \cr
&\int_{-\infty}^{\infty} dr' \int_{-\infty}^{\infty} dr''\; Q \; 
{\ln^{0,1} \Bigl[ H^2 \; (w^2,Y^2,z^2) \Bigr] \over 
(w^2)^{\ell} \; (Y^2)^m \; (z^2)^n} \;\; . &(3.37) \cr}$$
where $Q$ is a quadratic function of $r'$ and $r''$ whose precise form depends
upon what, if any, partial integrations were done to eliminate extra 
logarithms.

\noindent {\it 3.5 The Radial Integrations.} 

Since $w^2$, $z^2$ and $Y^2$ factorize into products of linear functions of
$r'$ and $r''$, the method of contours is especially effective. Note that the 
single possible logarithm can always be decomposed into a part which is 
analytic in the upper half plane and another part which is analytic in the 
lower half plane, for instance:
$$\ln (H^2 z^2) = 
\ln \Bigl[ H (r'' - d_z + i\epsilon) \Bigr] + 
\ln \Bigl[ H (r'' + d_z - i\epsilon) \Bigr] \;\; . \eqno(3.38)$$
The factor of $(2\pi)^2$ which comes from the two contour integrations is 
universal and is multiplied in with the rest.

As an example, consider (3.35b) with $\ell=n=1$ and $m=2$:
$$\eqalignno{ -\frac14 &\int_{-\infty}^{\infty} dr' \; r' \; 
\int_{-\infty}^{\infty} dr'' \; r'' \; 
{\ln(H^2 Y^2) + 1 \over w^2 \; z^2 \; Y^2} \cr
=& - {\pi^2 \over 4} {1 \over (d_w + dz + dy - 3i \epsilon) \;
(d_y - i\epsilon)} \cr
&+ {\pi^2 \over 4} \; {d_y \; \ln\Bigl[
H (d_w + d_z + d_y - 3i\epsilon) \Bigr] - 
(d_w + d_z) \; \ln\Bigl[2 H (d_y - i \epsilon)\Bigr] \over 
(d_w + d_z + d_y - 3i\epsilon) \; (d_w + d_z - d_y - i\epsilon) \;
(d_y - i\epsilon)} \cr
&+ {\pi^2 \over 4} \; {d_y \; \ln\Bigl[
-H (d_w + d_z + d_y - 3i\epsilon) \Bigr] -
(d_w + d_z) \; \ln\Bigl[-2 H (d_y - i \epsilon)\Bigr] \over 
(d_w + d_z + d_y - 3i\epsilon) \; (d_w + d_z - d_y - i\epsilon) \;
(d_y - i\epsilon)} \;\; . \qquad &(3.39) \cr}$$
Note that there is no pole for $d_y = d_w+d_z$; in this case the numerator
vanishes as well as the denominator and the residue is finite. Note also that
this integral can be differentiated with respect to $d_w$ and $d_z$ to give a
generating function for integrals with higher $\ell$ and $n$ values. In fact
only a handful of the radial integrations really need to be computed directly,
although this was done anyway as a check on accuracy.

For large exponents the results can be quite a bit longer than (3.39). In the
general case one also has to expect logarithms and inverse powers of $d_w-i
\epsilon$ and $d_z-i\epsilon$. The form we reach after doing the radial 
integrations is a long sum of conformal time integrations of the following 
type:
$$\eqalignno{ \# \; &{\kappa^4 H^2 \over 2^7 \pi^4} \; 
(\partial_u)^{0,1} \; \int_u^{H^{-1}} du' \int_{u'}^{H^{-1}} du'' \; 
{1 \over (u)^i (u')^j (u'')^k} \cr 
&{\ln^{0,1}\Bigl[\pm H \; (2d_w-2i\epsilon \; , \; 
2d_z-2i\epsilon \; , \; 2d_y-2i\epsilon \; , \; 
d_w+d_z+d_y-3i\epsilon) \Bigr] \over 
(d_w - i\epsilon)^{\zeta} \; (d_z - i\epsilon)^{\lambda} \; 
(d_y - i\epsilon)^{\xi} \; (d_w + d_z + d_y - 3i\epsilon)^{\sigma} \; 
(d_w + d_z - d_y - i \epsilon)^{\tau}} 
\;\; . \qquad\qquad &(3.40) \cr}$$
 
\noindent {\it 3.6 The Conformal Time Integrations.} 

It is at this stage that the ``$\pm$'' variations of the vertices become 
important. One has to sum (3.40) over the four possibilities, with the $d$'s 
assigned as in Table~1, and with a factor of $-1$ for each ``$-$'' vertex. One 
can see from Table~1 that the three $d$'s are always plus or minus the 
associated coordinate differences, for example, $d_z = \pm (u'-u)$. However, 
the two sums which can occur behave quite differently for the ``$++$'' and 
``$--$'' variations:
$$d_w + d_z + d_y = \pm 2 \; (u'' - u) \qquad , \qquad 
d_w + d_z - d_y = \pm 2 \; (u' - u) \;\; ; \eqno(3.41a)$$
than for the ``$+-$'' and ``$-+$'' variations:
$$d_w + d_z + d_y = \pm 2 \; (u'' - u') \qquad , \qquad 
d_w + d_z - d_y = 0 \;\; . \eqno(3.41b)$$
As noted above, apparent poles in at $d_y = d_w + d_z$ are always spurious. 
They are evaluated, for the ``$+-$'' and ``$-+$'' variations, by first taking
the limit $d_y \rightarrow d_w + d_z$, and then substituting for $d_w$ and
$d_z$ from Table~1.

The obvious strategy at this point is to decompose the integrands by partial 
fractions. Because only single powers of logarithms arise, this will always
suffice to reduce the result to a single integral.\footnote{*}{\tenpoint 
Although sometimes not in terms of either $u'$ or $u''$.} The remaining 
integral can also be decomposed by partial fractions, however, an integral of 
the form:
$$\int dx {\ln(x-a) \over x - b} \;\; , \eqno(3.42)$$
cannot be expressed in terms of elementary functions for $a \neq b$. When this 
occurs the strategy is to first extract any ultraviolet divergences by partial
integration, then expand the logarithm and integrate termwise. Since we only 
require the leading order form as $u \rightarrow 0^+$, this is 
straightforward. Consider, for example, the following:
$$\eqalignno{
\int_u^{H^{-1}} du' {\ln(Hu') \over u' - u - i \epsilon} &= 
\ln(Hu') \; \ln[H (u' - u - i\epsilon)] \Bigl \vert_u^{H^{-1}} - 
\int_u^{H^{-1}} {du' \over u'} \; \ln[H(u'-u)] \qquad \qquad &(3.43a) \cr
&= - \ln(Hu) \; \ln(-iH \epsilon) - \int_u^{H^{-1}} {du' \over u'} \; 
\Bigl[ \ln(Hu') - \sum_{n=1}^{\infty} \frac1{n} 
\Bigl(\frac{u}{u'}\Bigr)^n \Bigr] &(3.43b) \cr
&= - \ln(Hu) \; \ln(-iH \epsilon) + \frac12 \ln^2(Hu) + 
\sum_{n=1}^{\infty} \frac1{n^2} \Bigl[1 - (Hu)^n\Bigr] 
\;\; . &(3.43c) \cr}$$
Of course one can use these methods to perform the integrations in the original
order and with the original variables of expression (3.40). We did it both ways
and compared each term as a check on accuracy.

We have selected, as an example, the term which emerges from step 3 above in
the form:
$$\# {\kappa^4 H^2 \over 2^{12} \pi^8} \; \partial_u \int_u^{H^{-1}} du' \;
\int_{u'}^{H^{-1}} {du'' \over u^2 \; u''} \int d^3r' \int d^3r'' \; 
{\ln(H^2 y^2) \over w^2 y^4 z^2} \;\; . \eqno(3.44)$$
It has already been shown that the integrations over ${\vec r}\ '$ and 
${\vec r}\ ''$ give:
$$\eqalignno{ \# {1 \over 16} \; {\kappa^4 H^2 \over 2^7 \pi^4} \; 
&\partial_u \int_u^{H^{-1}} du' \; \int_{u'}^{H^{-1}} 
{du'' \over u^2 \; u''} \; \Biggl\{
-{1\over (d_w + dz + dy - 3i \epsilon) \; (d_y - i\epsilon)} \cr
&+ {d_y \; \ln\Bigl[H (d_w + d_z + d_y - 3i\epsilon)\Bigr] - 
(d_w + d_z) \; \ln\Bigl[2 H (d_y - i \epsilon)\Bigr] \over 
(d_w + d_z + d_y - 3i\epsilon) \; 
(d_w + d_z - d_y - i\epsilon) \; (d_y - i\epsilon)} \cr
&+ \Bigl( H \rightarrow -H \Bigr) \Biggr\} 
\;\; . \qquad &(3.45) \cr}$$
If we neglect terms of order $\epsilon$, the ``$++$'' and ``$+-$'' variations
give the following expressions:
$$\eqalignno{ \# {1 \over 16} \; {\kappa^4 H^2 \over 2^7 \pi^4} \; 
&\partial_u \int_u^{H^{-1}} du' \; \int_{u'}^{H^{-1}} 
{du'' \over u^2 \; u''} \; \Biggl\{
-{1\over 2} {1 \over (u''-u'-i \epsilon) \; (u''-u - i\epsilon)} \cr
&+ {1 \over 4} {\ln[2H (u''-u-i\epsilon)] \over 
(u''-u-i\epsilon) \; (u'-u-i\epsilon)} - 
{1 \over 4} {\ln[2H (u''-u'-i\epsilon)] \over 
(u''-u-i\epsilon) \; (u''-u'-i\epsilon)} \cr
&- {1 \over 4} {\ln[2H (u''-u'-i\epsilon)] \over 
(u''-u'-i\epsilon) \; (u'-u-i \epsilon)} + 
\Bigl( H \rightarrow -H \Bigr) \Biggr\} 
\;\; , \qquad &(3.46a) \cr}$$
$$\eqalignno{ -\# {1 \over 16} \; &{\kappa^4 H^2 \over 2^7 \pi^4}\; 
\partial_u \int_u^{H^{-1}} du' \; \int_{u'}^{H^{-1}} du'' \; 
{\ln[-2H (u''-u'+i\epsilon)] + (H \rightarrow -H) \over 
2 \; u^2 \; u'' \; (u''-u'+i\epsilon)^2} 
\;\; . \qquad &(3.46b) \cr}$$
The ``$--$'' and ``$-+$'' variations can be obtained by complex conjugation of
(3.46a) and (3.46b), respectively.

The first integral in the ``$++$'' variation (3.46a) has a simple result:
$$\eqalignno{ -{1\over 2 u^2} &\int_u^{H^{-1}} du' 
\int_{u'}^{H^{-1}} {du'' \over u'' (u'' - u' - i\epsilon) \; 
(u'' - u - i \epsilon)} = &(3.47a) \cr
&{1 \over 2 u^3} \sum_{n=1}^{\infty} {(Hu)^n  -1 \over n^2} 
+ {1 \over 4 u^3} \Bigl\{- \ln^2\Bigl({-i \epsilon \over u}\Bigr) 
+ 2 \ln(-iH \epsilon) \; \ln(1-Hu) - \ln^2(1-Hu)\Bigr\} \;\; . \cr}$$
Note how it illustrates the arguments given in sub-section 2.7 for the 
inevitability of infrared logarithms when an $\ell$-loop graph contains $\ell$
logarithmic ultraviolet divergences. Nice as it is to have exact expressions, 
what we really want is an asymptotic expansion for late times ($u \rightarrow 
0^+$):
$$-{1\over 2 u^2} \int_u^{H^{-1}} du' \int_{u'}^{H^{-1}} 
{du'' \over u'' (u'' - u' - i\epsilon) \; (u'' - u - i \epsilon)} 
= -{\ln^2(H u) \over 4 u^3} + O\Bigl({\ln(H u) \over u^3}\Bigr) 
\eqno(3.47b)$$
The asymptotic expansion for all of (3.46a) is:
$$\# \; {\kappa^4 H^2 \over 2^7 \pi^4} \; 
\partial_u \Biggl\{- {1 \over 96} {\ln^3(Hu) \over u^3} - 
{1\over 32} \Bigl[\ln(2) + {i\pi \over 2}\Bigr] \; {\ln^2(Hu) \over u^3}
+ O\Bigl({\ln(Hu) \over u^3}\Bigr) \Biggr\} 
\;\; . \eqno(3.48a)$$ 
The analogous expansion for the ``$+-$'' variation (3.46b) is:
$$\# \; {\kappa^4 H^2 \over 2^7 \pi^4} \; 
\partial_u \Biggl\{- {1\over 32} \; {\ln^2(Hu) \over u^3} + 
O\Bigl({\ln(Hu) \over u^3}\Bigr) \Biggr\} 
\;\; . \eqno(3.48b)$$
Summing these, along with the respective complex conjugates for the ``$--$''
and ``$-+$'' variations, gives:
$$\# \; {\kappa^4 H^2 \over 2^7 \pi^4} \; 
\Biggl\{ {1 \over 16} {\ln^3(Hu) \over u^4} + \
Bigl[{7 \over 32} + {3 \over 16} \; \ln(2)\Bigr] \; 
{\ln^2(Hu) \over u^4} + O\Bigl({\ln(Hu) \over u^4}\Bigr) 
\Biggr\} \;\; . \eqno(3.49)$$
Note the triple log terms. From the discussion of sub-section 2.8 we know that
these cannot appear in the full result,\footnote{*}{\tenpoint In fact the 
argument of 2.8 applies for the result from any triplet of vertex operators. 
We checked this explicitly for the triplet in which the $x^{\mu}$ vertex 
operator is \#10 and the vertex operators at ${x'}^{\mu}$ and ${x''}^{\mu}$ are
both \#41. Step 3 in the reduction of this triplet gives 117 terms containing a
factor of $\ln(H^2 y^2)$ and 77 terms with no logarithms.} however, they {\it 
do} appear in pieces of it such as this example. An important check on accuracy
is that the sum of all such triple log terms cancels.

Many, many terms such as (3.44) emerge from step 3. For example, the $\gamma$ 
contraction of diagram (4a) --- which is also diagram (2a) --- consists of 842 
terms whose integrands contain a factor of $\ln(H^2 y^2)$, 301 terms which 
contain $\ln(H^2 w^2)$, another 301 terms which contain $\ln(H^2 z^2)$, and 
1663 terms which contain no logarithm at all. When all the 3-3-3 integrals are 
computed and the results summed, the totals are:
$$\alpha(u) = {\kappa^4 H^2 \over 2^7 \pi^4} \; \Biggl\{ \; \Bigl[-492 
\;\; + \;\; 234 \;\; + \;\; 216\Bigr] \; {\ln^2(Hu) \over u^4} +
O\Bigl({\ln(Hu) \over u^4}\Bigr) \Biggr\} \;\; , \eqno(3.50a)$$ 
$$\gamma(u) = {\kappa^4 H^2 \over 2^7 \pi^4} \; \Biggl\{ \; \Bigl[+{1157 \over 
6} \;\; - \;\; {400 \over 3} \;\; - \;\; 56 \Bigr] \; {\ln^2(Hu) \over u^4} + 
O\Bigl({\ln(Hu) \over u^4}\Bigr) \Biggr\} \;\; . \eqno(3.50b)$$
where the three numbers between the square brackets refer to the contributions
from diagrams (2a), (2b) and (2c), respectively. Taking account of (3.27) this
gives the following 3-3-3 contributions to the coefficient functions $a(u)$ and
$c(u)$:
$$a_{333}(u) = {\kappa^4 H^2 \over 2^7 \pi^4} \; \Biggl\{ \; \Bigl[-{1795 \over
18} \;\; + \;\; {302 \over 9} \;\; + \;\; {160 \over 3} \Bigr] \; {\ln^2(Hu) 
\over u^4} + O\Bigl({\ln(Hu) \over u^4}\Bigr) \Biggr\} \;\; , \eqno(3.51a)$$
$$c_{333}(u) = {\kappa^4 H^2 \over 2^7 \pi^4} \; \Biggl\{ \; \Bigl[+{1157 \over
6} \;\; - \;\; {400 \over 3} \;\; - \;\; 56 \Bigr] \; {\ln^2(Hu) \over u^4} + 
O\Bigl({\ln(Hu) \over u^4}\Bigr) \Biggr\} \;\; . \eqno(3.51b)$$

The results quoted above contain the contributions of some degenerate terms 
whose reduction is actually much more akin to that of diagram (2d), to be
described in the next section. These terms come from the action of two 
$0$-component derivatives on the normal part (2.14a) of an outer leg 
propagator which connects two ``$+$'' vertices. The vertices can be either 
$x^{\mu}$ and ${x'}^{\mu}$ or $x^{\mu}$ and ${x''}^{\mu}$. What such a double
derivative really gives is:
$$\eqalignno{\partial_0 \; {\partial'}_0 \; i \Delta_N(x;x') =& 
{H^2 \over 8 \pi^2 {\Delta x}} \; \partial_0 \; {\partial'}_0 \; 
\Biggl\{ {u \; u' \over {\Delta x} - \vert {\Delta u} \vert + i \epsilon} 
+ { u \; u' \over {\Delta x} + \vert {\Delta u} \vert - i \epsilon}
\Biggr\} &(3.52a) \cr
=& {H^2 \over 8 \pi^2 {\Delta x}} \Biggl\{ 
{1 \over {\Delta x} - \vert {\Delta u}\vert + i \epsilon} + 
{1 \over {\Delta x} + \vert {\Delta u}\vert - i \epsilon} \cr
&+ \Bigl[\theta(\Delta u) - \theta(-\Delta u)\Bigr] \; \Biggl[ 
{\Delta u \over ({\Delta x} - \vert {\Delta u} \vert + i \epsilon)^2} -
{\Delta u \over ({\Delta x} + \vert {\Delta u}\vert - i \epsilon)^2}
\Biggr] \cr
&- {2 \; u \; u' \over ({\Delta x} - \vert {\Delta u} \vert + i \epsilon)^3} 
- {2 \; u \; u' \over ({\Delta x} + \vert {\Delta u}\vert - i \epsilon)^3}
\cr
&- 2 \; \delta(\Delta u) \; \Biggl[ 
{u^2 \over ({\Delta u} + i \epsilon)^2} -
{u^2 \over ({\Delta x} - i \epsilon)^2} \Biggr] \Biggr\} 
\;\; . &(3.52b) \cr}$$
(Recall that $\Delta u \equiv u' - u$ and $\Delta x \equiv \Vert {\vec x}\ ' -
{\vec x}\Vert$.) The reduction procedure outlined in this section accounts for 
all but the terms proportional to $\delta({\Delta u})$. They are of order
$\epsilon$ but sufficiently singular at ${\vec x} = {\vec x}\ '$ that the result
is a spacetime delta function:\footnote{*}{\tenpoint Similar terms of order
$\epsilon$ arise from the logarithm part of the propagator but they are not
singular enough to survive in the unregulated limit.}
$$\eqalignno{- {H^2 u^2 \over 4 \pi^2} \; 
{\delta(\Delta u) \over \Delta x} \; 
\Biggl[ {1 \over ({\Delta u} + i \epsilon)^2} - 
{1 \over ({\Delta x} - i \epsilon)^2} \Biggr] &= 
{H^2 u^2 \over \pi^2} \; {i \epsilon \; \delta(\Delta u) \over 
({\Delta x}^2 + \epsilon^2)^2} &(3.53a) \cr
&\longrightarrow \Omega^2 \; i \; \delta^4(x' - x) 
\;\; . \qquad &(3.53b) \cr}$$
Performing the now trivial integration over ${x'}^{\mu}$ causes the 3-3-3
diagram to degenerate to the 4-3 topology of diagram (2d). The reduction 
thereafter is the same as will be described in the next section. Of course 
delta function terms can happen whenever a propagator is doubly differentiated.
They give 4-3 diagrams when either of the two outer leg propagators is 
affected; when one of the two inner propagators is affected the degenerate
diagram has the topology of figure (2e) and fails to contribute to leading 
order for the same reason.

\vfill\eject

\centerline{\bf {4. The 4--3 Diagram}}

The 4-3 diagram consists of an outer 4-point vertex joined to a freely 
integrated 3-point vertex as shown in Fig.~5 below. From the discussion of 
sub-section 2.7 we see that this diagram can contribute at leading order when 
a factor of $\ln(H u)$ from integrating the free interaction vertex combines
with a factor of $\ln(H u)$ from an undifferentiated propagator logarithm. The
physical origin of the first logarithm is the growth in the invariant
volume of the past lightcone, while the second logarithm comes from the 
increasing correlation of the free graviton vacuum of an inflating universe.

\vskip 1cm

\centerline{\psfig{figure=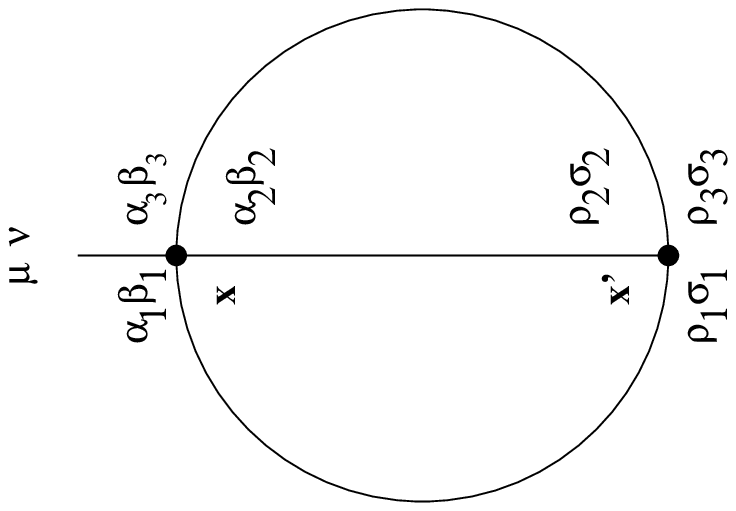,width=4.5truecm,angle=-90}}

\vskip -0.3truecm

{\bf Fig.~5:}{\ninepoint The tensor structure of the 4-3 diagram.} 

\vskip 0.3cm

The 4-3 diagram has an obvious 3-fold symmetry, corresponding to permutations 
of lines $1$, $2$ and $3$. In computing such a highly symmetric diagram it is 
not efficient to fully symmetrize the vertex operators and then divide by the 
symmetry factor of 6. The better strategy is to symmetrize the 4-point vertex 
operator only on the external line, and then contract this partially 
symmetrized vertex operator, through propagators, into the fully symmetrized 
3-point vertex operator. 

We can read off the graviton 4-point interaction from expression (2.4):

\vfill\eject

$$\eqalignno{ {\cal L}_{\rm inv}^{(4)} =  \; \kappa^2 \; \Omega^{2} &\; \Bigl\{
\frac{1}{32} \psi^2 \; \psi_{, \mu} \; \psi^{, \mu}  - \frac{1}{16} \psi^2 \; 
\psi_{, \mu} \; \psi^{\mu \nu }_{~~, \nu} - \frac{1}{32} \psi^2 \; \psi_{\mu 
\nu , \alpha} \; \psi^{\mu \nu , \alpha} + \frac{1}{16} \psi^2 \; \psi_{\alpha 
\mu , \nu} \; \psi^{\alpha \nu , \mu} \cr 
&- \frac18 \psi \; \psi^{\alpha \beta } \; \psi_{, \alpha} \; \psi_{, \beta} 
+ \frac18 \psi \; \psi^{\alpha \beta } \; \psi^{\mu \nu}_{~~, \alpha} \; 
\psi_{\mu \nu , \beta} + \frac14 \psi \; \psi^{\alpha \beta } \; \psi_{\alpha 
\beta , \mu} \; \psi^{\mu \nu}_{~~, \nu} \cr 
&- \frac14 \psi \; \psi^{\alpha \beta } \; \psi_{\alpha \beta , \nu} \; 
\psi^{, \nu} + \frac14 \psi \; \psi^{\alpha \beta } \; \psi_{\alpha \mu , \nu} 
\; \psi^{~\mu , \nu}_{\beta}  - \frac14 \psi \; \psi^{\alpha \beta } \; 
\psi^{~~, \nu}_{\alpha \mu} \; \psi^{~~, \mu}_{\beta \nu} \cr 
&+ \frac14 \psi \; \psi^{\alpha \beta } \; \psi_{\mu \alpha , \beta} \; 
\psi^{, \mu} + \frac14 \psi \; \psi^{\alpha \beta } \; \psi_{, \alpha} \; 
\psi^{~~, \nu}_{\beta \nu} - \frac12 \psi \; \psi^{\alpha \beta } \; 
\psi_{\alpha \mu , \nu} \; \psi^{\mu \nu}_{~~, \beta} \cr 
&- \frac{1}{16} \psi^{\alpha}_{~\beta} \; \psi^{\beta}_{~\alpha} \; 
\psi^{\mu }_{~\mu , \gamma} \; \psi^{, \gamma} 
+ \frac18 \psi^{\alpha}_{~\beta} \; \psi^{\beta}_{~\alpha} \; 
\psi_{, \mu} \; \psi^{\mu \nu}_{~~, \nu}  
+ \frac{1}{16} \psi^{\alpha}_{~\beta} \; \psi^{\beta}_{~\alpha} \; 
\psi_{\mu \nu , \gamma} \; \psi^{\mu \nu , \gamma} \cr 
&- \frac18 \psi^{\alpha}_{~\beta} \; \psi^{\beta}_{~\alpha} \; 
\psi_{\gamma \nu , \mu} \; \psi^{\gamma \mu , \nu}  
+ \frac14 \psi^{\alpha}_{~\beta} \; \psi^{\beta \gamma} \; 
\psi_{, \alpha} \; \psi_{, \gamma} 
- \frac14 \psi^{\alpha}_{~\beta} \; \psi^{\beta \gamma} \; 
\psi_{\mu \nu , \alpha} \; \psi^{\mu \nu}_{~~, \gamma} \cr 
&+ \frac12 \psi^{\alpha}_{~\beta} \; \psi^{\beta \gamma} \; 
\psi_{\alpha \gamma , \mu} \; \psi^{, \mu}  
- \frac12 \psi^{\alpha}_{~\beta} \; \psi^{\beta \gamma} \; 
\psi_{\alpha \gamma , \mu} \; \psi^{\mu \nu}_{~~, \nu} 
+ \frac12 \psi^{\alpha}_{~\beta} \; \psi^{\beta \gamma} \; 
\psi^{~~, \nu}_{\alpha \mu} \; \psi^{~~, \mu}_{\gamma \nu} \cr 
&- \frac12 \psi^{\alpha}_{~\beta} \; \psi^{\beta \gamma} \; 
\psi_{\alpha \mu , \nu} \; \psi^{~\mu , \nu}_{\gamma}  
+ \psi^{\alpha}_{~\beta} \; \psi^{\beta \gamma} \; 
\psi_{\alpha \mu , \nu} \; \psi^{\mu \nu}_{~~, \gamma} 
- \frac12 \psi^{\alpha}_{~\beta} \; \psi^{\beta \gamma} \; 
\psi_{, \alpha} \; \psi^{~~, \nu}_{\gamma \nu} \cr 
&- \frac12 \psi^{\alpha}_{~\beta} \; \psi^{\beta \gamma} \; 
\psi^{, \nu} \; \psi_{\alpha \nu , \gamma}  
+ \frac14 \psi^{\alpha \beta} \; \psi^{\mu \nu} \; 
\psi_{\alpha \beta , \gamma} \; \psi^{~~, \gamma}_{\mu \nu} 
- \frac12 \psi^{\alpha \beta} \; \psi^{\mu \nu} \; 
\psi_{\alpha \beta , \mu} \; \psi^{~~, \gamma}_{\nu \gamma} \cr 
&+ \frac12 \psi^{\alpha \beta} \; \psi^{\mu \nu} \; 
\psi_{\alpha \beta , \mu} \; \psi_{, \nu}  
- \frac12 \psi^{\alpha \beta} \; \psi^{\mu \nu} \; 
\psi_{\alpha \mu , \nu} \; \psi_{, \beta} 
- \frac12 \psi^{\alpha \beta} \; \psi^{\mu \nu} \; 
\psi_{\alpha \gamma , \mu} \; \psi^{\gamma}_{~\beta , \nu} \cr 
&- \frac12 \psi^{\alpha \beta} \; \psi^{\mu \nu} \; 
\psi_{\alpha \beta , \gamma} \; \psi^{\gamma}_{~\mu , \nu}  
- \frac14 \psi^{\alpha \beta} \; \psi^{\mu \nu} \; 
\psi_{\alpha \mu , \gamma} \; \psi^{~~, \gamma}_{\beta \nu} 
+ \psi^{\alpha \beta} \; \psi^{\mu \nu} \; 
\psi_{\alpha \mu , \gamma} \; \psi^{\gamma}_{~\beta , \nu} \cr 
&+ \frac12 \psi^{\alpha \beta} \; \psi^{\mu \nu} \; 
\psi_{\alpha \gamma , \mu} \; \psi^{\gamma}_{~\nu , \beta} 
- \frac{1}{8u} \psi_{, \mu} \; \psi^{2} \; 
\psi^{\mu \nu} \; t_{\nu} 
+ \frac{1}{4u} \psi_{, \mu} \; \psi^{\alpha \beta} \; 
\psi_{\alpha \beta} \; \psi^{\mu \nu} \; t_{\nu} \cr 
&- \frac1u \psi^{\alpha \beta}_{~~, \mu} \; \psi_{\alpha \gamma} \; 
\psi^{~\gamma}_{\beta} \; \psi^{\mu \nu} \; t_{\nu} 
- \frac1u \psi^{~\alpha}_{\mu} \; \psi_{\alpha \beta} \; 
\psi^{, \beta} \; \psi^{\mu \nu} \; t_{\nu} 
+ \frac{1}{2u} \psi^{\alpha \beta}_{~~, \mu} \; \psi_{\alpha \beta} \; 
\psi \; \psi^{\mu \nu} \; t_{\nu} \cr 
&+ \frac{1}{2u} \psi_{\alpha \mu} \; \psi^{, \alpha} \; \psi \; 
\psi^{\mu \nu} \; t_{\nu} 
- \frac{1}{u} \psi_{\gamma \mu} \; \psi_{\alpha \beta} \; 
\psi^{\alpha \beta , \gamma} \; \psi^{\mu \nu} \; t_{\nu} 
\Bigr\} \;\; . &(4.1) \cr}$$ 
(Recall that $\psi \equiv \psi^{\mu}_{~\mu}$ and $t_{\mu} \equiv 
\eta_{\mu 0}$.) Almost all of this ungainly expression can be checked against
published results [11] by taking the flat space limit: $u = H^{-1} - t$ and
$H \rightarrow 0$. The procedure for partially symmetrizing a vertex was 
described at the beginning of Section 3. As an example consider the final term 
in (4.1):
$$-\kappa^2 \Omega^2 \; \frac1{u} \psi_{\gamma\mu} \; \psi_{\alpha\beta} \; 
\psi^{\alpha\beta, \gamma} \; \psi^{\mu\nu} \; t_{\nu} \;\; . \eqno(4.2)$$
If the indices $\mu$ and $\nu$ and the derivative $\partial_u$ represent the 
distinguished line, a valid partial symmetrization of this interaction gives 
the following vertex operators:
$$- \kappa^2 \Omega^2 \; \frac1{u} \; \eta^{\mu (\alpha_2} \; \eta^{\beta_2) 
\nu} \; \partial_2^{(\alpha_1} \; \eta^{\beta_1) (\alpha_3} \; t^{\beta_3)} 
\;\; , \eqno(4.3a)$$
$$- \kappa^2 \Omega^2 \; \frac1{u} \; \partial_2^{(\mu} \; \eta^{\nu) 
(\alpha_3} \; t^{\beta_3)} \; \eta^{\alpha_1 (\alpha_2} \; \eta^{\beta_2) 
\beta_1} \;\; , \eqno(4.3b)$$
$$- \kappa^2 \Omega^2 \; \frac1{u} \; \partial_u \; \eta^{\mu (\alpha_1} \; 
\eta^{\beta_1) \nu} \; t^{(\alpha_2} \; \eta^{\beta_2) (\alpha_3} \; 
t^{\beta_3)} \;\; , \eqno(4.3c)$$
$$- \kappa^2 \Omega^2 \; \frac1{u} \; t^{(\mu} \; \eta^{\nu) (\alpha_1} \; 
\partial_3^{\beta_1)} \; \eta^{\alpha_2 (\alpha_3} \; \eta^{\beta_3) \beta_2} 
\;\; . \eqno(4.3d)$$
These are vertex operators $i=127,...,130$ in Table~4. The fully symmetrized 
cubic vertex operators are given in Table~5. They were obtained by 
interchanging legs $2$ and $3$ from the Table~3, but we have taken account of 
symmetries to reduce the $2 \times 43 = 86$ terms to only 75.

The unprimed vertex at $x^{\mu}$ is ``$+$'' type and we must sum the primed 
vertex over both ``$+$'' and ``$-$'' variations. Since the $(++)$ and 
$(+-)$ propagators are identical for $u' < u$, the relative minus sign 
between the two vertex assignments allows us to restrict the range of 
integration to $u \leq u' \leq H^{-1}$. In this region the $(+-)$ 
propagator is the complex conjugate of the $(++)$ one, so we can write
the total contribution of the 4-3 diagram as twice the real part of
the $(++)$ term:
$$\eqalignno{
{\cal T}_{43}^{\mu \nu} & \equiv
a_{43}(u) \;\; {\overline \eta}^{\mu\nu} \;
+ \; c_{43}(u) \;\; t^{\mu} \; t^{\nu} \cr
&=2{\rm Re}\Biggl\{ 
-i \kappa \int_u^{H^{-1}} du' \int d^3x' \; \sum_{i=1}^{130} \; 
V_i^{\mu\nu \alpha_1 \beta_1 \alpha_2 \beta_2 \alpha_3 \beta_3}
(x;\partial_u,\partial_1, \partial_2,\partial_3) \cr
& \qquad 
i\Bigl[ {_{\alpha_1\beta_1}} \Delta_{\rho_1 \sigma_1} \Bigr](x;x') \;\; 
i\Bigl[ {_{\alpha_2\beta_2}} \Delta_{\rho_2 \sigma_2} \Bigr](x;x') \;\;
i\Bigl[ {_{\alpha_3\beta_3}} \Delta_{\rho_3 \sigma_3} \Bigr](x;x') \cr 
& \qquad \sum_{j=1}^{75} \; 
V_j^{\rho_1\sigma_1\rho_2\sigma_2\rho_3\sigma_3}
(x';\partial_1',\partial_2',\partial_3')\Biggr\} 
\;\; . &(4.4) \cr}$$
The subscripts $i$ and $j$ refer to Tables~4 and 5 respectively. 
Recall that where each derivative acts is indicated by primes and 
subscripts. For example, the derivative $\partial_3$ in the 4-point
vertex operator acts on the first argument of the propagator 
$i[ {_3}\Delta_3 ](x;x')$. The derivative $\partial_u$ acts on 
all $u$'s in the vertex and the three propagators.

The entire calculation was performed by computer using Mathematica 
[12] and FeynCalc [13]. The first step was to contract each pair of 
vertex operators into the three internal propagator and write the 
results onto a file. The next step was to act the internal derivatives 
($\partial_{1-3}$ and $\partial_{ 1-3}'$) and store the results for 
each pair of vertex operators. Selected vertex pairs were computed by 
hand to check the procedure. 

\vskip 1cm

\vbox{\tabskip=0pt \offinterlineskip
\def\tablerule{\noalign{\hrule}}
\halign to505pt {\strut#& \vrule#\tabskip=1em plus2em& 
\hfil#& \vrule#& \hfil#\hfil& \vrule#& \hfil#& \vrule#& \hfil#\hfil& 
\vrule#\tabskip=0pt\cr
\tablerule
\omit&height4pt&\omit&&\omit&&\omit&&\omit&\cr
&&\omit\hidewidth $i$ &&
\omit\hidewidth {\rm Vertex Operator}\hidewidth&& 
\omit\hidewidth $i$\hidewidth&& \omit\hidewidth {\rm Vertex Operator}
\hidewidth&\cr
\omit&height4pt&\omit&&\omit&&\omit&&\omit&\cr
\tablerule
\omit&height1.8pt&\omit&&\omit&&\omit&&\omit&\cr
&& 1 && $\frac18 \eta^{\mu\nu} \; \eta^{\alpha_1\beta_1} \;
\partial_3^{(\alpha_2} \; \eta^{\beta_2) (\alpha_3} \; 
\partial_2^{\beta_3)}$ && 66 && $\frac1{16} \partial_u \; \eta^{\mu\nu}
\; \eta^{\alpha_1 \beta_1} \; \eta^{\alpha_2 \beta_2} 
\; \eta^{\alpha_3 \beta_3} \; t \cdot \partial_3$ &\cr
\omit&height1.8pt&\omit&&\omit&&\omit&&\omit&\cr
\tablerule
\omit&height1.8pt&\omit&&\omit&&\omit&&\omit&\cr
&& 2 && $\frac18 \partial_u \; \partial_3^{(\mu} \; \eta^{\nu)
(\alpha_3} \; t^{\beta_3)} \; \eta^{\alpha_1 \beta_1} 
\; \eta^{\alpha_2 \beta_2}$ && 67 && $-\frac18 \eta^{\mu (\alpha_1}
\; \eta^{\beta_1) \nu} \; \eta^{\alpha_2 \beta_2} \;
\eta^{\alpha_3 \beta_3} \; \partial_2 \cdot \partial_3$ &\cr
\omit&height1.8pt&\omit&&\omit&&\omit&&\omit&\cr
\tablerule
\omit&height1.8pt&\omit&&\omit&&\omit&&\omit&\cr
&& 3 && $-\frac14 \eta^{\mu (\alpha_1} \; \eta^{\beta_1) \nu} 
\; \partial_3^{(\alpha_2} \; \eta^{\beta_2) (\alpha_3} 
\; \partial_2^{\beta_3)}$ && 68 && $-\frac18 \partial_u \;
\eta^{\mu \nu} \; \eta^{\alpha_1 (\alpha_2} \; \eta^{\beta_2)
\beta_1} \; \eta^{\alpha_3 \beta_3} \; t \cdot \partial_3$ &\cr
\omit&height1.8pt&\omit&&\omit&&\omit&&\omit&\cr
\tablerule
\omit&height1.8pt&\omit&&\omit&&\omit&&\omit&\cr
&& 4 && $-\frac14 \partial_u \; \partial_3^{(\mu} \; \eta^{\nu) 
(\alpha_3} \; t^{\beta_3)} \; \eta^{\alpha_1 (\alpha_2} 
\; \eta^{\beta_2) \beta_1}$ && 69 && $\eta^{\mu) (\alpha_1} \;
\eta^{\beta_1) (\alpha_2} \; \eta^{\beta_2) (\nu} \; \eta^{
\alpha_3 \beta_3} \; \partial_2 \cdot \partial_3$ &\cr
\omit&height1.8pt&\omit&&\omit&&\omit&&\omit&\cr
\tablerule
\omit&height1.8pt&\omit&&\omit&&\omit&&\omit&\cr
&& 5 && $\eta^{\mu) (\alpha_1} \; \partial_3^{\beta_1)} \;
\partial_2^{(\alpha_3} \; \eta^{\beta_3) (\alpha_2} \; \eta^{
\beta_2) (\nu}$ && 70 && $\frac12 \partial_u \; \eta^{\mu) (\alpha_1}
\; \eta^{\beta_1) (\alpha_2} \; \eta^{\beta_2) (\nu} \;
\eta^{\alpha_3 \beta_3} \; t \cdot \partial_3$ &\cr
\omit&height1.8pt&\omit&&\omit&&\omit&&\omit&\cr
\tablerule
\omit&height1.8pt&\omit&&\omit&&\omit&&\omit&\cr
&& 6 && $\partial_2^{(\mu} \; \eta^{\nu) (\alpha_1} \; \eta^{
\beta_1) (\alpha_3} \; \eta^{\beta_3) (\alpha_2} \; 
\partial_3^{\beta_2)}$ && 71 && $\frac12 \partial_u \; \eta^{\mu\nu}
\; \eta^{\alpha_1) (\alpha_2} \; \eta^{\beta_2) (\alpha_3}
\; \eta^{\beta_3) (\beta_1} \; t\cdot \partial_3$ &\cr
\omit&height1.8pt&\omit&&\omit&&\omit&&\omit&\cr
\tablerule
\omit&height1.8pt&\omit&&\omit&&\omit&&\omit&\cr
&& 7 && $\partial_u \; \eta^{\mu) (\alpha_2} \; \eta^{\beta_2)
(\alpha_1} \; \partial_3^{\beta_1)} \; t^{(\alpha_3} \;
\eta^{\beta_3) (\nu}$ && 72 && $\frac12 \partial_2^{(\mu} \; \eta^{\nu)
(\alpha_1} \; \partial_3^{\beta_1)} \; \eta^{\alpha_2 \beta_2}
\; \eta^{\alpha_3 \beta_3}$ &\cr
\omit&height1.8pt&\omit&&\omit&&\omit&&\omit&\cr
\tablerule
\omit&height1.8pt&\omit&&\omit&&\omit&&\omit&\cr
&& 8 && $\partial_u \; \partial_3^{(\mu} \; \eta^{\nu) 
(\alpha_3} \; \eta^{\beta_3) (\alpha_1} \; \eta^{\beta_1) 
(\alpha_2} \; t^{\beta_2)}$ && 73 && $\frac12 \partial_u \;
\eta^{\mu\nu} \; t^{(\alpha_1} \; \eta^{\beta_1) (\alpha_2}
\; \partial_3^{\beta_2)} \; \eta^{\alpha_3 \beta_3}$ &\cr
\omit&height1.8pt&\omit&&\omit&&\omit&&\omit&\cr
\tablerule
\omit&height1.8pt&\omit&&\omit&&\omit&&\omit&\cr
&& 9 && $\eta^{\mu) (\alpha_1} \; \eta^{\beta_1) (\alpha_3} \;
\partial_2^{\beta_3)} \; \partial_3^{(\alpha_2} \; \eta^{
\beta_2) (\nu}$ && 74 && $-\frac14 \eta^{\mu \nu} \; \eta^{\alpha_1 
(\alpha_2} \; \eta^{\beta_2) \beta_1} \; \eta^{\alpha_3 
\beta_3} \; \partial_2 \cdot \partial_3$ &\cr
\omit&height1.8pt&\omit&&\omit&&\omit&&\omit&\cr
\tablerule
\omit&height1.8pt&\omit&&\omit&&\omit&&\omit&\cr
&& 10 && $\partial_u \; \partial_3^{(\mu} \; \eta^{\nu) 
(\alpha_2} \; \eta^{\beta_2) (\alpha_1} \; \eta^{\beta_1) 
(\alpha_3} \; t^{\beta_3)}$ && 75 && $-\frac14 \eta^{\mu (\alpha_2}
\; \eta^{\beta_2) \nu} \; \eta^{\alpha_1 \beta_1} \;
\eta^{\alpha_3 \beta_3} \; \partial_2 \cdot \partial_3$ &\cr
\omit&height1.8pt&\omit&&\omit&&\omit&&\omit&\cr
\tablerule
\omit&height1.8pt&\omit&&\omit&&\omit&&\omit&\cr
&& 11 && $-\frac12 \eta^{\mu \nu} \; \partial_3^{(\alpha_1} \;
\eta^{\beta_1) (\alpha_2} \; \eta^{\beta_2) (\alpha_3} \;
\partial_2^{\beta_3)}$ && 76 && $-\frac14 \partial_u \; \eta^{\mu 
(\alpha_2} \; \eta^{\beta_2) \nu} \; \eta^{\alpha_1 \beta_1}
\; \eta^{\alpha_3 \beta_3} \; t \cdot \partial_3$ &\cr
\omit&height1.8pt&\omit&&\omit&&\omit&&\omit&\cr
\tablerule
\omit&height1.8pt&\omit&&\omit&&\omit&&\omit&\cr
&& 12 && $-\frac12 \partial_2^{(\mu} \; \eta^{\nu) (\alpha_3} 
\; \eta^{\beta_3) (\alpha_2} \; \partial_3^{\beta_2)} 
\; \eta^{\alpha_1 \beta_1}$ && 77 && $-\frac14 \partial_u \;
\eta^{\mu \nu} \; \eta^{\alpha_1 \beta_1} \; \eta^{\alpha_2
(\alpha_3} \; \eta^{\beta_3) \beta_2} \; t \cdot \partial_3$ 
&\cr
\omit&height1.8pt&\omit&&\omit&&\omit&&\omit&\cr
\tablerule
\omit&height1.8pt&\omit&&\omit&&\omit&&\omit&\cr
&& 13 && $-\frac12 \partial_u \; \eta^{\mu) (\alpha_2} \; 
\partial_3^{\beta_2)} \; t^{(\alpha_3} \; \eta^{\beta_3) (\nu}
\; \eta^{\alpha_1 \beta_1}$ && 78 && $-\frac18 \eta^{\mu\nu} 
\partial_2^{(\alpha_1} \; \partial_3^{\beta_1)} \; \eta^{
\alpha_2 \beta_2} \; \eta^{\alpha_3 \beta_3}$ &\cr
\omit&height1.8pt&\omit&&\omit&&\omit&&\omit&\cr
\tablerule
\omit&height1.8pt&\omit&&\omit&&\omit&&\omit&\cr
&& 14 && $-\frac12 \partial_u \; \partial_3^{(\mu} \; \eta^{
\nu) (\alpha_3} \; \eta^{\beta_3) (\alpha_2} \; t^{\beta_2)}
\; \eta^{\alpha_1 \beta_1}$ && 79 && $-\frac18 \partial_2^{(\mu} 
\; \partial_3^{\nu)} \; \eta^{\alpha_1 \beta_1} \;
\eta^{\alpha_2 \beta_2} \; \eta^{\alpha_3 \beta_3}$ &\cr
\omit&height1.8pt&\omit&&\omit&&\omit&&\omit&\cr
\tablerule
\omit&height1.8pt&\omit&&\omit&&\omit&&\omit&\cr
&& 15 && $-\frac14 \eta^{\mu \nu} \; \partial_3^{(\alpha_2} \;
\eta^{\beta_2) (\alpha_1} \; \eta^{\beta_1) (\alpha_3} \;
\partial_2^{\beta_3)}$ && 80 && $-\frac14 \partial_u \; \eta^{\mu \nu}
\; \eta^{\alpha_1 \beta_1} \; \eta^{\alpha_3 \beta_3} 
\; \partial_3^{(\alpha_2} \; t^{\beta_2)}$ &\cr
\omit&height1.8pt&\omit&&\omit&&\omit&&\omit&\cr
\tablerule
\omit&height1.8pt&\omit&&\omit&&\omit&&\omit&\cr
&& 16 && $-\frac14 \eta^{\mu) (\alpha_2} \; \partial_3^{\beta_2)}
\; \eta^{\alpha_1 \beta_1} \; \partial_2^{(\alpha_3} \;
\eta^{\beta_3) (\nu}$ && 81 && $\frac12 \eta^{\mu (\alpha_2} \; \eta^{
\beta_2) \nu} \; \partial_2^{(\alpha_1} \; \partial_3^{
\beta_1)} \; \eta^{\alpha_3 \beta_3}$ &\cr
\omit&height1.8pt&\omit&&\omit&&\omit&&\omit&\cr
\tablerule
\omit&height1.8pt&\omit&&\omit&&\omit&&\omit&\cr
&& 17 && $-\frac12 \partial_u \; \partial_3^{(\mu} \; \eta^{
\nu) (\alpha_2} \; \eta^{\beta_2) (\alpha_3} \; t^{\beta_3)} 
\; \eta^{\alpha_1 \beta_1}$ && 82 && $\frac12 \partial_2^{(\mu} 
\; \partial_3^{\nu)} \; \eta^{\alpha_1 (\alpha_2} \;
\eta^{\beta_2) \beta_1} \; \eta^{\alpha_3 \beta_3}$ &\cr
\omit&height1.8pt&\omit&&\omit&&\omit&&\omit&\cr
\tablerule
\omit&height1.8pt&\omit&&\omit&&\omit&&\omit&\cr
&& 18 && $\partial_2^{(\mu} \; \eta^{\mu) (\alpha_3} \; \eta^{
\beta_3) (\alpha_1} \; \eta^{\beta_1) (\alpha_2} \; 
\partial_3^{\beta_2)}$ && 83 && $\frac12 \partial_u \eta^{\mu (\alpha_1}
\; \eta^{\beta_1) \nu} \; \partial_3^{(\alpha_2} \;
t^{\beta_2)} \; \eta^{\alpha_3 \beta_3}$ &\cr
\omit&height1.8pt&\omit&&\omit&&\omit&&\omit&\cr
\tablerule
\omit&height1.8pt&\omit&&\omit&&\omit&&\omit&\cr
&& 19 && $\eta^{\mu) (\alpha_2} \; \eta^{\beta_2) (\alpha_1} \;
\partial_3^{\beta_1)} \; \partial_2^{(\alpha_3} \; \eta^{
\beta_3) (\nu}$ && 84 && $\frac12 \partial_u \; \eta^{\mu\nu} 
\; \partial_3^{(\alpha_1} \; t^{\beta_1)} \; \eta^{
\alpha_2 (\alpha_3} \; \eta^{\beta_3) \beta_2}$ &\cr
\omit&height1.8pt&\omit&&\omit&&\omit&&\omit&\cr
\tablerule
\omit&height1.8pt&\omit&&\omit&&\omit&&\omit&\cr
&& 20 && $\partial_u \; \eta^{\mu) (\alpha_1} \; \eta^{\beta_1)
(\alpha_3} \; t^{\beta_3)} \; \partial_3^{(\alpha_2} \;
\eta^{\beta_2) (\nu}$ && 85 && $\frac12 \eta^{\mu (\alpha_3} \; \eta^{
\beta_3) \nu} \; \eta^{\alpha_1 (\alpha_2} \; \eta^{\beta_2)
\beta_1} \; \partial_2 \cdot \partial_3$ &\cr
\omit&height1.8pt&\omit&&\omit&&\omit&&\omit&\cr
\tablerule
\omit&height1.8pt&\omit&&\omit&&\omit&&\omit&\cr
&& 21 && $\partial_u \; \partial_3^{(\mu} \; \eta^{\nu) 
(\alpha_2} \; \eta^{\beta_2) (\alpha_3} \; \eta^{\beta_3) 
(\alpha_1} \; t^{\beta_1)}$ && 86 && $\frac12 \partial_u \; 
\eta^{\mu (\alpha_2} \; \eta^{\beta_2) \nu} \; \eta^{\alpha_1
(\alpha_3} \; \eta^{\beta_3) \beta_1} \; t \cdot \partial_3$ 
&\cr
\omit&height1.8pt&\omit&&\omit&&\omit&&\omit&\cr
\tablerule
\omit&height1.8pt&\omit&&\omit&&\omit&&\omit&\cr
&& 22 && $\partial_2^{(\mu} \; \eta^{\nu) (\alpha_3} \; \eta^{
\beta_3) (\alpha_2} \; \eta^{\beta_2) (\alpha_1} \; 
\partial_3^{\beta_1)}$ && 87 && $-\frac1{16} \eta^{\mu \nu} \; \eta^{
\alpha_1 \beta_1} \; \eta^{\alpha_2 (\alpha_3} \; \eta^{
\beta_3) \beta_2} \; \partial_2 \cdot \partial_3$ &\cr
\omit&height1.8pt&\omit&&\omit&&\omit&&\omit&\cr
\tablerule}}

{\bf Table~4:} {\ninepoint The partially symmetrized quartic 
pseudo-graviton vertex operators $V_i^{\mu \nu \alpha_1 \beta_1
\alpha_2 \beta_2 \alpha_3 \beta_3}$ without 
\vskip -0.6truecm \noindent \hglue 2.8truecm
the factor of $\kappa^2 \Omega^2$.}

\vfil \eject

\vbox{\tabskip=0pt \offinterlineskip
\def\tablerule{\noalign{\hrule}}
\halign to505pt {\strut#& \vrule#\tabskip=1em plus2em& \hfil#& \vrule#& 
\hfil#\hfil& \vrule#& \hfil#& \vrule#& \hfil#\hfil& \vrule#\tabskip=0pt\cr
\tablerule
\omit&height4pt&\omit&&\omit&&\omit&&\omit&\cr
&&\omit\hidewidth $i$ 
&&\omit\hidewidth {\rm Vertex Operator}\hidewidth&& 
\omit\hidewidth $i$\hidewidth&& 
\omit\hidewidth {\rm Vertex Operator}
\hidewidth&\cr
\omit&height4pt&\omit&&\omit&&\omit&&\omit&\cr
\tablerule
\omit&height1.8pt&\omit&&\omit&&\omit&&\omit&\cr
&& 23 && $\partial_u \; \eta^{\mu) (\alpha_2} \; \partial_3^{
\beta_2)} \; t^{(\alpha_1} \; \eta^{\beta_1) (\alpha_3} 
\; \eta^{\beta_3) (\nu}$ && 88 && \hidewidth $-\frac1{16} \partial_u 
\; \eta^{\mu (\alpha_3} \; \eta^{\beta_3) \nu} \; 
\eta^{\alpha_1 \beta_1} \; \eta^{\alpha_2 \beta_2} \; t \cdot 
\partial_3$ \hidewidth &\cr
\omit&height2pt&\omit&&\omit&&\omit&&\omit&\cr
\tablerule
\omit&height2pt&\omit&&\omit&&\omit&&\omit&\cr
&& 24 && $-\frac18 \eta^{\mu \nu} \; \eta^{\alpha_1 \beta_1} \;
\eta^{\alpha_2 \beta_2} \; \partial_2^{(\alpha_3} \; 
\partial_3^{\beta_3)}$ && 89 && $\frac18 \eta^{\mu (\alpha_1} \; \eta^{
\beta_1) \nu} \; \eta^{\alpha_2 (\alpha_3} \; \eta^{\beta_3)
\beta_2} \; \partial_2 \cdot \partial_3$ &\cr
\omit&height2pt&\omit&&\omit&&\omit&&\omit&\cr
\tablerule
\omit&height2pt&\omit&&\omit&&\omit&&\omit&\cr
&& 25 && $-\frac1{16} \partial_u \; \eta^{\mu \nu} \; \eta^{
\alpha_1 \beta_1} \; \eta^{\alpha_2 \beta_2} \; \partial_3^{(
\alpha_3} \; t^{\beta_3)}$ && 90 && $\frac18 \partial_u \; 
\eta^{\mu (\alpha_3} \; \eta^{\beta_3) \nu} \; \eta^{\alpha_1 
(\alpha_2} \; \eta^{\beta_2) \beta_1} \; t \cdot \partial_3$ 
&\cr
\omit&height2pt&\omit&&\omit&&\omit&&\omit&\cr
\tablerule
\omit&height2pt&\omit&&\omit&&\omit&&\omit&\cr
&& 26 && $-\frac1{16} \partial_u \; \partial_3^{(\mu} \; t^{
\nu)} \; \eta^{\alpha_1 \beta_1} \; \eta^{\alpha_2 \beta_2}
\; \eta^{\alpha_3 \beta_3}$ && 91 && \hidewidth $-\eta^{\mu) (\alpha_1}
\; \eta^{\beta_1) (\alpha_3} \; \eta^{\beta_3) (\alpha_2} 
\; \eta^{\beta_2) (\nu} \; \partial_2 \cdot \partial_3$ 
\hidewidth &\cr
\omit&height2pt&\omit&&\omit&&\omit&&\omit&\cr
\tablerule
\omit&height2pt&\omit&&\omit&&\omit&&\omit&\cr
&& 27 && $\frac14 \eta^{\mu (\alpha_1} \; \eta^{\beta_1) \nu} 
\; \eta^{\alpha_2 \beta_2} \; \partial_2^{(\alpha_3} \;
\partial_3^{\beta_3)}$ && 92 && \hidewidth $-\partial_u \; \eta^{\mu) 
(\alpha_2} \; \eta^{\beta_2) (\alpha_1} \; \eta^{\beta_1) 
(\alpha_3} \; \eta^{\beta_3) (\nu} \; t \cdot \partial_3$ 
\hidewidth &\cr
\omit&height2pt&\omit&&\omit&&\omit&&\omit&\cr
\tablerule
\omit&height2pt&\omit&&\omit&&\omit&&\omit&\cr
&& 28 && $\frac18 \partial_u \; \eta^{\mu \nu} \; \eta^{
\alpha_1 (\alpha_2} \; \eta^{\beta_2) \beta_1} \; \partial_3^{
(\alpha_3} \; t^{\beta_3)}$ && 93 && $-\frac12 \partial_2^{(\mu} 
\; \eta^{\nu) (\alpha_1} \; \partial_3^{\beta_1)} \;
\eta^{\alpha_2 (\alpha_3} \; \eta^{\beta_3) \beta_2}$ &\cr
\omit&height2pt&\omit&&\omit&&\omit&&\omit&\cr
\tablerule
\omit&height2pt&\omit&&\omit&&\omit&&\omit&\cr
&& 29 && $\frac18 \partial_u \partial_3^{(\mu} \; t^{\nu)} \;
\eta^{\alpha_1 (\alpha_2} \; \eta^{\beta_2) \beta_1} \; \eta^{
\alpha_3 \beta_3}$ && 94 && $-\frac12 \partial_u \; \eta^{\mu 
(\alpha_3} \; \eta^{\beta_3) \nu} \; t^{(\alpha_1} \;
\eta^{\beta_1) (\alpha_2} \; \partial_3^{\beta_2)}$ &\cr
\omit&height2pt&\omit&&\omit&&\omit&&\omit&\cr
\tablerule
\omit&height2pt&\omit&&\omit&&\omit&&\omit&\cr
&& 30 && $-\eta^{\mu) (\alpha_1} \; \eta^{\beta_1) (\alpha_2} 
\; \eta^{\beta_2) (\nu} \; \partial_2^{(\alpha_3} \;
\partial_3^{\beta_3)}$ && 95 && $\frac14 \eta^{\mu \nu} \; \eta^{
\alpha_1 (\alpha_2} \; \eta^{\beta_2) (\alpha_3} \; \eta^{
\beta_3) (\beta_1} \; \partial_2 \cdot \partial_3$ &\cr
\omit&height2pt&\omit&&\omit&&\omit&&\omit&\cr
\tablerule
\omit&height2pt&\omit&&\omit&&\omit&&\omit&\cr
&& 31 && \hidewidth $-\frac12 \partial_u \; \eta^{\mu) (\alpha_1} 
\; \eta^{\beta_1) (\alpha_2} \; \eta^{\beta_2) (\nu} \; 
\partial_3^{(\alpha_3} \; t^{\beta_3)}$ \hidewidth && 96 && $\frac14 
\eta^{\mu) (\alpha_2} \; \eta^{\beta_2) (\alpha_3} \; \eta^{
\beta_3) (\nu} \; \eta^{\alpha_1 \beta_1} \; \partial_2 \cdot 
\partial_3$ &\cr
\omit&height2pt&\omit&&\omit&&\omit&&\omit&\cr
\tablerule
\omit&height2pt&\omit&&\omit&&\omit&&\omit&\cr
&& 32 && \hidewidth $-\frac12 \partial_u \; \partial_3^{(\mu} 
\; t^{\nu)} \; \eta^{\alpha_1) (\alpha_2} \; \eta^{
\beta_2) (\alpha_3} \; \eta^{\beta_3) (\beta_1}$ \hidewidth && 97 && 
\hidewidth $\frac12 \partial_u \; \eta^{\mu) (\alpha_2} \; 
\eta^{\beta_2) (\alpha_3} \; \eta^{\beta_3) (\nu} \; \eta^{
\alpha_1 \beta_1} \; t \cdot \partial_3$ \hidewidth &\cr
\omit&height2pt&\omit&&\omit&&\omit&&\omit&\cr
\tablerule
\omit&height2pt&\omit&&\omit&&\omit&&\omit&\cr
&& 33 && $-\frac12 \partial_2^{(\mu} \; \eta^{\nu) (\alpha_1} 
\; \eta^{\beta_1) (\alpha_3} \; \partial_3^{\beta_3)} 
\; \eta^{\alpha_2 \beta_2}$ && 98 && $\frac18 \eta^{\mu \nu} \;
\partial_2^{(\alpha_1} \; \partial_3^{\beta_1)} \; \eta^{
\alpha_2 (\alpha_3} \; \eta^{\beta_3) \beta_2}$ &\cr
\omit&height2pt&\omit&&\omit&&\omit&&\omit&\cr
\tablerule
\omit&height2pt&\omit&&\omit&&\omit&&\omit&\cr
&& 34 && $-\frac12 \eta^{\mu) (\alpha_1} \; \partial_3^{\beta_1)} 
\; \partial_2^{(\alpha_2} \; \eta^{\beta_2) (\nu} \;
\eta^{\alpha_3 \beta_3}$ && 99 && $\frac18 \partial_2^{(\mu} \; 
\partial_3^{\nu)} \; \eta^{\alpha_1 \beta_1} \; \eta^{\alpha_2 
(\alpha_3} \; \eta^{\beta_3) \beta_2}$ &\cr
\omit&height2pt&\omit&&\omit&&\omit&&\omit&\cr
\tablerule
\omit&height2pt&\omit&&\omit&&\omit&&\omit&\cr
&& 35 && $-\frac12 \partial_u \eta^{\mu \nu} \; t^{(\alpha_1} 
\; \eta^{\beta_1) (\alpha_2} \; \eta^{\beta_2) (\alpha_3}
\; \partial_3^{\beta_3)}$ && 100 && $\frac14 \partial_u \; 
\eta^{\mu (\alpha_3} \; \eta^{\beta_3) \nu} \; \partial_3^{
(\alpha_2} \; t^{\beta_2)} \; \eta^{\alpha_1 \beta_1}$ &\cr
\omit&height2pt&\omit&&\omit&&\omit&&\omit&\cr
\tablerule
\omit&height2pt&\omit&&\omit&&\omit&&\omit&\cr
&& 36 && $-\frac12 \partial_u \; t^{(\mu} \; \eta^{\nu) 
(\alpha_2} \; \eta^{\beta_2) (\alpha_1} \; \partial_3^{\beta_1
)} \; \eta^{\alpha_3 \beta_3}$ && 101 && \hidewidth$-\frac12 \eta^{\mu) 
(\alpha_2} \; \eta^{\beta_2) (\alpha_1} \; \eta^{\beta_1) 
(\alpha_3} \; \eta^{\beta_3) (\nu} \; \partial_2 \cdot 
\partial_3$ \hidewidth &\cr
\omit&height2pt&\omit&&\omit&&\omit&&\omit&\cr
\tablerule
\omit&height2pt&\omit&&\omit&&\omit&&\omit&\cr
&& 37 && $-\frac12 \eta^{\mu) (\alpha_1} \; \partial_2^{\beta_1)}
\; \partial_3^{(\alpha_2} \; \eta^{\beta_2) (\nu} \;
\eta^{\alpha_3 \beta_3}$ && 102 && \hidewidth $-\frac12 \partial_u \; 
\eta^{\mu) (\alpha_1} \; \eta^{\beta_1) (\alpha_3} \; \eta^{
\beta_3) (\alpha_2} \; \eta^{\beta_2) (\nu} \; t \cdot 
\partial_3$ \hidewidth &\cr
\omit&height2pt&\omit&&\omit&&\omit&&\omit&\cr
\tablerule
\omit&height2pt&\omit&&\omit&&\omit&&\omit&\cr
&& 38 && $-\frac12 \partial_2^{(\mu} \; \eta^{\nu) (\alpha_1} 
\; \eta^{\beta_1) (\alpha_2} \; \partial_3^{\beta_2)}
\; \eta^{\alpha_3 \beta_3}$ && 103 && $-\frac12 \eta^{\mu) (\alpha_2}
\; \eta^{\beta_2) (\alpha_3} \; \eta^{\beta_3) (\nu} \;
\partial_2^{(\alpha_1} \; \partial_3^{\beta_1)}$ &\cr
\omit&height2pt&\omit&&\omit&&\omit&&\omit&\cr
\tablerule
\omit&height2pt&\omit&&\omit&&\omit&&\omit&\cr
&& 39 && $-\frac12 \partial_u \; \eta^{\mu \nu} \; \partial_3^{
(\alpha_2} \; \eta^{\beta_2) (\alpha_1} \; \eta^{\beta_1) 
(\alpha_3} \; t^{\beta_3)}$ && 104 && $-\frac12 \partial_2^{(\mu}
\; \partial_3^{\nu)} \; \eta^{\alpha_1) (\alpha_2} \;
\eta^{\beta_2) (\alpha_3} \; \eta^{\beta_3) (\beta_1}$ &\cr
\omit&height2pt&\omit&&\omit&&\omit&&\omit&\cr
\tablerule
\omit&height2pt&\omit&&\omit&&\omit&&\omit&\cr
&& 40 && $-\frac12 \partial_u \; \partial_1^{(\mu} \; \eta^{
\nu) (\alpha_2} \; \eta^{\beta_2) (\alpha_3} \; t^{\beta_3)}
\; \eta^{\alpha_1 \beta_1}$ && 105 && $-\partial_u \; \eta^{
\mu) (\alpha_2} \; \eta^{\beta_2) (\alpha_3} \; \eta^{\beta_3)
(\nu} \; \partial_3^{(\alpha_1} \; t^{\beta_1)}$ &\cr
\omit&height2pt&\omit&&\omit&&\omit&&\omit&\cr
\tablerule
\omit&height2pt&\omit&&\omit&&\omit&&\omit&\cr
&& 41 && $\frac14 \eta^{\mu \nu} \; \eta^{\alpha_1 (\alpha_2} 
\; \eta^{\beta_2) \beta_1} \; \partial_2^{(\alpha_3} \;
\partial_3^{\beta_3)}$ && 106 && $-\frac1{4 u} \eta^{\mu \nu} \; \eta^{
\alpha_1 \beta_1} \; \eta^{\alpha_2 \beta_2} \; \partial_2^{(
\alpha_3} \; t^{\beta_3)}$ &\cr
\omit&height2pt&\omit&&\omit&&\omit&&\omit&\cr
\tablerule
\omit&height2pt&\omit&&\omit&&\omit&&\omit&\cr
&& 42 && $\frac14 \eta^{\mu (\alpha_2} \; \eta^{\beta_2) \nu} 
\; \eta^{\alpha_1 \beta_1} \; \partial_2^{(\alpha_3} \;
\partial_3^{\beta_3)}$ && 107 && $-\frac1{8 u} \partial_u \; \eta^{\mu
\nu} \; \eta^{\alpha_1 \beta_1} \; \eta^{\alpha_2 \beta_2}
\; t^{\alpha_3} \; t^{\beta_3}$ &\cr
\omit&height2pt&\omit&&\omit&&\omit&&\omit&\cr
\tablerule
\omit&height2pt&\omit&&\omit&&\omit&&\omit&\cr
&& 43 && $\frac14 \partial_u \; \eta^{\mu (\alpha_2} \; \eta^{
\beta_2) \nu} \; \eta^{\alpha_1 \beta_1} \; \partial_3^{
(\alpha_3} \; t^{\beta_3)}$ && 108 && $-\frac1{8 u} t^{(\mu} \;
\partial_3^{\nu)} \; \eta^{\alpha_1 \beta_1} \; \eta^{\alpha_2
\beta_2} \; \eta^{\alpha_3 \beta_3}$ &\cr
\omit&height2pt&\omit&&\omit&&\omit&&\omit&\cr
\tablerule
\omit&height2pt&\omit&&\omit&&\omit&&\omit&\cr
&& 44 && $\frac14 \partial_u \; \partial_3^{(\mu} \; t^{\nu)}
\; \eta^{\alpha_1 \beta_1} \; \eta^{\alpha_2 (\alpha_3} 
\; \eta^{\beta_3) \beta_2}$ && 109 && $\frac1{2 u} \eta^{\mu (\alpha_1}
\; \eta^{\beta_1) \nu} \; \eta^{\alpha_2 \beta_2} \;
\partial_2^{(\alpha_3} \; t^{\beta_3)}$ &\cr
\omit&height2pt&\omit&&\omit&&\omit&&\omit&\cr
\tablerule}}

\vskip 0.5cm

{\bf Table~4:} {\ninepoint $(continued \; from \; previous \; page)$} 

\vfill\eject

\vbox{\tabskip=0pt \offinterlineskip
\def\tablerule{\noalign{\hrule}}
\halign to505pt {\strut#& \vrule#\tabskip=1em plus2em& \hfil#& \vrule#& 
\hfil#\hfil& \vrule#& \hfil#& \vrule#& \hfil#\hfil& \vrule#\tabskip=0pt\cr
\tablerule
\omit&height4pt&\omit&&\omit&&\omit&&\omit&\cr
&&\omit\hidewidth $i$ &&
\omit\hidewidth {\rm Vertex Operator}\hidewidth&& 
\omit\hidewidth $i$\hidewidth&&
\omit\hidewidth {\rm Vertex Operator}
\hidewidth&\cr
\omit&height4pt&\omit&&\omit&&\omit&&\omit&\cr
\tablerule
\omit&height2pt&\omit&&\omit&&\omit&&\omit&\cr
&& 45 && $\frac14 \eta^{\mu \nu} \; \partial_2^{(\alpha_1} \;
\eta^{\beta_1) (\alpha_3} \; \partial_3^{\beta_3)} \; \eta^{
\alpha_2 \beta_2}$ && 110 && $\frac1{4 u} \eta^{\mu \nu} \; \eta^{
\alpha_1 (\alpha_2} \; \eta^{\beta_2) \beta_1} \; t^{\alpha_3}
\; t^{\beta_3}$ &\cr
\omit&height2pt&\omit&&\omit&&\omit&&\omit&\cr
\tablerule
\omit&height2pt&\omit&&\omit&&\omit&&\omit&\cr
&& 46 && $\frac14 \partial_2^{(\mu} \; \eta^{\nu) (\alpha_3} \;
\partial_3^{\beta_3)} \; \eta^{\alpha_1 \beta_1} \; \eta^{
\alpha_2 \beta_2}$ && 111 && $\frac1{4 u} t^{(\mu} \; \partial_3^{\nu)}
\; \eta^{\alpha_1 (\alpha_2} \; \eta^{\beta_2) \beta_1}
\; \eta^{\alpha_3 \beta_3}$ &\cr
\omit&height2pt&\omit&&\omit&&\omit&&\omit&\cr
\tablerule
\omit&height2pt&\omit&&\omit&&\omit&&\omit&\cr
&& 47 && $\frac14 \partial_u \; \eta^{\mu \nu} \; \eta^{
\alpha_1 \beta_1} \; t^{(\alpha_2} \; \eta^{\beta_2) (\alpha_3}
\partial_3^{\beta_3)}$ && 112 && $-\frac2{u} \eta^{\mu) (\alpha_1} \;
\eta^{\beta_1) (\alpha_2} \; \eta^{\beta_2) (\nu} \; 
\partial_2^{(\alpha_3} \; t^{\beta_3)}$ &\cr
\omit&height2pt&\omit&&\omit&&\omit&&\omit&\cr
\tablerule
\omit&height2pt&\omit&&\omit&&\omit&&\omit&\cr
&& 48 && $\frac14 \partial_u \; t^{(\mu} \; \eta^{\nu) 
(\alpha_2} \; \partial_3^{\beta_2)} \; \eta^{\alpha_1 \beta_1}
\; \eta^{\alpha_3 \beta_3}$ && 113 && $-\frac1{u} \partial_u 
\; \eta^{\mu) (\alpha_1} \; \eta^{\beta_1) (\alpha_2} 
\; \eta^{\beta_2) (\nu} \; t^{\alpha_3} \; t^{\beta_3}$
&\cr
\omit&height2pt&\omit&&\omit&&\omit&&\omit&\cr
\tablerule
\omit&height2pt&\omit&&\omit&&\omit&&\omit&\cr
&& 49 && $\frac14 \eta^{\mu \nu} \; \partial_3^{(\alpha_1} \;
\eta^{\beta_1) (\alpha_3} \; \partial_2^{\beta_3)} \; \eta^{
\alpha_2 \beta_2}$ && 114 && $-\frac1{u} t^{(\mu} \; \partial_3^{\nu)}
\; \eta^{\alpha_1) (\alpha_2} \; \eta^{\beta_2) (\alpha_3}
\; \eta^{\beta_3) (\beta_1}$ &\cr
\omit&height2pt&\omit&&\omit&&\omit&&\omit&\cr
\tablerule
\omit&height2pt&\omit&&\omit&&\omit&&\omit&\cr
&& 50 && $\frac14 \partial_3^{(\mu} \; \eta^{\nu) (\alpha_3} \;
\partial_2^{\beta_3)} \; \eta^{\alpha_1 \beta_1} \; \eta^{
\alpha_2 \beta_2}$ && 115 && $-\frac1{u} \partial_2^{(\mu} \; \eta^{
\nu) (\alpha_1} \; \eta^{\beta_1) (\alpha_3} \; t^{\beta_3)}
\; \eta^{\alpha_2 \beta_2}$ &\cr
\omit&height2pt&\omit&&\omit&&\omit&&\omit&\cr
\tablerule
\omit&height2pt&\omit&&\omit&&\omit&&\omit&\cr
&& 51 && $\frac14 \partial_u \; \eta^{\mu \nu} \; \eta^{
\alpha_1 \beta_1} \; \partial_3^{(\alpha_2} \; \eta^{\beta_2) 
(\alpha_3} \; t^{\beta_3)}$ && 116 && $-\frac1{u} \eta^{\mu) (\alpha_1}
\; \partial_2^{\beta_1)} \; t^{(\alpha_3} \; \eta^{
\beta_3) (\nu} \; \eta^{\alpha_2 \beta_2}$ &\cr
\omit&height2pt&\omit&&\omit&&\omit&&\omit&\cr
\tablerule
\omit&height2pt&\omit&&\omit&&\omit&&\omit&\cr
&& 52 && $\frac14 \partial_u \; \partial_3^{(\mu} \; \eta^{\nu)
(\alpha_2} \; t^{\beta_2)} \; \eta^{\alpha_1 \beta_1} 
\; \eta^{\alpha_3 \beta_3}$ && 117 && $-\frac1{u} \; \partial_u
\; \eta^{\mu \nu} \; t^{(\alpha_1} \; \eta^{\beta_1)
(\alpha_2} \; \eta^{\beta_2) (\alpha_3} \; t^{\beta_3)}$ &\cr
\omit&height2pt&\omit&&\omit&&\omit&&\omit&\cr
\tablerule
\omit&height2pt&\omit&&\omit&&\omit&&\omit&\cr
&& 53 && $-\frac12 \eta^{\mu (\alpha_2} \; \eta^{\beta_2) \nu} 
\; \partial_2^{(\alpha_1} \; \eta^{\beta_1) (\alpha_3} 
\; \partial_3^{\beta_3)}$ && 118 && $-\frac1{u} t^{(\mu} \;
\eta^{\nu) (\alpha_1} \; \eta^{\beta_1) (\alpha_2} \; 
\partial_3^{\beta_2)} \; \eta^{\alpha_3 \beta_3}$ &\cr
\omit&height2pt&\omit&&\omit&&\omit&&\omit&\cr
\tablerule
\omit&height2pt&\omit&&\omit&&\omit&&\omit&\cr
&& 54 && $-\frac12 \partial_2^{(\mu} \; \eta^{\nu) (\alpha_3} 
\; \partial_3^{\beta_3)} \; \eta^{\alpha_1 (\alpha_2} 
\; \eta^{\beta_2) \beta_1}$ && 119 && $\frac1{2 u} \eta^{\mu \nu} 
\; \eta^{\alpha_1 (\alpha_2} \; \eta^{\beta_2) \alpha_1}
\; \partial_2^{(\alpha_3} \; t^{\beta_3)}$ &\cr
\omit&height2pt&\omit&&\omit&&\omit&&\omit&\cr
\tablerule
\omit&height2pt&\omit&&\omit&&\omit&&\omit&\cr
&& 55 && $-\frac12 \partial_u \; \eta^{\mu (\alpha_1} \; \eta^{
\beta_1) \nu} \; t^{(\alpha_2} \; \eta^{\beta_2) (\alpha_3}
\; \partial_3^{\beta_3)}$ && 120 && $\frac1{2 u} \eta^{\mu (\alpha_2} 
\; \eta^{\beta_2) \nu} \; \eta^{\alpha_1 \beta_1} \; 
\partial_2^{(\alpha_3} \; t^{\beta_3)}$ &\cr
\omit&height2pt&\omit&&\omit&&\omit&&\omit&\cr
\tablerule
\omit&height2pt&\omit&&\omit&&\omit&&\omit&\cr
&& 56 && $-\frac12 \partial_u \; t^{(\mu} \; \eta^{\nu) 
(\alpha_2} \partial_3^{\beta_2)} \; \eta^{\alpha_1 (\alpha_3} 
\; \eta^{\beta_3) \beta_1}$ && 121 && $\frac1{2 u} \partial_u 
\; \eta^{\mu (\alpha_2} \; \eta^{\beta_2) \nu} \; 
\eta^{\alpha_1 \beta_1} \; t^{\alpha_3} \; t^{\beta_3}$ &\cr
\omit&height2pt&\omit&&\omit&&\omit&&\omit&\cr
\tablerule
\omit&height2pt&\omit&&\omit&&\omit&&\omit&\cr
&& 57 && $-\frac12 \eta^{\mu (\alpha_2} \; \eta^{\beta_2) \nu} 
\; \partial_3^{(\alpha_1} \; \eta^{\beta_1) (\alpha_3} 
\; \partial_2^{\beta_3)}$ && 122 && $\frac1{2 u} t^{(\mu} \;
\partial_3^{\nu)} \; \eta^{\alpha_1 \beta_1} \; \eta^{\alpha_2
(\alpha_3} \; \eta^{\beta_3) \beta_2}$ &\cr
\omit&height2pt&\omit&&\omit&&\omit&&\omit&\cr
\tablerule
\omit&height2pt&\omit&&\omit&&\omit&&\omit&\cr
&& 58 && $-\frac12 \partial_3^{(\mu} \; \eta^{\nu) (\alpha_3} 
\; \partial_2^{\beta_3)} \; \eta^{\alpha_1 (\alpha_2}
\; \eta^{\beta_2) \beta_1}$ && 123 && $\frac1{2 u} \eta^{\mu \nu} 
\; \partial_2^{(\alpha_1} \; \eta^{\beta_1) (\alpha_3} 
\; t^{\beta_3)} \; \eta^{\alpha_2 \beta_2}$ &\cr
\omit&height2pt&\omit&&\omit&&\omit&&\omit&\cr
\tablerule
\omit&height2pt&\omit&&\omit&&\omit&&\omit&\cr
&& 59 && $-\frac12 \partial_u \; \eta^{\mu (\alpha_1} \; \eta^{
\beta_1) \nu} \; \partial_3^{(\alpha_2} \; \eta^{\beta_2) 
(\alpha_3} \; t^{\beta_3)}$ && 124 && $\frac1{2 u} \partial_2^{(\mu}
\; \eta^{\nu) (\alpha_3} \; t^{\beta_3)} \; \eta^{
\alpha_1 \beta_1} \; \eta^{\alpha_2 \beta_2}$ &\cr
\omit&height2pt&\omit&&\omit&&\omit&&\omit&\cr
\tablerule
\omit&height2pt&\omit&&\omit&&\omit&&\omit&\cr
&& 60 && $-\frac12 \partial_u \; \partial_2^{(\mu} \; \eta^{
\nu) (\alpha_3} \; t^{\beta_3)} \; \eta^{\alpha_1 (\alpha_2}
\; \eta^{\beta_2) \beta_1}$ && 125 && $\frac1{2 u} \partial_u 
\; \eta^{\mu \nu} \; \eta^{\alpha_1 \beta_1} \; t^{(
\alpha_2} \; \eta^{\beta_2) (\alpha_3} \; t^{\beta_3)}$ &\cr
\omit&height2pt&\omit&&\omit&&\omit&&\omit&\cr
\tablerule
\omit&height2pt&\omit&&\omit&&\omit&&\omit&\cr
&& 61 && $-\frac12 \partial_2^{(\mu} \; \eta^{\nu) (\alpha_3} 
\; \eta^{\beta_3) (\alpha_1} \; \partial_3^{\beta_1)} \eta^{
\alpha_2 \beta_2}$ && 126 && $\frac1{2 u} t^{(\mu} \; \eta^{\nu) 
(\alpha_2} \; \partial_3^{\beta_2)} \; \eta^{\alpha_1 \beta_1} 
\; \eta^{\alpha_3 \beta_3}$ &\cr
\omit&height2pt&\omit&&\omit&&\omit&&\omit&\cr
\tablerule
\omit&height2pt&\omit&&\omit&&\omit&&\omit&\cr
&& 62 && $-\frac12 \partial_3^{(\mu} \; \eta^{\nu) (\alpha_3} 
\; \eta^{\beta_3) (\alpha_1} \; \partial_2^{\beta_1)}
\; \eta^{\alpha_2 \beta_2}$ && 127 && $-\frac1{u} \eta^{\mu (\alpha_2}
\; \eta^{\beta_2) \nu} \; \partial_2^{(\alpha_1} \;
\eta^{\beta_1) (\alpha_3} \; t^{\beta_3)}$ &\cr
\omit&height2pt&\omit&&\omit&&\omit&&\omit&\cr
\tablerule
\omit&height2pt&\omit&&\omit&&\omit&&\omit&\cr
&& 63 && $-\frac12 \partial_u \; \eta^{\mu \nu} \; t^{
(\alpha_1} \; \eta^{\beta_1) (\alpha_3} \; \eta^{\beta_3)
(\alpha_2} \partial_3^{\beta_2)}$ && 128 && $-\frac1{u} \partial_2^{(\mu}
\; \eta^{\nu) (\alpha_3} \; t^{\beta_3)} \; \eta^{
\alpha_1 (\alpha_2} \; \eta^{\beta_2) \beta_1}$ &\cr
\omit&height2pt&\omit&&\omit&&\omit&&\omit&\cr
\tablerule
\omit&height2pt&\omit&&\omit&&\omit&&\omit&\cr
&& 64 && $-\frac12 \partial_u \; \eta^{\mu) (\alpha_1} \; t^{
\beta_1)} \; \partial_3^{(\alpha_2} \; \eta^{\beta_2) (\nu}
\eta^{\alpha_3 \beta_3}$ && 129 && $-\frac1{u} \partial_u \; \eta^{\mu
(\alpha_1} \; \eta^{\beta_1) \nu} \; t^{(\alpha_2} \;
\eta^{\beta_2) (\alpha_3} \; t^{\beta_3)}$ &\cr
\omit&height2pt&\omit&&\omit&&\omit&&\omit&\cr
\tablerule
\omit&height2pt&\omit&&\omit&&\omit&&\omit&\cr
&& 65 && $\frac1{16} \eta^{\mu \nu} \; \eta^{\alpha_1 \beta_1} 
\; \eta^{\alpha_2 \beta_2} \; \eta^{\alpha_3 \beta_3} 
\; \partial_2 \cdot \partial_3$ && 130 && $-\frac1{u} t^{(\mu} 
\; \eta^{\nu) (\alpha_1} \; \partial_3^{\beta_1)} \;
\eta^{\alpha_2 (\alpha_3} \; \eta^{\beta_3) \beta_2}$ &\cr
\omit&height2pt&\omit&&\omit&&\omit&&\omit&\cr
\tablerule}}

\vskip 0.5cm

{\bf Table~4:} {\ninepoint $(continued \; from \; previous \; page)$} 

\vfill\eject

\vbox{\tabskip=0pt \offinterlineskip
\def\tablerule{\noalign{\hrule}}
\halign to450pt {\strut#& \vrule#\tabskip=1em plus2em& \hfil#& \vrule#& 
\hfil#\hfil& \vrule#& \hfil#& \vrule#& \hfil#\hfil& \vrule#\tabskip=0pt\cr
\tablerule
\omit&height4pt&\omit&&\omit&&\omit&&\omit&\cr
&&\omit\hidewidth $j$ &&
\omit\hidewidth {\rm Vertex Operator}\hidewidth&& 
\omit\hidewidth $j$\hidewidth&& 
\omit\hidewidth {\rm Vertex Operator}
\hidewidth&\cr
\omit&height4pt&\omit&&\omit&&\omit&&\omit&\cr
\tablerule
\omit&height1.8pt&\omit&&\omit&&\omit&&\omit&\cr
&& 1 && $-\frac1{2u'} \eta^{\rho_1 \sigma_1} \; \eta^{\rho_2 \sigma_2}
\; \partial_2^{(\rho_3} \; t^{\sigma_3)}$ && 
39 && $-\frac14 \eta^{\rho_2 \sigma_2} \; \eta^{\rho_3 (\rho_1} 
\; \eta^{\sigma_1) \sigma_3} \; \partial_3 \cdot \partial_1$ &\cr
\omit&height1.8pt&\omit&&\omit&&\omit&&\omit&\cr
\tablerule
\omit&height1.8pt&\omit&&\omit&&\omit&&\omit&\cr
&& 2 && $-\frac1{2u'} \eta^{\rho_2 \sigma_2} \; \eta^{\rho_3 \sigma_3}
\; \partial_3^{(\rho_1} \; t^{\sigma_1)}$ && 
40 && $\eta^{\rho_1) (\rho_2} \; \eta^{\sigma_2) (\rho_3} \;
\eta^{\sigma_3) (\sigma_1} \; \partial_2 \cdot \partial_3$ &\cr
\omit&height1.8pt&\omit&&\omit&&\omit&&\omit&\cr
\tablerule
\omit&height1.8pt&\omit&&\omit&&\omit&&\omit&\cr
&& 3 && $-\frac1{2u'} \eta^{\rho_3 \sigma_3} \; \eta^{\rho_1 \sigma_1}
\; \partial_1^{(\rho_2} \; t^{\sigma_2)}$ && 
41 && $\eta^{\rho_1) (\rho_2} \; \eta^{\sigma_2) (\rho_3} 
\; \eta^{\sigma_3) (\sigma_1} \; \partial_3 \cdot \partial_1$ &\cr
\omit&height1.8pt&\omit&&\omit&&\omit&&\omit&\cr
\tablerule
\omit&height1.8pt&\omit&&\omit&&\omit&&\omit&\cr
&& 4 && $\frac1{u'} \eta^{\rho_1 (\rho_2} \; \eta^{\sigma_2) \sigma_1}
\; \partial_2^{(\rho_3} \; t^{\sigma_3)}$ && 
42 && $\frac12 \partial_2^{(\rho_1} \; \partial_3^{\sigma_1)} \; \eta^{
\rho_2 (\rho_3} \; \eta^{\sigma_3) \sigma_2}$ &\cr 
\omit&height1.8pt&\omit&&\omit&&\omit&&\omit&\cr
\tablerule
\omit&height1.8pt&\omit&&\omit&&\omit&&\omit&\cr
&& 5 && $\frac1{u'} \eta^{\rho_2 (\rho_3} \; \eta^{\sigma_3) \sigma_2}
\; \partial_3^{(\rho_1} \; t^{\sigma_1)}$ && 
43 && $\frac12 \partial_3^{(\rho_2} \; \partial_1^{\sigma_2)} \; \eta^{
\rho_3 (\rho_1} \; \eta^{\sigma_1) \sigma_3}$ &\cr 
\omit&height1.8pt&\omit&&\omit&&\omit&&\omit&\cr
\tablerule
\omit&height1.8pt&\omit&&\omit&&\omit&&\omit&\cr
&& 6 && $\frac1{u'} \eta^{\rho_3 (\rho_1} \; \eta^{\sigma_1) \sigma_3}
\; \partial_1^{(\rho_2} \; t^{\sigma_2)}$ && 
44 && $- \frac1{2u'} \eta^{\rho_1 \sigma_1} \; \eta^{\rho_3 \sigma_3} 
\; \partial_3^{(\rho_2} \; t^{\sigma_2)}$ &\cr
\omit&height1.8pt&\omit&&\omit&&\omit&&\omit&\cr
\tablerule
\omit&height1.8pt&\omit&&\omit&&\omit&&\omit&\cr
&& 7 && $\frac1{u'} t^{(\rho_3} \; \eta^{\sigma_3) (\rho_1}
\; \partial_2^{\sigma_1)} \; \eta^{\rho_2 \sigma_2}$ && 
45 && $- \frac1{2u'} \eta^{\rho_3 \sigma_3} \; \eta^{\rho_2 \sigma_2} 
\; \partial_2^{(\rho_1} \; t^{\sigma_1)}$ &\cr
\omit&height1.8pt&\omit&&\omit&&\omit&&\omit&\cr
\tablerule
\omit&height1.8pt&\omit&&\omit&&\omit&&\omit&\cr
&& 8 && $\frac1{u'} t^{(\rho_1} \; \eta^{\sigma_1) (\rho_2} 
\; \partial_3^{\sigma_2)} \; \eta^{\rho_3 \sigma_3}$ && 
46 && $- \frac1{2u'} \eta^{\rho_2 \sigma_2} \; \eta^{\rho_1 \sigma_1} 
\; \partial_1^{(\rho_3} \; t^{\sigma_3)}$ &\cr
\omit&height1.8pt&\omit&&\omit&&\omit&&\omit&\cr
\tablerule
\omit&height1.8pt&\omit&&\omit&&\omit&&\omit&\cr
&& 9 && $\frac1{u'} t^{(\rho_2} \; \eta^{\sigma_2) (\rho_3} 
\; \partial_1^{\sigma_3)} \; \eta^{\rho_1 \sigma_1}$ && 
47 && $\frac1{u'} \eta^{\rho_1 (\rho_3} \; \eta^{\sigma_3) \sigma_1}
\; \partial_3^{(\rho_2} \; t^{\sigma_2)}$ &\cr
\omit&height1.8pt&\omit&&\omit&&\omit&&\omit&\cr
\tablerule
\omit&height1.8pt&\omit&&\omit&&\omit&&\omit&\cr
&& 10 && $\frac12 \eta^{\rho_1 \sigma_1} \; \partial_3^{(\rho_2} 
\; \eta^{\sigma_2) (\rho_3} \; \partial_2^{\sigma_3)}$ && 
48 && $\frac1{u'} \eta^{\rho_3 (\rho_2} \; \eta^{\sigma_2) \sigma_3}
\; \partial_2^{(\rho_1} \; t^{\sigma_1)}$ &\cr
\omit&height1.8pt&\omit&&\omit&&\omit&&\omit&\cr
\tablerule
\omit&height1.8pt&\omit&&\omit&&\omit&&\omit&\cr
&& 11 && $\frac12 \eta^{\rho_2 \sigma_2} \; \partial_1^{(\rho_3} 
\; \eta^{\sigma_3) (\rho_1} \; \partial_3^{\sigma_1)}$ && 
49 && $\frac1{u'} \eta^{\rho_2 (\rho_1} \; \eta^{\sigma_1) \sigma_2}
\; \partial_1^{(\rho_3} \; t^{\sigma_3)}$ &\cr
\omit&height1.8pt&\omit&&\omit&&\omit&&\omit&\cr
\tablerule
\omit&height1.8pt&\omit&&\omit&&\omit&&\omit&\cr
&& 12 && $\frac12 \eta^{\rho_3 \sigma_3} \; \partial_2^{(\rho_1} 
\; \eta^{\sigma_1) (\rho_2} \; \partial_1^{\sigma_2)}$ && 
50 && $\frac1{u'} \partial_3^{(\rho_1} \; \eta^{\sigma_1) (\rho_2} 
\; t^{\sigma_2)} \eta^{\rho_3 \sigma_3}$ &\cr
\omit&height1.8pt&\omit&&\omit&&\omit&&\omit&\cr
\tablerule
\omit&height1.8pt&\omit&&\omit&&\omit&&\omit&\cr
&& 13 && $-\partial_3^{(\rho_1} \; \eta^{\sigma_1) (\rho_2} 
\; \eta^{\sigma_2) (\rho_3} \; \partial_2^{\sigma_3)}$ && 
51 && $\frac1{u'} \partial_2^{(\rho_3} \; \eta^{\sigma_3) (\rho_1}
\; t^{\sigma_1)} \eta^{\rho_2 \sigma_2}$ &\cr
\omit&height1.8pt&\omit&&\omit&&\omit&&\omit&\cr
\tablerule
\omit&height1.8pt&\omit&&\omit&&\omit&&\omit&\cr
&& 14 && $-\partial_1^{(\rho_2} \; \eta^{\sigma_2) (\rho_3} 
\; \eta^{\sigma_3) (\rho_1} \; \partial_3^{\sigma_1)}$ && 
52 && $\frac1{u'} \partial_1^{(\rho_2} \; \eta^{\sigma_2) (\rho_3} 
\; t^{\sigma_3)} \eta^{\rho_1 \sigma_1}$ &\cr
\omit&height1.8pt&\omit&&\omit&&\omit&&\omit&\cr
\tablerule
\omit&height1.8pt&\omit&&\omit&&\omit&&\omit&\cr
&& 15 && $-\partial_2^{(\rho_3} \; \eta^{\sigma_3) (\rho_1} 
\; \eta^{\sigma_1) (\rho_2} \; \partial_1^{\sigma_2)}$ && 
53 && $-\partial_2^{(\rho_1} \; \eta^{\sigma_1) (\rho_3} \; \eta^{
\sigma_3) (\rho_2} \; \partial_3^{\sigma_2)}$ &\cr
\omit&height1.8pt&\omit&&\omit&&\omit&&\omit&\cr
\tablerule
\omit&height1.8pt&\omit&&\omit&&\omit&&\omit&\cr
&& 16 && $- \partial_3^{(\rho_2} \; \eta^{\sigma_2) (\rho_1} 
\; \eta^{\sigma_1) (\rho_3} \; \partial_2^{\sigma_3)}$ && 
54 && $-\partial_1^{(\rho_3} \; \eta^{\sigma_3) (\rho_2} \; \eta^{
\sigma_2) (\rho_1} \; \partial_2^{\sigma_1)}$ &\cr
\omit&height1.8pt&\omit&&\omit&&\omit&&\omit&\cr
\tablerule
\omit&height1.8pt&\omit&&\omit&&\omit&&\omit&\cr
&& 17 && $- \partial_1^{(\rho_3} \; \eta^{\sigma_3) (\rho_2} 
\; \eta^{\sigma_2) (\rho_1} \; \partial_3^{\sigma_1)}$ && 
55 && $-\partial_3^{(\rho_2} \; \eta^{\sigma_2) (\rho_1} \; \eta^{
\sigma_1) (\rho_3} \; \partial_1^{\sigma_3)}$ &\cr
\omit&height1.8pt&\omit&&\omit&&\omit&&\omit&\cr
\tablerule
\omit&height1.8pt&\omit&&\omit&&\omit&&\omit&\cr
&& 18 && $- \partial_2^{(\rho_1} \; \eta^{\sigma_1) (\rho_3} 
\; \eta^{\sigma_3) (\rho_2} \; \partial_1^{\sigma_2)}$ && 
56 && $-\frac14 \eta^{\rho_1 \sigma_1} \; \eta^{\rho_3 \sigma_3} \;
\partial_2^{(\rho_2} \; \partial_3^{\sigma_2)}$ &\cr
\omit&height1.8pt&\omit&&\omit&&\omit&&\omit&\cr
\tablerule
\omit&height1.8pt&\omit&&\omit&&\omit&&\omit&\cr
&& 19 && $-\frac14 \eta^{\rho_1 \sigma_1} \; \eta^{\rho_2 \sigma_2}
\; \partial_2^{(\rho_3} \; \partial_3^{\sigma_3)}$ && 
57 && $-\frac14 \eta^{\rho_3 \sigma_3} \; \eta^{\rho_2 \sigma_2} \;
\partial_1^{(\rho_1} \; \partial_2^{\sigma_1)}$ &\cr
\omit&height1.8pt&\omit&&\omit&&\omit&&\omit&\cr
\tablerule}}

\vskip 0.5cm

{\bf Table~5:} {\ninepoint The fully symmetrized cubic pseudo-graviton 
vertex operators $V_j^{\rho_1 \sigma_1 \rho_2 \sigma_2 \rho_3 \sigma_3}$ 
without 
\vskip -0.6truecm \noindent \hglue 3.05truecm
the factor of $\kappa {\Omega'}^2$.}

\vfill\eject

\vbox{\tabskip=0pt \offinterlineskip
\def\tablerule{\noalign{\hrule}}
\halign to450pt {\strut#& \vrule#\tabskip=1em plus2em& \hfil#& \vrule#& 
\hfil#\hfil& \vrule#& \hfil#& \vrule#& \hfil#\hfil& \vrule#\tabskip=0pt\cr
\tablerule
\omit&height4pt&\omit&&\omit&&\omit&&\omit&\cr
&&\omit\hidewidth $j$ &&
\omit\hidewidth {\rm Vertex Operator}\hidewidth&& 
\omit\hidewidth $j$\hidewidth&& 
\omit\hidewidth {\rm Vertex Operator}
\hidewidth&\cr
\omit&height4pt&\omit&&\omit&&\omit&&\omit&\cr
\tablerule
\omit&height1.8pt&\omit&&\omit&&\omit&&\omit&\cr
&& 20 && $-\frac14 \eta^{\rho_2 \sigma_2} \; \eta^{\rho_3 \sigma_3} 
\; \partial_3^{(\rho_1} \; \partial_1^{\sigma_1)}$ && 
58 && $-\frac14 \eta^{\rho_2 \sigma_2} \; \eta^{\rho_1 \sigma_1} \;
\partial_3^{(\rho_3} \; \partial_1^{\sigma_3)}$ &\cr
\omit&height1.8pt&\omit&&\omit&&\omit&&\omit&\cr
\tablerule
\omit&height1.8pt&\omit&&\omit&&\omit&&\omit&\cr
&& 21 && $-\frac14 \eta^{\rho_3 \sigma_3} \; \eta^{\rho_1 \sigma_1} 
\; \partial_1^{(\rho_2} \; \partial_2^{\sigma_2)}$ && 
59 && $\frac12 \eta^{\rho_1 (\rho_3} \; \eta^{\sigma_3) \sigma_1}
\; \partial_2^{(\rho_2} \; \partial_3^{\sigma_2)}$ &\cr
\omit&height1.8pt&\omit&&\omit&&\omit&&\omit&\cr
\tablerule
\omit&height1.8pt&\omit&&\omit&&\omit&&\omit&\cr
&& 22 && $\frac12 \eta^{\rho_1 (\rho_2} \; \eta^{\sigma_2) \sigma_1} 
\; \partial_2^{(\rho_3} \; \partial_3^{\sigma_3)}$ && 
60 && $\frac12 \eta^{\rho_3 (\rho_2} \; \eta^{\sigma_2) \sigma_3}
\; \partial_1^{(\rho_1} \; \partial_2^{\sigma_1)}$ &\cr
\omit&height1.8pt&\omit&&\omit&&\omit&&\omit&\cr
\tablerule
\omit&height1.8pt&\omit&&\omit&&\omit&&\omit&\cr
&& 23 && $\frac12 \eta^{\rho_2 (\rho_3} \; \eta^{\sigma_3) \sigma_2} 
\; \partial_3^{(\rho_1} \; \partial_1^{\sigma_1)}$ && 
61 && $\frac12 \eta^{\rho_2 (\rho_1} \; \eta^{\sigma_1) \sigma_2}
\; \partial_3^{(\rho_3} \; \partial_1^{\sigma_3)}$ &\cr
\omit&height2pt&\omit&&\omit&&\omit&&\omit&\cr
\tablerule
\omit&height2pt&\omit&&\omit&&\omit&&\omit&\cr
&& 24 && $\frac12 \eta^{\rho_3 (\rho_1} \; \eta^{\sigma_1) \sigma_3} 
\; \partial_1^{(\rho_2} \; \partial_2^{\sigma_2)}$ && 
62 && $\frac12 \partial_3^{(\rho_1} \; \eta^{\sigma_1) (\rho_2} \;
\partial_2^{\sigma_2)} \; \eta^{\rho_3 \sigma_3}$ &\cr
\omit&height2pt&\omit&&\omit&&\omit&&\omit&\cr
\tablerule
\omit&height2pt&\omit&&\omit&&\omit&&\omit&\cr
&& 25 && $\frac12 \partial_2^{(\rho_1} \; \eta^{\sigma_1) (\rho_3} 
\; \partial_3^{\sigma_3)} \; \eta^{\rho_2 \sigma_2}$ && 
63 && $\frac12 \partial_2^{(\rho_3} \; \eta^{\sigma_3) (\rho_1} \;
\partial_1^{\sigma_1)} \; \eta^{\rho_2 \sigma_2}$ &\cr
\omit&height2pt&\omit&&\omit&&\omit&&\omit&\cr
\tablerule
\omit&height2pt&\omit&&\omit&&\omit&&\omit&\cr
&& 26 && $\frac12 \partial_3^{(\rho_2} \; \eta^{\sigma_2) (\rho_1} 
\; \partial_1^{\sigma_1)} \; \eta^{\rho_3 \sigma_3}$ && 
64 && $\frac12 \partial_1^{(\rho_2} \; \eta^{\sigma_2) (\rho_3} \;
\partial_3^{\sigma_3)} \; \eta^{\rho_1 \sigma_1}$ &\cr
\omit&height2pt&\omit&&\omit&&\omit&&\omit&\cr
\tablerule
\omit&height2pt&\omit&&\omit&&\omit&&\omit&\cr
&& 27 && $\frac12 \partial_1^{(\rho_3} \; \eta^{\sigma_3) (\rho_2} 
\; \partial_2^{\sigma_2)} \; \eta^{\rho_1 \sigma_1}$ && 
65 && $\frac12 \partial_2^{(\rho_3} \; \eta^{\sigma_3) (\rho_1} \;
\partial_3^{\sigma_1)} \; \eta^{\rho_2 \sigma_2}$ &\cr
\omit&height2pt&\omit&&\omit&&\omit&&\omit&\cr
\tablerule
\omit&height2pt&\omit&&\omit&&\omit&&\omit&\cr
&& 28 && $\frac12 \partial_2^{(\rho_1} \; \eta^{\sigma_1) (\rho_2} 
\; \partial_3^{\sigma_2)} \; \eta^{\rho_3 \sigma_3}$ && 
66 && $\frac12 \partial_1^{(\rho_2} \; \eta^{\sigma_2) (\rho_3} \;
\partial_2^{\sigma_3)} \; \eta^{\rho_1 \sigma_1}$ &\cr
\omit&height2pt&\omit&&\omit&&\omit&&\omit&\cr
\tablerule
\omit&height2pt&\omit&&\omit&&\omit&&\omit&\cr
&& 29 && $\frac12 \partial_3^{(\rho_2} \; \eta^{\sigma_2) (\rho_3} 
\; \partial_1^{\sigma_3)} \; \eta^{\rho_1 \sigma_1}$ && 
67 && $\frac12 \partial_3^{(\rho_1} \; \eta^{\sigma_1) (\rho_2} 
\; \partial_1^{\sigma_2)} \; \eta^{\rho_3 \sigma_3}$ &\cr
\omit&height2pt&\omit&&\omit&&\omit&&\omit&\cr
\tablerule
\omit&height2pt&\omit&&\omit&&\omit&&\omit&\cr
&& 30 && $\frac12 \partial_1^{(\rho_3} \; \eta^{\sigma_3) (\rho_1} 
\; \partial_2^{\sigma_1)} \; \eta^{\rho_2 \sigma_2}$ && 
68 && $\frac14 \eta^{\rho_1 \sigma_1} \; \eta^{\rho_2 \sigma_2}
\; \eta^{\rho_3 \sigma_3} \; \partial_1 \cdot \partial_2$ &\cr
\omit&height2pt&\omit&&\omit&&\omit&&\omit&\cr
\tablerule
\omit&height2pt&\omit&&\omit&&\omit&&\omit&\cr
&& 31 && $\frac14 \eta^{\rho_1 \sigma_1} \; \eta^{\rho_2 \sigma_2} 
\; \eta^{\rho_3 \sigma_3} \; \partial_2 \cdot \partial_3$ && 
69 && $-\frac12 \eta^{\rho_1 (\rho_3} \; \eta^{\sigma_3) \sigma_1}
\; \eta^{\rho_2 \sigma_2} \; \partial_2 \cdot \partial_3$ &\cr
\omit&height2pt&\omit&&\omit&&\omit&&\omit&\cr
\tablerule
\omit&height2pt&\omit&&\omit&&\omit&&\omit&\cr
&& 32 && $\frac14 \eta^{\rho_1 \sigma_1} \; \eta^{\rho_2 \sigma_2} 
\; \eta^{\rho_3 \sigma_3} \; \partial_3 \cdot \partial_1$ && 
70 && $-\frac12 \eta^{\rho_3 (\rho_2} \; \eta^{\sigma_2) \sigma_3}
\; \eta^{\rho_1 \sigma_1} \; \partial_1 \cdot \partial_2$ &\cr
\omit&height2pt&\omit&&\omit&&\omit&&\omit&\cr
\tablerule
\omit&height2pt&\omit&&\omit&&\omit&&\omit&\cr
&& 33 && $-\frac12 \eta^{\rho_1 (\rho_2} \; \eta^{\sigma_2) \sigma_1}
\; \eta^{\rho_3 \sigma_3} \; \partial_2 \cdot \partial_3$ && 
71 && $-\frac12 \eta^{\rho_1 (\rho_2} \; \eta^{\sigma_2) \sigma_1}
\; \eta^{\rho_3 \sigma_3} \; \partial_3 \cdot \partial_1$ &\cr
\omit&height2pt&\omit&&\omit&&\omit&&\omit&\cr
\tablerule
\omit&height2pt&\omit&&\omit&&\omit&&\omit&\cr
&& 34 && $-\frac12 \eta^{\rho_2 (\rho_3} \; \eta^{\sigma_3) \sigma_2}
\; \eta^{\rho_1 \sigma_1} \; \partial_3 \cdot \partial_1$ && 
72 && $-\frac12 \eta^{\rho_1 \sigma_1} \; \eta^{\rho_2 \sigma_2} \;
\partial_1^{(\rho_3} \; \partial_2^{\sigma_3)}$ &\cr
\omit&height2pt&\omit&&\omit&&\omit&&\omit&\cr
\tablerule
\omit&height2pt&\omit&&\omit&&\omit&&\omit&\cr
&& 35 && $-\frac12 \eta^{\rho_3 (\rho_1} \; \eta^{\sigma_1) \sigma_3}
\; \eta^{\rho_2 \sigma_2} \; \partial_1 \cdot \partial_2$ && 
73 && $-\frac14 \eta^{\rho_1 (\rho_2} \; \eta^{\sigma_2) \sigma_1}
\; \eta^{\rho_3 \sigma_3} \; \partial_1 \cdot \partial_2 $ &\cr
\omit&height2pt&\omit&&\omit&&\omit&&\omit&\cr
\tablerule
\omit&height2pt&\omit&&\omit&&\omit&&\omit&\cr
&& 36 && $-\frac12 \partial_2^{(\rho_1} \; \partial_3^{\sigma_1)} 
\; \eta^{\rho_2 \sigma_2} \; \eta^{\rho_3 \sigma_3}$ && 
74 && $\eta^{\rho_1) (\rho_2} \; \eta^{\sigma_2) (\rho_3} \; 
\eta^{\sigma_3) (\sigma_1} \; \partial_1 \cdot \partial_2$ &\cr
\omit&height2pt&\omit&&\omit&&\omit&&\omit&\cr
\tablerule
\omit&height2pt&\omit&&\omit&&\omit&&\omit&\cr
&& 37 && $-\frac12 \partial_3^{(\rho_2} \; \partial_1^{\sigma_2)} 
\; \eta^{\rho_3 \sigma_3} \; \eta^{\rho_1 \sigma_1}$ && 
75 && $\frac12 \partial_1^{(\rho_3} \; \partial_2^{\sigma_3)} 
\; \eta^{\rho_1 (\rho_2} \; \eta^{\sigma_2) \sigma_1}$ &\cr
\omit&height2pt&\omit&&\omit&&\omit&&\omit&\cr
\tablerule
\omit&height2pt&\omit&&\omit&&\omit&&\omit&\cr
&& 38 && $-\frac14 \eta^{\rho_1 \sigma_1} \; \eta^{\rho_2 (\rho_3} 
\; \eta^{\sigma_3) \sigma_2} \; \partial_2 \cdot \partial_3$ && 
\omit && \omit &\cr
\omit&height2pt&\omit&&\omit&&\omit&&\omit&\cr
\tablerule}}

\vskip 0.5cm

{\bf Table~5:} {\ninepoint $(continued \; from \; previous \; page)$}

\vfill\eject

As an example, consider the contraction of the $i=1$ and $j=10$ vertex 
operators:
$$\eqalignno{ &-i \kappa \; 
V_1^{\mu\nu\alpha_1\beta_1\alpha_2\beta_2\alpha_3\beta_3}
(x;\partial_u,\partial_1,\partial_2,\partial_3) \;\;
i\Bigl[ {_{\alpha_1\beta_1}} \Delta_{\rho_1\sigma_1} \Bigr](x;x') \;\;
i\Bigl[ {_{\alpha_2\beta_2}} \Delta_{\rho_2\sigma_2} \Bigr](x;x') \cr
& \qquad
i\Bigl[ {_{\alpha_3\beta_3}} \Delta_{\rho_3\sigma_3} \Bigr](x;x') \;\; 
V_{10}^{\rho_1\sigma_1\rho_2\sigma_2\rho_3\sigma_3}
(x';\partial_1',\partial_2',\partial_3') \cr
&= -\frac{i}{16} \; \kappa^4 \; \Omega^2 \; {\Omega'}^2 \; 
\eta^{\mu\nu} \; \eta^{\alpha_1\beta_1} \; 
\partial_3^{(\alpha_2} \; \eta^{\beta_2) (\alpha_3} \; 
\partial_2^{\beta_3)} \; \eta^{\rho_1\sigma_1} \; 
{\partial_3'}^{(\rho_2} \; \eta^{\sigma_2) (\rho_3} \; 
{\partial_2'}^{ \sigma_3)} \cr
&\qquad 
\Biggl\{ i\Delta_{1N} \Bigl[
2 \eta_{\alpha_1( \rho_1} \; \eta_{\sigma_1) \beta_1} 
- \eta_{\alpha_1 \beta_1} \; \eta_{\rho_1 \sigma_1}
\Bigr] - i\Delta_{1L} \Bigl[
2 {\overline \eta}_{\alpha_1 (\rho_1} \; 
{\overline \eta}_{\sigma_1 \beta_1} 
- \; {\overline \eta}_{\alpha_1 \beta_1} \;
{\overline \eta}_{\rho_1 \sigma_1}
\Bigr] \Biggr\} \qquad \cr
&\qquad 
\Biggl\{ i\Delta_{2N} \Bigl[
2 \eta_{\alpha_2( \rho_2} \; \eta_{\sigma_2) \beta_2} 
- \eta_{\alpha_2 \beta_2} \; \eta_{\rho_2 \sigma_2}
\Bigr] - i\Delta_{2L} \Bigl[
2 {\overline \eta}_{\alpha_2 (\rho_2} \; 
{\overline \eta}_{\sigma_2 \beta_2} 
- 2 {\overline \eta}_{\alpha_2 \beta_2} \;
{\overline \eta}_{\rho_2 \sigma_2}
\Bigr] \Biggr\} \qquad \cr
&\qquad 
\Biggl\{ i\Delta_{3N} \Bigl[
2 \eta_{ \alpha_3( \rho_3} \; \eta_{\sigma_3) \beta_3} 
- \eta_{\alpha_3 \beta_3} \; \eta_{\rho_3 \sigma_3}
\Bigr] - i\Delta_{3L} \Bigl[
2 {\overline \eta}_{\alpha_3 (\rho_3} \; 
{\overline \eta}_{\sigma_3 \beta_3} 
- 2 {\overline \eta}_{\alpha_3 \beta_3} \;
{\overline \eta}_{\rho_3 \sigma_3}
\Bigr] \Biggr\} \quad &(4.5a) \cr
&= - \frac{i}4 \; \kappa^4 \; \Omega^2 \; {\Omega'}^2 \; 
\eta^{\mu\nu} \; \Bigl(
-2 \; i\Delta_{1N} + 3 \; i\Delta_{1L} \Bigr) \cr
&\qquad \Biggl\{\Bigl[
4 \partial_2\cdot\partial_2' \;
\partial_3\cdot\partial_3' 
- 2 \partial_2\cdot\partial_3' \;
\partial_3\cdot\partial_2' 
+ 2 \partial_2\cdot\partial_3 \;
\partial_2'\cdot\partial_3'
\Bigr] \; {i\Delta_{2N}} \; {i \Delta_{3N}} \cr
&\qquad\qquad + \Bigl[
- 2 {\overline \partial}_2\cdot{\overline \partial}_2' \; 
\partial_3\cdot\partial_3' 
+ 3 {\overline \partial}_2\cdot{\overline \partial}_3' \; 
{\overline \partial}_3\cdot{\overline \partial}_2' 
- 3 {\overline \partial}_2\cdot{\overline \partial}_3 \; 
{\overline \partial}_2'\cdot {\overline \partial}_3'
\Bigr] \; {i\Delta_{2N}} \; {i\Delta_{3L}} \cr
&\qquad\qquad + \Bigl[
- 2 \partial_2\cdot\partial_2' \; 
{\overline \partial}_3\cdot{\overline \partial}_3' 
+ 3 {\overline \partial}_2\cdot{\overline \partial}_3' \; 
{\overline \partial}_3\cdot{\overline \partial}_2' 
- 3 {\overline \partial}_2\cdot{\overline \partial}_3 \; 
{\overline \partial}_2'\cdot {\overline \partial}_3'
\Bigr] \; {i\Delta_{2L}} \; {i\Delta_{3N}} \cr
&\qquad\qquad + \Bigl[
{\overline \partial}_2\cdot{\overline\partial}_2' \; 
{\overline \partial}_3\cdot{\overline \partial}_3' 
- 4 {\overline \partial}_2\cdot{\overline \partial}_3' \; 
{\overline \partial}_3\cdot{\overline \partial}_2' 
+ 5 {\overline \partial}_2\cdot{\overline \partial}_3 \; 
{\overline \partial}_2'\cdot {\overline \partial}_3' 
\Bigr] \; {i\Delta_{2L}} \; {i\Delta_{3L}} 
\Biggr\} \quad &(4.5b) \cr
&= {-i \kappa^4 H^2 \over 2^9 \pi^6 u^2 {u'}^2} \; 
\eta^{\mu\nu} \; \Biggl\{ \Bigl[
{\scriptstyle 24} \frac{t\cdot w^4}{w^8} + 
{\scriptstyle 24} \frac{u^2 \; t\cdot w^2}{w^8} - 
{\scriptstyle 336} \frac{u u' \; t\cdot w^2}{w^8} + 
{\scriptstyle 24} \frac{{u'}^2 \; t\cdot w^2}{w^8} - 
{\scriptstyle 48} \frac{u \; t\cdot w}{w^6} - 
{\scriptstyle 24} \frac{u^2}{w^6} \cr
&\qquad\qquad\qquad\qquad + 
{\scriptstyle 48} \frac{u' \; t\cdot w}{w^6} - 
{\scriptstyle 96} \frac{u u'}{w^6} - 
{\scriptstyle 24} \frac{{u'}^2}{w^6} + 
{\scriptstyle 24} \frac1{w^4}
\Bigr] \; \ln(H^2 w^2) \cr 
&\qquad\qquad\qquad\qquad + \Bigl[ -
{\scriptstyle 32} \frac{u u' \; t\cdot w^4}{w^{10}} - 
{\scriptstyle 32} \frac{u^3 u' \; t\cdot w^2}{w^{10}} + 
{\scriptstyle 448} \frac{u^2 {u'}^2 \; t\cdot w^2}{w^{10}} - 
{\scriptstyle 32} \frac{u {u'}^3 \; t\cdot w^2}{w^{10}} + 
{\scriptstyle 64} \frac{u^2 u' \; t\cdot w}{w^8} \cr 
&\qquad\qquad\qquad\qquad + 
{\scriptstyle 32} \frac{u^3 u'}{w^8} - 
{\scriptstyle 64} \frac{u {u'}^2 \; t\cdot w}{w^8} + 
{\scriptstyle 128} \frac{u^2 {u'}^2}{w^8} + 
{\scriptstyle 32} \frac{u{u'}^3}{w^8} - 
{\scriptstyle 32} \frac{u u'}{w^6}
\Bigr] \Biggr\} \;\; . \quad &(4.5c) \cr}$$
Note that we have dropped the order $\epsilon$ terms in derivatives
of $w^2$:
$${\partial \over \partial x^{\mu}} \; w^2 
\; = \;
2 \; w_{\mu} - 2 i \; \epsilon \; t_{\mu} 
\; \longrightarrow \; 
2 \; w_{\mu} \;\; . \eqno(4.6)$$
The neglected terms affect ultraviolet divergences but not the 
infrared terms of interest.

The next step is to perform the spatial integrations. The angular 
integrations give a factor of $4\pi$ and the radial integral was 
done by the method of contours. When no logarithm is present we 
obtain:
$$\eqalignno{ \int d^3x' \; {1 \over w^{2N}} & = 
2 \pi \int_{-\infty}^{\infty} dr \; 
{r^2 \over (r-w_0+i\epsilon)^N \; (r+w_0-i\epsilon)^N} \cr
&= 8\pi^2 i \; {(-1)^{N-1} \; (2N-5)!! \over 2^{N+1} \; (N-1)!} \;
\Bigl( {1 \over w_0 - i\epsilon} \Bigr)^{2N-3} \;\; . &(4.7a) \cr}$$
The logarithm gives:
$$\eqalignno{ \int& d^3x' \; {\ln (H^2 w^2) \over w^{2N}} = 
2\pi^2 \int_{-\infty}^{\infty} dr \; 
{r^2 \; \ln \Bigl[ H^2 (r-w_0+i\epsilon) \; (r+w_0-i\epsilon) \Bigr] 
\over (r-w_0+i\epsilon)^N \; (r+w_0-i\epsilon)^N} &(4.7b) \cr
&= 8\pi^2 i \; {(-1)^{N-1} \; (2N-5)!! \over 2^{N+1} \; (N-1)!} \; 
\Bigl\{ \ln \Bigl[ 2H (w_0 - i\epsilon) \Bigr] + 
\ln \Bigl[ -2H (w_0 - i\epsilon) \Bigr] \Bigr\} \; 
\Bigl( {1 \over w_0 - i\epsilon} \Bigr)^{2N-3} \cr
&\quad + \; {8\pi^2 i \; (-1)^N \over 2^{2N-1} \; (N-1)!} \; 
\sum_{k=1}^{N-1} {(2N-2-k)! \; [2N - 2 - k - k^2] \over 
k \; (N-1-k)! \; (2N - 2 - k) (2N - 3 - k)} \; 
\Bigl( {1 \over w_0 - i \epsilon} \Bigr)^{2N-3} 
\;\; . \cr}$$
The case of $N = 2$ has to be treated specially:
$$\int d^3x' \; {1 \over w^4} = -8 \pi^2 i \; {1 \over 8} \; 
{1 \over w_0 - i\epsilon} \;\; , \eqno(4.8a)$$
$$\int d^3x' \; {\ln (H^2 w^2) \over w^4} = 
-8 \pi^2 i \; {1 \over 8} \Bigr( \; 
\frac{\ln[ 2H(w_0 - i\epsilon)] \; + \; \ln[ -2H(w_0 - i\epsilon)]}
{w_0 - i\epsilon} + \frac1{w_0 - i\epsilon} 
\; \Bigr) \;\; . \eqno(4.8b)$$ 
Performing the spatial integrations for our $i=1$, $j=10$ vertex 
pair gives:
$$\eqalignno{ &-i \kappa \int d^3x' \; 
V_1^{\mu\nu\alpha_1\beta_1\alpha_2\beta_2\alpha_3\beta_3}
(x;\partial_u,\partial_1,\partial_2,\partial_3) \;\;
i\Bigl[ {_{\alpha_1\beta_1}} \Delta_{\rho_1\sigma_1} \Bigr](x;x') \;\;
i\Bigl[ {_{\alpha_2\beta_2}} \Delta_{\rho_2\sigma_2} \Bigr](x;x') \cr
& \qquad\qquad\qquad
i\Bigl[ {_{\alpha_3\beta_3}} \Delta_{\rho_3\sigma_3} \Bigr](x;x') \;\; 
V_{10}^{\rho_1\sigma_1\rho_2\sigma_2\rho_3\sigma_3}
(x';\partial_1',\partial_2',\partial_3') \cr
&= {\kappa^4 H^2 \over 2^6 \pi^4 u^2 {u'}^2} \; \eta^{\mu\nu} \; 
\Biggl\{ - {\scriptstyle 3} 
\frac{\ln[ 2H(u'-u-i\epsilon)] \; + \; \ln[ -2H(u'-u-i\epsilon)]}
{u'-u-i\epsilon} 
- \frac72 \frac1{u'-u-i\epsilon} \cr 
&\qquad\qquad\qquad\qquad\;\; 
+ \frac{15}8 \frac{u' \; u}{(u'-u- i \epsilon)^3} 
- \frac7{16} \frac{{u'}^2 \; u^2}{(u'-u-i \epsilon)^5}
\Biggr\} \;\; . &(4.9) \cr}$$

The final step is to integrate over the conformal time of the free 
vertex. The integrand always consists of a numerator --- which may 
contain a pair of logarithms --- and a denominator --- which contains 
up to two powers of $u'$, and up to seven powers of $u'-u-i\epsilon$. 
Decomposing the denominator by partial fractions results in four 
distinct terms requiring special attention: $(u'-u-i\epsilon)$, $u'$, 
${u'}^2$ and $(u'-u-i\epsilon)^k$ for $k > 1$. When the pair of 
logarithms is present the four denominators give:
$$\eqalignno{ &{\rm (i)} \;\;
\int_u^{H^{-1}} \frac{du'}{u'-u-i\epsilon} \; 
\Bigl\{ \ln \Bigl[ 2H(u'-u-i\epsilon) \Bigr] + 
\ln \Bigl[ -2H(u'-u-i \epsilon) \Bigr] \Bigr\} &(4.10) \cr
&= \;
\frac12 \Bigl\{ \ln^2 \Bigl[ 2H (u'-u-i\epsilon) \Bigr] + 
\ln^2 \Bigl[ -2H (u'-u-i\epsilon \Bigr]
\Bigr\} \Biggl\vert_u^{H^{-1}} \cr
&\longrightarrow \; 
\ln^2 \Bigl[ 2(1-Hu) \Bigr] - 
\ln^2 (2H\epsilon) + 
i \pi \ln \Bigl[ 2(1-Hu) \Bigr] -
\frac14 \pi^2 \;\; , \cr}$$
$$\eqalignno{ &{\rm (ii)} \;\;
\int_u^{H^{-1}} \frac{du'}{u'} \; 
\Bigl\{ \ln \Bigl[ 2H (u'-u-i\epsilon) \Bigr] + 
\ln \Bigl[ -2H(u'-u-i \epsilon) \Bigr] \Bigr\} &(4.11) \cr
&\longrightarrow \; 
\Bigl\{ \ln^2(2 H u') + i \pi \ln(H u') + 
2 \sum_{k=1}^{\infty} \frac1{k^2} \Bigl(\frac{u}{u'}\Bigr)^k
\Bigr\} \Biggr\vert_u^{H^{-1}} \cr
&= \;
-2 \ln(2) \ln(Hu) - \ln^2(Hu) - i \pi \ln(Hu) + 
2 \Bigl[ H u \Phi(Hu,2,1) - \zeta(2) \Bigr] 
\;\; , \cr}$$
$$\eqalignno{ &{\rm (iii)} \;\;
\int_u^{H^{-1}} \frac{du'}{{u'}^2} \; 
\Bigl\{ \ln \Bigl[ 2H(u'-u-i\epsilon) \Bigr] +  
\ln \Bigl[ -2H(u'-u-i \epsilon) \Bigr] \Bigr\} &(4.12) \cr
&\longrightarrow \;
\Bigl\{ -\frac2{u'} \ln(2 H u') -\frac{i\pi}{u'} + 
2 \frac{u'-u}{u' u} \; \ln \Bigl[ 1- \frac{u}{u'} \Bigr]
\Bigr\} \Biggr\vert_u^{H^{-1}} \cr
&= \; 
\frac2{u} \ln(Hu) + \frac2{u} (1-Hu) 
\Bigl\{ \frac{i}2 \pi + 
\ln\Bigl[2 (1-Hu)\Bigr]
\Bigr\} \;\; , \cr}$$
$$\eqalignno{ &{\rm (iv)} \;\;
\int_u^{H^{-1}} \frac{du'}{(u'-u-i\epsilon)^k} \; 
\Bigl\{ \ln \Bigl[ 2H(u'-u-i\epsilon) \Bigr] + 
\ln \Bigl[ -2H(u'-u-i \epsilon) \Bigr] \Bigr\} &(4.13) \cr
&= \;
- \frac1{k-1} \frac1{(u'-u-i\epsilon)^{k-1}} \Bigl\{ 
\ln \Bigl[ 2 H (u'-u-i\epsilon) \Bigr] + 
\ln \Bigl[ -2H (u'-u-i\epsilon) \Bigr] +
\frac2{k-1}
\Bigr\} \Biggr\vert_u^{H^{-1}} \cr
&\longrightarrow \;
\frac2{k-1} \Bigl( \frac{i}{\epsilon} \Bigr)^{k-1} \Bigl[ 
\ln (2H\epsilon) + \frac1{k-1} \Bigr] 
- \frac2{k-1} \Bigl( \frac{H}{1-Hu}\Bigr)^{k-1} \Bigl\{ 
\ln \Bigl[ 2(1-Hu) \Bigr] + \frac{i}2 \pi + \frac1{k-1} 
\Bigr\} \;\; , \cr}$$
where $\Phi(z,s,v)$ is the special function [14]:
$$\Phi(z,s,v) \equiv \sum_{n=0}^{\infty} \;
{z^n \over (v + n)^s} \;\; . \eqno(4.14)$$
and the arrow indicates the leading infinitesimal $\epsilon$ 
contributions. When no logarithms are present the integrals are 
simple enough that we give only the small $\epsilon$ forms:
$${\rm (i)} \;\;
\int_u^{H^{-1}} {du' \over u' - u -i\epsilon} 
\; \longrightarrow \;
\ln(1-Hu) - \ln(H\epsilon) + \frac{i}2 \pi 
\;\; , \eqno(4.15)$$
$${\rm (ii)} \;\; 
\int_u^{H^{-1}} {du' \over u'} \; = \; - \ln(Hu) 
\;\; , \eqno(4.16)$$
$${\rm (iii)} \;\;
\int_u^{H^{-1}} {du' \over {u'}^2} \; = \; \frac1{u} (1-Hu) 
\;\; , \eqno(4.17)$$
$${\rm (iv)} \;\;
\int_u^{H^{-1}} {du' \over (u'-u-i\epsilon)^k} 
\; \longrightarrow \; 
\frac1{k-1} \; \Bigl\{ 
\Bigl( \frac{i}{\epsilon} \Bigr)^{k-1} - 
\Bigl( \frac{H}{1-Hu} \Bigr)^{k-1}
\Bigr\} \;\; . \eqno(4.18)$$

We can now complete the reduction of the contribution from the 
$i=1$, $j=10$ vertex pair:
$$\eqalignno{ {\cal T}_{1,10}^{\mu \nu} & = 
{\kappa^4 H^2 \over 2^6 \pi^4} \; \eta^{\mu\nu} \; \Biggl\{ \; 
\Bigl\{ -6 \ln^2(Hu) + \Bigl[ \frac{35}4 - 12 \ln2 \Bigr] \ln(Hu) + 
19 (1 -Hu) + \frac32 \pi^2 \cr
&\qquad\qquad\qquad + 12 \Bigl[ \zeta(2) - Hu \Phi(Hu,2,1) \Bigr] 
- \frac{13}4 \ln \Bigl[ 2(1-Hu) \Bigr] - 6 \ln^2 \Bigl[ 2(1-Hu) \Bigr] \cr 
&\qquad\qquad\qquad + \frac{13}4 \ln (2 H \epsilon) + 
6 \ln^2 (2 H \epsilon) \Bigr\} {1 \over u^4} \; + \; 
\frac{15}4 \Bigl( \frac{H}{1-Hu} \Bigr) {1 \over u^3} \cr 
&\qquad\qquad\qquad - \frac{15}8 \Bigl[ \Bigl( \frac{H}{1-Hu} \Bigr)^2 + 
\frac1{\epsilon^2} \Bigr] {1 \over u^2} \; + \; 
\frac38 \Bigl[ -\Bigl( \frac{H}{1-Hu} \Bigr)^4 + \frac1{\epsilon^4} 
\Bigr] \; \Biggr\} \;\; . &(4.19) \cr}$$
It is worth noting that the coefficient of the leading term for late times
--- $-6 \; u^{-4} \ln^2(H u)$ --- is opposite to that of the double logarithmic
ultraviolet divergence --- $+ 6 \; u^{-4} \ln^2(2 H \epsilon)$. This is in
sharp contrast to 3-3-3 contributions such as (3.47a) where an undifferentiated
propagator logarithm is not involved. Then the leading infrared contribution
has {\it the same} sign as the double logarithmic ultraviolet divergence. 

The same sign phenomenon of (3.47a) was explained in sub-section 2.7. The 
parameter $\epsilon$ has dimensions of length so it can only appear in the
combinations $H \epsilon$ or $\epsilon u^{-1}$. The dimensionality of the 
conformal time integrands shows that they converge even if the upper limit, 
$H^{-1}$, diverges, so the only factors of $\ln(H)$ can come from the single 
possible undifferentiated propagator logarithm. When no such logarithm is 
present any double logarithmic ultraviolet divergence must take the form:
$$\ln^2(\epsilon u^{-1}) = \ln^2(H \epsilon) - 2 \; \ln(H \epsilon) \; 
\ln(H u) + \ln^2(H u) \;\; . \eqno(4.20)$$
One can understand why the opposite relation prevails in all 4-3 contributions
by the need to avoid overlapping divergences. The 4-3 diagram contains only
two vertices so its ultraviolet divergences must be primitive; there cannot be
any non-local sub-divergences such as $\ln(H \epsilon) \; \ln(H u)$. The 
possibility of $\ln^2(H \epsilon)$ is ruled out because no factors of $\ln(H)$
can come from the limits of integration; the only $\ln(H)$ can come from the
single possible undifferentiated propagator logarithm. So we are left with
just one dimensionally consistent double ultraviolet logarithm:
$$\ln^2(H \epsilon) - \ln^2(H u) = \ln^2(\epsilon) + 2 \; \ln(\epsilon) \;
\ln(H) - 2 \; \ln(H) \; \ln(u) - \ln^2(u) \;\; . \eqno(4.21)$$

Although the tensor algebra and derivatives of the 4-3 diagram are only 
marginally simpler than that of the most complicated 3-3-3 diagram (2a), the
number of terms they produce is much smaller. There are just 39 integrands
which contain $\ln(H^2 w^2)$ and only 57 which have no logarithm. We have
also seen that the integrals are simple enough for exact results to be 
obtained. Although we did this, we report just the leading results here:
$$a_{43}(u) \; = \; {\kappa^4 H^2 \over 2^6 \pi^4} \; \Biggl\{-{13 \over 3}
\; {\ln^2(H u) \over u^4} + O\Bigl({\ln(H u) \over u^4}\Bigr) \Biggr\} 
\;\; ,\eqno(4.22a)$$
$$c_{43}(u) \; = \; {\kappa^4 H^2 \over 2^6 \pi^4} \; \Biggl\{+ 2
\; {\ln^2(H u) \over u^4} + O\Bigl({\ln(H u) \over u^4}\Bigr) \Biggr\} 
\;\; . \eqno(4.22b)$$

\vskip 1cm
\centerline{\bf 5. Epilogue}

We can now assemble the results from the various diagrams of Fig.~2. 
Combining (3.51) with (4.22) gives the following coefficient functions for
the amputated 1-point function (2.19):
$$a(u) = H^{-2} \; \Bigl({\kappa H \over 4 \pi u}\Bigr)^4 \; \Biggl\{
\Bigl( -{1795 \over 9} + {604 \over 9} + {320 \over 3} - {52 \over 3} \Bigr)
\; \ln^2(H u) + O\Bigl(\ln(H u) \Bigr) \; \Biggr\} + O(\kappa^6) 
\eqno(5.1a)$$
$$c(u) = H^{-2} \; \Bigl({\kappa H \over 4 \pi u}\Bigr)^4 \; \Biggl\{
\Bigl( +{1157 \over 3} - {800 \over 3} - 112 + 8 \Bigr) 
\; \ln^2(H u) + O\Bigl(\ln(H u) \Bigr) \; \Biggr\} + O(\kappa^6) 
\eqno(5.1b)$$
The four numbers in each of these expressions represent the respective 
contributions from diagrams (2a), (2b), (2c) and (2d). As explained in 
sub-section 2.9, diagram (2e) can only contribute a single factor of 
$\ln(H u)$ and diagram (2f) cannot contribute any.

Attaching the retarded Green's functions discussed in sub-section 2.5 gives
the coefficient functions for the full 1-point function (2.20):
$$A(u) = \Bigl({\kappa H \over 4 \pi u}\Bigr)^4 \; 
\Biggl\{ \Bigl( +{7180 \over 81} - {2416 \over 81} - 
{1280 \over 27} + {208 \over 27} \Bigr) \; \ln^3(H u) 
+ O\Bigl(\ln^2(H u) \Bigr) \; \Biggr\} + O(\kappa^6) 
\eqno(5.2a)$$
$$C(u) = \Bigl({\kappa H \over 4 \pi u}\Bigr)^4 \; 
\Biggl\{ \Bigl( +{319 \over 6} + {49 \over 3} - 52 + 11 \Bigr) \; 
\ln^2(H u) + O\Bigl(\ln(H u) \Bigr) \; \Biggr\} + O(\kappa^6) 
\;\; . \qquad \eqno(5.2b)$$
None of the four coefficient functions given above is free of gauge 
dependence. The physical, gauge independent observable is the combination
(2.24) which gives the effective Hubble constant. Our result for it is:
$$\eqalignno{H_{\rm eff}(t) &= 
H \; \Biggl\{1 - \Bigl({\kappa H \over 4 \pi}\Bigr)^4 \; \Biggl[ \; 
\Bigl(+ {4309 \over 54} - {1649 \over 27} - {172 \over 9} + 
{5 \over 9} \Bigr) \; (H t)^2 + O\Bigl( H t\Bigr) \; \Biggr] \cr
&\qquad\quad + O(\kappa^6)\Biggr\} &(5.3a) \cr
&= H \; \Biggl\{ 1 - \Bigl( {\kappa H \over 4 \pi}\Bigr)^4 \; 
\Biggl[ \; {1 \over 6} \; (H t)^2 + O\Bigl( H t\Bigr) \; \Biggr] + 
O(\kappa^6)\Biggr\} 
\;\; . &(5.3b) \cr}$$
Although we have given the contribution from each diagram in (5.3a), only
the sum (5.3b) is really physical and gauge independent. Still, it is worth
pointing out that the pure graviton diagrams --- (2a) and (2d) --- act to slow
inflation while the ghost diagrams --- (2b) and (2c) --- contribute in the 
opposite sense. 

The physical significance of our result has been discussed elsewhere [4]
but two points should be mentioned here. First, we get a {\it reduction}
in the effective Hubble constant due to the negative energy of the 
gravitational interaction between the zero point motions of 
gravitons.\footnote{*}{\tenpoint The induced stress-energy is that of 
negative vacuum energy, to leading order, because causality restricts the
gravitational interaction to the constant Hubble volume. The induced 
energy density is therefore independent of the true volume of the inflating
universe, so the total energy is just $E = V \rho$ and the pressure is
$p = -{\partial E}/{\partial V} = - \rho$.} This tendency ought to persist 
to all orders in perturbation theory. Second, recall from sub-section 2.8 
that at most $\ell$ infrared logarithms can appear at $\ell$ loops. This 
means that the $\ell$-loop contribution to the bracketed term in (5.3b) is 
at most:
$$- \# \; (\kappa H)^{2 \ell} \; (H t)^{\ell} \;\; . \eqno(5.4)$$
The bound is quite likely to be saturated because one can form higher loop
graphs by attaching the 2-loop tadpole we have computed. But one in any case
obtains the following estimate for the number of e-foldings needed to end 
rapid inflation:
$$N = H t \sim (\kappa H)^{-2} \;\; . \eqno(5.5)$$
This is $\gtwid 10^{12}$ for inflation at the GUT scale or below, which is 
more than enough to explain the homogeneity and isotropy of the observed 
universe. If the unknown numerical coefficients in (5.4) fall off less rapidly
than $\ell^{-1}$ then the end of rapid inflation is likely to be quite abrupt 
because all orders will become strong at once. This augurs well for reheating.

We turn now to a discussion of accuracy. The strongest checks on the 
consistency of our basic formalism and the accuracy of our implementation
are provided by the one-loop self-energy. This quantity was relatively simple
to obtain [15] because technical considerations compelled us to compute the 
inner loops of diagrams (2a) and (2b) before doing the outer loop contractions 
and acting the outer loop derivatives. We checked that the one-loop self-energy
has the appropriate reflection symmetry, that it obeys the Ward identity, and 
that it agrees, for $H \rightarrow 0$ at fixed $t$, with the flat space 
self-energy obtained by Capper [16] in the same gauge. Since we used only 
partially symmetrized vertices and then interchanged lines where necessary, 
reflection symmetry is a non-trivial test of our programs for the tensor 
algebra. It also checks the programs for taking derivatives because the order
of differentiation breaks manifest reflection symmetry. The Ward identity 
tests the apparatus of gauge fixing, our solutions for the ghost and graviton 
propagators, and the 3-point interaction vertices. It also provides a powerful 
independent check of the tensor algebra and derivative programs. In addition 
to giving further independent tests for all these things, the flat space limit 
checks the overall factor and the sign. 

One of the major complications in computing diagram (2a) was that the number 
of intermediate expressions becomes prohibitive if one attempts to perform 
the tensor algebra all at once, even for a single ``trio'' of vertex 
operators.\footnote{*}{\tenpoint The problem is that expanding the four 
propagators gives $6^4 = 1296$ terms, each of which involves a contraction 
over 16 indices.} This is why we first did the inner loop tensor algebra and 
acted the inner loop derivatives for each pair of inner loop vertices, then 
summed the results and projected the lengthy total onto a basis formed from 
products of the 79 independent 4-index objects and the 10 possible contracted 
outer leg derivatives. Each basis element with a non-zero coefficient was 
contracted into the outer loop propagators and the outer vertex operators, 
and the various outer loop derivatives were acted. Then the inner loop 
coefficients were multiplied and the results summed. 

The cumbersome nature of this procedure caused us much anxiety and we devised 
a number of ways to check it. First, a program was written to check the inner 
loop projection by simply summing the product of each basis element with its 
coefficient and then comparing with the original expression. Second, the 
outer loop tensor algebra and derivative programs were based on the same 
scheme as those which were so effectively checked by the Ward identity, and 
of course we used the same stored expressions for the vertices and 
propagators. Third, it was not too time consuming to perform all the tensor 
algebra at once for a single ``trio'' of vertex operators. This was done 
for the case where the vertex operator at $x^{\mu}$ is \#10 in Table~3, 
and where those at ${x'}^{\mu}$ and ${x''}^{\mu}$ are both \#41. The 
derivatives were then acted (using a different program from the usual one) 
and the result compared with what our standard programs give. The same 
calculation was performed by hand as an additional check.

Diagram (2b) was computed using the same programs as for (2a), so its accuracy
is checked by that of (2a). An additional test was provided by the fact that 
no undifferentiated propagator logarithms can come from ghost 
lines.\footnote{*}{\tenpoint To see this note that the 10 vertices of Table~2 
have either a a factor of $\partial_2$ or a factor of $t^{\alpha_2}$ --- which 
accesses only the normal part of the propagator.} We checked that the 
integrands for (2b) are indeed free of $\ln(H^2 y^2)$, and that those of 
diagram (2c) are free of either $\ln(H^2 w^2)$ or $\ln(H^2 z^2)$.

Our greatest worry was the complicated integrations of 3-3-3 diagrams. We 
checked these a number of ways. First, we did a large number of examples by 
hand. Second, the two authors wrote independent integration schemes. The 
first of these did each radial integration independently by the method of 
contours and changed variables as necessary so that only the final conformal 
time integration needed to be expanded. This program also computed the four 
``$\pm$'' variations separately. The second program did the radial integrals 
by differentiating generating functions. The conformal time integrations were 
carried out in the standard order, expanding whenever necessary. This second 
program also added the ``$\pm$'' variations of the integrands before 
evaluating the integral. We ran the two programs for each of the thousands of 
terms and then scanned the results to make sure they agreed.

One of the strongest checks on the accuracy of the 3-3-3 diagrams derives 
from our proof in sub-section 2.8 that each ``trio'' of vertices for each 
of the 3-3-3 diagrams must be free of triple log terms --- $u^{-4} \; 
\ln^3(H u)$. As mentioned there, and as we saw explicitly in (3.49), this
is {\it not} true for each of the integrands which emerge from step 3 of our
reduction procedure. Yet when the results from integrating the thousands
of distinct integrands were summed, the triple logs cancelled for each 
diagram. We also checked that our programs show this cancellation separately
for the previously mentioned 10-41-41 vertex triad.

We fretted long about how to check the 4-3 diagram (2d). Of course it is 
partially checked by the 3-3-3 diagrams by virtue of the fact that all use 
the same stored expression for the graviton propagator, and since their
tensor algebra and derivative action programs are based on the same scheme. 
One of our big worries was the different vertices. We checked the fully
symmetrized 3-point vertex by having the computer symmetrize the partially 
symmetrized vertex of Table~3. We also wrote a program in which the computer 
produces fully symmetrized 3-point and 4-point vertex operators from the 
respective interaction Lagrangians (3.6) and (4.1). These were compared 
directly with our stored expression for the symmetrized 3-point vertex; of 
course we had to (computer) symmetrize our 4-point vertex before comparing it.

As an additional check on the 4-3 diagram, we compared the output from our
programs with a hand computation of the result for vertex operator \#1 at 
$x^{\mu}$ and \#10 at ${x'}^{\mu}$. It is also worth pointing out that the
4-3 integrations are vastly simpler than those of the 3-3-3 diagrams. In
fact only the single term (4.11) can contribute at leading order.

Finally, there is the fact that the result is plausible. Quantum gravity is 
not on-shell finite at one loop when the cosmological constant is non-zero
[17], so one has to expect ultraviolet divergences of the form $\ln^2(
\epsilon)$ at two loops. Since $\epsilon$ has dimensions of length, logarithms 
of it must come in the form $\ln(\epsilon u^{-1})$ and $\ln(H \epsilon)$. We 
showed in sub-section 2.7 that only a single factor of $\ln(H)$ can occur, so
there must be at least one $\ln(u)$, and we can think of no reason why two
should not occur. The reason why these infrared logarithms act to slow 
inflation is that they represent the negative gravitational interaction
energy between the zero point motions of gravitons. At the risk of putting
too much faith in gauge-dependent results one can even understand that pure 
graviton diagrams should act to slow inflation, and that the ghost diagrams
should diminish this effect.\footnote{*}{\tenpoint The near cancellation 
between the two classes is perhaps explicable from the fact that the gauge 
fixed graviton field carries eight unphysical modes and only two physical
ones. At two loops this means roughly $10^2 = 100$ mode pairs of which the 
ghost loops must remove all but $2^2 = 4$.}

There is also a good physical interpretation for infrared logarithms. The
up to two powers of $\ln(H u)$ that derive from integrating factors of 
$u'^{-1}$ and $u''^{-1}$ represent the invariant volume of the past lightcone 
as viewed from the observation point [2]. The single $\ln(H u)$ which can come
from an undifferentiated propagator logarithm represents the ever-increasing 
correlation of the free graviton vacuum in an inflating universe. It has long 
been known that the assumption of correlated de Sitter vacuum over an infinite 
surface of simultaneity leads to a divergence in the propagator. (This was
first proved by Allen and Folacci [18]. See also [7] and the references cited 
therein.) Infrared logarithms are the causal manifestation of this effect 
when one approaches an infinite surface of simultaneity by inflating a finite
patch of correlated de Sitter vacuum.

In sum, the procedure we used should work, it gives a reasonable result, and
every care has been taken to ensure accuracy. However, we do not wish this 
discussion to convey a false sense of infallibility. We have been schooled in 
humility by the disconcerting experience of detecting errors even after the 
final one seemed to have been expunged. It must be stressed that this was an 
enormously complicated piece of work. Only one other two-loop result has ever 
been obtained in quantum gravity [19], and it was confined to the ultraviolet 
divergent part of the on-shell effective action for zero cosmological 
constant. We feel very strongly that computer calculations on this scale 
should be regarded as experiments whose results require independent 
verification before they can be completely trusted. We would be happy to 
cooperate with anyone wishing to undertake even a partial check.

We conclude by briefly discussing a perturbative issue which demands further 
study: the case of a negative cosmological constant. This is interesting in
its own right and because it may have relevance to the period after rapid
inflation has ended, when an energetically favored phase transition would
be expected to generate a negative effective cosmological constant. Two
qualitative questions are of great importance: are there strong infrared
effects from quantum gravity? and, do they tend to resist the contraction of 
spacetime?

Careful consideration of the effect for positive cosmological constant leads
to the conclusion that it derives from the combination of three features:
\item{(1)} Propagators which do not oscillate or fall off over large temporal
separations;
\item{(2)} An interaction of dimension three;
\item{(3)} The fact that the invariant volume of the past lightcone increases
without bound.

\noindent The last two are certainly true as well for a negative cosmological
constant. In fact the causal structure of anti-de Sitter space allows one to
access spatial infinity after only a finite amount of time. The non-trivial
issue is the propagator. The usual prescription for defining what happens at
spatial infinity is to impose reflective boundary conditions [20]. These are
enforced mathematically by a negative image source at the antipodal point, so
they make propagators fall off too rapidly to give a big infrared effect.

We believe that reflective boundary conditions are reasonable for Euclidean
anti-de Sitter space because its antipodal points are not part of the 
manifold. However, the antipodal points of the Minkowski signature formalism 
are on the manifold. In this case the use of reflective boundary conditions 
implies the physical absurdity that every source of stress energy has an 
antipodal anti-source. It is not reasonable to suppose a man is forbidden to
shake his fist (which generates gravitational radiation) without the 
cooperation of someone of the other side of the Universe. 

The more sensible boundary conditions seem to be transmissive, in which 
information is allowed to flow to spatial infinity without hindrance. These
do not lead to a globally well-posed initial value problem, but they do make 
physical sense locally. The propagator associated with this condition does
not fall off rapidly enough to preclude a large infrared effect. It should
also be noted that arguments in the literature about the stability of 
supergravity or superstrings on an anti de Sitter background are based on 
reflective boundary conditions.\footnote{*}{\tenpoint We are indebted to I.
Antoniadis for bringing this point to our attention.}

It is very much more difficult to determine whether quantum gravity makes a
negative $\Lambda$ universe collapse slower or faster. Local considerations 
compel us to the view that gravitational interaction energy should still be 
negative. However, the fact that the sign of the dimension three coupling 
changes makes it hard to say what this does. (Note that the 1-point function 
is odd in the 3-point vertex.) An additional complication is that we can no 
longer count on the induced stress tensor to be that of pure vacuum energy. 
That it came out this way for the case of positive cosmological constant 
derives from the finite causal horizon of de Sitter space, which means that 
the gravitational interaction energy density must be independent of the true 
volume. This is not true for anti-de Sitter space. And it is worth recalling 
that inflation is driven by the negative pressure of a positive cosmological
constant; the positive energy density serves as a drag on expansion. So it
is conceivable that negative $\Lambda$ quantum gravity induces a negative 
energy density, while still resisting contraction by virtue of generating
less than an equal and opposite pressure.

\vskip 1cm
\centerline{ACKNOWLEDGEMENTS}

We wish to give special thanks to T. J. M. Zouros for the use of his SUN SPARC 
station, and to S. Deser for his encouragement and support during this project.
We have profited from conversations with I. Antoniadis, C. Bachas and
J. Iliopoulos. One of us (RPW) thanks the University of Crete and the Theory 
Group of FO.R.T.H. for their hospitality during the execution of this project. 
This work was partially supported by DOE contract 86-ER40272, by NSF grant 
94092715 and by EEC grant 933582.

\vfill\eject

\centerline{REFERENCES}

\item{[1]} N. C. Tsamis and R. P. Woodard, {\sl Phys. Lett.} {\bf B1993} (1993)
351.

\item{[2]} N. C. Tsamis and R. P. Woodard, {\sl Ann. Phys.} {\bf 238} (1995) 1. 

\item{[3]} S. Deser and L. F. Abbott, {\sl Nucl. Phys.} {\bf B195} (1982) 
76.\hfill\break
P. Ginsparg and M. J. Perry, {\sl Nucl. Phys.} {\bf B222} (1983) 245.

\item{[4]} N. C. Tsamis and R. P. Woodard, {\sl Nucl. Phys.} {\bf B474} (1996)
235.

\item{[5]} N. C. Tsamis and R. P. Woodard, {\sl Commun. Math. Phys.} {\bf 162} 
(1994) 217.

\item{[6]} N. C. Tsamis and R. P. Woodard, {\sl Phys. Lett.} {\bf B269} (1992)
269.

\item{[7]} N. C. Tsamis and R. P. Woodard, {\sl Class. Quantum Grav.} {\bf 11} 
(1994) 2969

\item{[8]} J. Schwinger, {\sl J. Math. Phys.} {\bf 2} (1961) 407; {\it 
Particles, Sources and Fields} (Addison-Wesley, Reading, MA, 1970).

\item{[9]} L. H. Ford, {\sl Phys. Rev.} {\bf D31} (1985) 710.

\item{[10]} A. D. Dolgov, M. B. Einhorn and V. I. Zakharov, {\sl Phys. Rev.} 
{\bf D52} (1995) 717.

\item{[11]} B. S. DeWitt, {\sl Phys. Rev.} {\bf 162} (1967) 1239.
\hfill\break
F. A. Berends and R. Gastmans, {\sl Nucl. Phys.} {\bf B88} (1975) 99.
 
\item{[12]} S. Wolfram, {\it Mathematica, 2nd edition} (Addison-Wesley, 
Redwood City, CA, 1991). 

\item{[13]} R. Mertig, {\it Guide to FeynCalc 1.0}, University of 
W\"urzburg preprint, March 1992.

\item{[14]} I. S. Gradshteyn and I. M. Ryzhik, {\it Table of Integrals,
Series, and Products, 4th edition} (Academic Press, New York, 1965)
pp. 1075 -- 1076.

\item{[15]} N. C. Tsamis and R. P. Woodard, {\sl Phys. Rev.} {\bf D15} (1996)
2621.

\item{[16]} D. M. Capper, {\sl J. Phys.} {\bf A13} (1980) 199.

\item{[17]} S. Deser and P. van Nieuwenhuizen, {\sl Phys. Rev.} {\bf D10}
(1974) 401.

\item{[18]} B. Allen and A. Folacci, {\sl J. Math, Phys.} {\bf 32} (1991)
2828.

\item{[19]} M. Goroff and A. Sagnotti, {\sl Phys. Lett.} {\bf B160} (1986) 
81; {\sl Nucl. Phys.} {\bf B266} (1986) 709.

\item{[20]} S. J. Avis, C. J. Isham and D. Storey, {\sl Phys. Rev.} {\bf D18}
(1978) 3565.

\bye